%% file: main.tex
\documentclass[lettersize,journal]{IEEEtran}

\input{preamble}
\input{preamble1}
\input{preamble_xr}

\usepackage{titling}

\begin{document}

\title{Compressive Recovery of Signals Defined on Perturbed Graphs}

\author{
\IEEEauthorblockN
{
Sabyasachi Ghosh,
Ajit Rajwade} \\
\IEEEauthorblockA{
\{ssgosh,ajitvr.cse.iitb\}@gmail.com\\
Dept. of Computer Science and Engineering,\\
Indian Institute of Technology Bombay, Mumbai, India.}
}
\date{}

\maketitle

\begin{abstract}
Recovery of signals with elements defined on the nodes of a graph, from compressive measurements is an important problem, which can arise in various domains such as sensor networks, image reconstruction and group testing.
In some scenarios, the graph may not be accurately known, and there may exist a few edge additions or deletions relative to a ground truth graph. Such perturbations, even if small in number, significantly affect the Graph Fourier Transform (GFT). This impedes recovery of signals which may have sparse representations in the GFT bases of the ground truth graph. We present an algorithm which simultaneously recovers the signal from the compressive measurements and also corrects the graph perturbations. We analyze some important theoretical properties of the algorithm. Our approach to correction for graph perturbations is based on model selection techniques such as cross-validation in compressed sensing. We validate our algorithm on signals which have a sparse representation in the GFT bases of many commonly used graphs in the network science literature. An application to compressive image reconstruction is also presented, where graph perturbations are modeled as undesirable graph edges linking pixels with significant intensity difference. In all experiments, our algorithm clearly outperforms baseline techniques which either ignore the perturbations or use first order approximations to the perturbations in the GFT bases.
\end{abstract}



\section{Introduction}
\label{sec:introduction}

Efficient acquisition of data arranged on the nodes of a graph (i.e.\ a graph signal) may be performed via the technique of Compressed Sensing (CS) \cite{Candes2008a}. Instead of recording the data at each node of the graph separately, a small number of random linear measurements of the graph signal may be acquired in a domain-specific manner. CS techniques allow accurate recovery of a high-dimensional vector from a small number of linear measurements, given sufficient conditions such as sparsity/compressibility of the signal representation in a known orthonormal basis, and measurement matrix characteristics such as the restricted isometry property (RIP) \cite{Candes2008a}.
In some applications, graph signals defined on the nodes of an undirected, unweighted graph, possess a sparse representation in terms of the eigenvectors of the Laplacian matrix of the graph, which are also known as a Graph Fourier Transform (GFT) basis for such graphs in the Graph Signal Processing (GSP) literature \cite{Ortega2018}. Given the sparsity of this representation, such graph signals may be recovered from the compressive linear measurements using CS decoding algorithms. Compressive graph signal measurements naturally arise in sensor networks \cite{zhu2012graph}, group testing with side information such as contact tracing graphs \cite{goenka2021contact}, or in image acquisition if the image is regarded as a graph.  

In this work, we consider the case when a graph signal has been compressively acquired, but there is some uncertainty in the knowledge of the graph on which the signal was defined. A \emph{nominal} graph is known, which has the same nodes as the \emph{actual} (unknown) graph, but has a small number of edge perturbations (edge additions and/or deletions) relative to the actual graph. Due to this, significant uncertainty is induced in the GFT basis of the graph, since the perturbation of a few edges of the graph will perturb its Laplacian matrix, and correspondingly its eigenvectors.
Hence the graph signal will not be recovered accurately by a CS decoding algorithm if the GFT basis of the nominal graph is used, and some alternative approach is needed for recovery of the graph signal.
In some applications, it may be desirable to recover the actual graph as well.
We refer to the problem of recovery of the actual graph and the graph signal from compressive measurements as the \emph{Compressive Perturbed Graph Recovery} problem.

We present a method called Greedy Edge Selection (\textsc{Ges}), which solves this problem by refining the nominal graph one edge at a time, based on the cross-validation (CV) errors of the signals recovered using these graphs on a held-out set of measurements.
The algorithm keeps proceeding as long as the CV errors of the successive graphs keep decreasing.
Finally, the algorithm outputs the refined graph and the graph signal defined using the GFT basis of the refined graph. To summarize, our main contributions in this work are:
\begin{itemize}[leftmargin=1mm]
    \item We present the novel (to our best knowledge) problem of Compressive Perturbed Graph Recovery (CPGR), in which there is uncertainty in the orthonormal basis used for CS recovery, induced via edge perturbations in the underlying graph.
    
    \item We present the Greedy Edge Selection (\textsc{Ges}) method (Sec.~\ref{subsec:greedy_edge_selection}) for solving the CPGR problem, using an approach based on CV errors. 
    
    \item We frame the problem of edge-aware recovery of patch-wise compressively acquired images (such as in \cite{kulkarni2016reconnet}) as a CPGR problem.
    Based on the idea of using CV errors, we propose Inferred Linear-Edge Compressive Image Recovery (\textsc{Ilecir}, Sec.~\ref{subsec:compressive_image_acquisition}), an algorithm which performs the recovery of such images along with inferring linear image edges in the patches of the image via structured perturbation of the edges of a 2-D lattice graph (unlike unstructured edge perturbations done in \textsc{Ges}).
    
    \item Using CV theory for compressed sensing \cite{Zhang2014}, we prove high-probability signal and graph recovery guarantees for a brute-force version of our algorithms. We also provide similar guarantees for solution improvement in \textsc{Ilecir} and at each step of the \textsc{Ges}.
    
    \item We empirically validate the \textsc{Ges} algorithm on signals on a variety of graphs commonly used in the Network Science literature \cite{networksciencewikipedia},
    and also validate the \textsc{Ilecir} algorithm on $40$ images of various kinds -- natural images, synthetically generated piece-wise smooth images, cartoon images, and depth-maps of indoor scenes (Sec.~\ref{sec:gsp_results}).
    Our algorithms outperform the baseline method of using the GFT basis of the nominal graph or the GFT basis produced by first-order perturbations, in a standard CS decoding algorithm .
\end{itemize}

The remainder of this paper is organised as follows:
Sec.~\ref{subsec:problem_statement} defines the CPGR problem formally and gives an overview of the literature tackling similar problems.
Sec.~\ref{sec:method} describes the methods presented in this paper in detail.
Details and results of an empirical evaluation of our methods are presented in Sec.~\ref{sec:gsp_results}.
Finally, conclusions and future work are discussed in Sec.~\ref{sec:gsp_conclusion}.

\section{Problem Statement}
\label{subsec:problem_statement}
Consider that we are given linear, and possibly noisy, compressive measurements $\bb y \in \mathbb{R}^m$ of a graph signal defined on the vertices of an undirected, unweighted graph $\mathcal{G_{\text{actual}}} := (\mathcal{V}, \mathcal{E_{\text{actual}}})$ where the set of the nodes is denoted by $\mathcal{V} \triangleq \{1, \dots, n\}$ (and thus $|\mathcal{V}| = n$), and the set of edges is denoted by $\mathcal{E_{\text{actual}}}$. The graph signal is given by a vector $\bb x^* \in \mathbb{R}^n$ with each entry representing the value on the corresponding graph node.
Hence, we have
\begin{equation}
\label{eq:cs_measurement}
    \bb y = \bb \Phi \bb x^*+ \bb \eta,
\end{equation}
where $\bb \Phi$ is a $m\times n$ `measurement matrix' with $m \ll n$ (since we are in the compressive regime), and $\bb \eta $ is a noise vector, each of whose entries is assumed to be i.i.d. Gaussian with mean $0$ and variance $\sigma^2$.
In our setting, the graph $\mathcal{G_{\text{actual}}}$ is not known with full accuracy.
Instead, we have access to the graph $\mathcal{G_\text{nominal}} = (\mathcal{V}, \mathcal{E}_\text{nominal})$, which has the same nodes as $\act{\mc G}$, but the set of edges $\nom{\mc E}$ contains a few edge perturbations (additions or deletions) relative to $\act{\mc E}$.
We assume that an upper bound $d_0$ on the number of perturbations is known, i.e.\ $|\mathcal{E}_\text{actual} - \mathcal{E}_\text{nominal}| + |\mathcal{E}_\text{nominal} - \mathcal{E}_\text{actual}| \leq d_0$.
For any undirected graph with adjacency matrix $\boldsymbol{W}$, its Laplacian matrix is given by $\boldsymbol{L} = \boldsymbol{D} - \boldsymbol{W}$, where $\boldsymbol{D}$ is the diagonal matrix whose $i^\textrm{th}$ diagonal entry contains the degree of the $i^\text{th}$ node. For simplicity, we assume unweighted graphs. For undirected graphs, the Laplacian matrix is positive semi-definite and has real-valued eigenvectors and eigenvalues.
Let $\boldsymbol{L}_{\text{actual}}$ be the (unknown) Laplacian matrix of $\mathcal{G}_{\text{actual}}$, and let $\boldsymbol{L}_{\text{actual}} = \boldsymbol{V}_{\text{actual}} \boldsymbol{\Lambda}_{\text{actual}} \boldsymbol{V}^T_{\text{actual}}$ be its eigen-decomposition.
From the GSP literature, we know that the eigenvectors $\boldsymbol{V}_{\text{actual}}$ form a GFT basis, with each eigenvector having an associated `frequency' -- the eigenvalue -- which is a measure of its variation across the neighbouring nodes of the graph $\act{\mc G}$ \cite[Sec.~II.E]{Ortega2018}.
We assume that $\bb x^*$ is \emph{sparse-spectrum}, i.e.\ its GFT \cite[Eqn.~14]{Ortega2018} given by $\bb \theta^* = \boldsymbol{V}^T_{\text{actual}} \bb x^*$ has very few non-zero entries.
The goal is to recover the original signal $\bb x^*$, as well as the actual graph $\act{\mc G}$ given just $\boldsymbol{y}, \boldsymbol{\Phi}$ and $\mathcal{G}_{\text{nominal}}$.
This is formally defined as follows:
\begin{problem}[Compressive Perturbed Graph Recovery (CPGR) Problem]
    \label{eqn:main_problem}
    Given $\bb y$, $\bb \Phi$, $\nom{\mc E}$ and $d_0$, find $\bb x^*$ and $\act{\mc E}$.
\end{problem}
To our best knowledge, this is a novel computational problem, and has not been explored in the literature. While we are mainly concerned with sparse-spectrum signals as defined earlier, the model of \emph{band-limited} signals -- a subset of sparse-spectrum signals -- has been widely  considered in the literature (albeit not in conjunction with incorrectly defined graphs) \cite{zhu2012graph,Ortega2018}. In this model, the signals are linear combinations of the first $s$ eigenvectors with the smallest frequencies, with $s$ being the band limit. In the following section, we present a literature review of closely related work.

\begingroup
\subsection{Related Work}
\noindent\textbf{Graph Spectral Compressed Sensing:}
Recovery of graph signals from compressive measurements while making use of the graph structure via their GFT is sometimes called graph spectral compressed sensing \cite{zhu2012graph, Jung2018}.
While these and other works in GSP (e.g.\ see references in \cite{Ortega2018}) focus on band-limited signals, we mainly focus on the recovery of sparse-spectrum signals, i.e. signals with arbitrary sparse support in the GFT domain.
In \cite{zhu2012graph,Jung2018,zou2020compressive}, the measurements are the values of the graph signals on a subset of the nodes, and the purpose of compressive recovery is to fill in the missing data at other nodes,
whereas our work focuses on compressive measurements which are random linear combinations of the values at the different nodes, more in line with traditional CS and with applications such as group testing \cite{goenka2021contact} or image reconstruction.

\noindent\textbf{Compressed Sensing of Signals over Graphs with Partly Erroneous Topology:}
Our work is different as compared to a large body of work in graph learning (\textit{eg}, \cite{segarra2017network} or the references in \cite{Dong2019}), which typically assumes \emph{no} knowledge of the graph, but an availability of many graph signals defined over the nodes.
Instead, we assume that the graph is known up to a few edge perturbations, and that only a single graph signal is indirectly observed via compressive measurements.
The approach of correcting a small number of edges has been followed in a small number of papers, albeit in the non-compressive regime given a single signal vector -- \cite{barbarossa2020, Ceci2020total}.
The work in \cite{barbarossa2020} gives an approximate formula to compute eigenvectors of a perturbed graph from those of the original graph, and uses it in a brute-force algorithm which performs corrections to a graph topology given a band-limited graph signal with known band limit.
A total least squares approach to robust graph signal recovery given structural equation models is presented in \cite{Ceci2020total}.
The work in \cite{miettinen2021modelling} develops models for perturbations to edges in a graph and examines their effect on graph signal filtering and independent components analysis.
As opposed to this, in our work, we present a method for robust graph signal recovery from \emph{compressive measurements} given a small number of perturbations in the graph topology, and assuming the signal has a \emph{sparse}, but not necessarily band-limited, representation in the underlying GFT basis. The work in \cite{mahmood2018adaptive} performs joint recovery of an undirected, weighted, data-dependent graph with nodes representing overlapping patches of a 2-D image along with the image itself (i.e.,the graph signal) from compressive measurements. This is done by alternately re-computing the graph from the image, and then invoking graph total variation (GTV) and wavelet-sparsity prior for tomographic reconstruction. 
We note that their method is not applicable to our setting, where the graph is unweighted, and not data-dependent - i.e., it cannot be computed directly from the graph signal. Also, our technique focuses on recovery of a small number graph perturbations and not the full graph, and is not restricted to just graphs associated with images. In our setting, the signal is a linear combination of a small number of arbitrary GFT basis vectors, and hence the signals need not be smooth over the graph, whereas in \cite{mahmood2018adaptive}, the GTV prior assumes piece-wise smoothness of the graph signal. Unlike \cite{mahmood2018adaptive}, we also provide sufficient conditions for successful signal and graph recovery using one of our methods in Sec.~\ref{subsec:proof_brute_force}.

\noindent\textbf{Compressed Sensing with Perturbed Models:}
We now give an overview of techniques in general CS that deal with model mismatches. Referring to Eqn.~\ref{eq:cs_measurement}, we could have perturbations in either the sensing matrix $\boldsymbol{\Phi}$ or the representation matrix $\boldsymbol{\Psi}$ or both.
The former problem has been explored in terms of perturbations to specified Fourier frequencies in Fourier sensing matrices in magnetic resonance imaging in work such as \cite{Pandotra2019,Ianni2016}.
More unstructured and dense perturbations to $\boldsymbol{\Phi}$ are considered in \cite{Zhu2011,Parker2011,Fosson2020}.
The focus of the work in this paper, however, is related to perturbations in $\boldsymbol{\Psi}$, because perturbations to the graph topology induces changes in the GFT matrix, which is the signal \emph{representation} matrix (and not the sensing matrix which is completely independent).
The problem of perturbations in $\boldsymbol{\Psi}$ has been extensively explored in the context of off-the-grid signal representation in sinusoidal bases such as the Fourier or discrete cosine transform in  \cite{Chi2011,Nichols2014,nehorai2014structured}, using approaches such as alternating minimization \cite{Nichols2014}, modified greedy algorithms like orthogonal matching pursuit (OMP) \cite{Teke2013}, or structured sparsity \cite{Zhu2011,nehorai2014structured}. Applications to problems such as direction of arrival estimation have been explored.
In contrast to this, in this work, we explore perturbations in the representation matrix in an \emph{indirect} manner. That is, we explore methods to correct for perturbations in \emph{edge specifications} in the adjacency matrix within our optimization framework.
Given changes to the adjacency matrix, the Laplacian matrix and its eigenvectors are recomputed. We also note that small perturbations in the Laplacian matrix may lead to large perturbations of some of its eigenvectors -- especially in those with high frequency. Hence the methods which make the assumption of small perturbation in $\bb \Psi$ \cite{Zhu2011} are not directly applicable to the problem considered in our work.
\endgroup

\section{Method}
\label{sec:method}

\subsection{Recovery using standard CS decoding algorithms}
\label{subsec:standard_cs_recovery}
It is well known that a signal $\bb x^*$ acquired compressively as in Eqn.~\ref{eq:cs_measurement} and sparse in some known orthonormal basis $\bb \Psi$ may be recovered via \textsc{Lasso} (Least Absolute Shrinkage and Selection Operator) \cite{THW2015} as:
\begin{align}
    \label{eqn:lasso}
    &\textsc{Lasso:}&\hat{\bb x}_{\text{lasso}} = \underset{\bb x}{\arg\min} \: \|\bb y - \bb \Phi \bb x \|_2^2 + \mu \|\bb \Psi^T \bb x\|_1,
\end{align}
where $\mu$ is a regularization parameter.
Suitable choices of matrix $\bb \Phi$, such as those whose entries are drawn independently from a zero-mean Gaussian, satisfy with high probability the Restricted Eigenvalue Condition \cite[Sec.~11.2.2]{THW2015} if $m = O(s\log n)$ or other similar properties sufficient for recovery of $\bb x^*$ using \textsc{Lasso} \cite[Thm. 11.1]{THW2015} or other CS decoding algorithms.
The regularization parameter $\mu$ is typically chosen via cross-validation (CV) \cite{Zhang2014}, by holding out some $m_\text{cv} < m$ measurements, performing recovery using only the remaining $m_\text{r} = m - m_\text{cv}$ measurements, and choosing the value of $\mu$ for which the CV error is minimized. As shown in \cite[Theorem 1]{Zhang2014}, the CV error acts as a data-driven proxy for the (unobservable) mean-squared error. 
\begin{align}
\mathtt{CVE}(\bb y, &\bb \Phi, \bb \Psi, \Gamma, m_\text{cv}, \mu):\nonumber\\
   &\hat{\bb x}_\text{r}^\mu = \underset{\bb x}{\arg\min} \: \|\bb y_\text{r} - \bb \Phi_\text{r} \bb x \|_2^2 + \mu \|\bb \Psi^T \bb x\|_1, \\
   &\epsilon_\text{cv}^\mu = \|y_\text{cv} - \bb \Phi_\text{cv} \hat{\bb x}_\text{r}^\mu\|_2^2, \label{eqn:cv_error_formula} \\\nonumber\\
\textsc{Lasso}&\text{-Cv}(\bb y, \bb \Phi, \bb \Psi, \Gamma, m_\text{cv}):\nonumber\\
   & \hat{\mu} = \underset{\mu \in \Gamma}{\arg\min} \: \epsilon_\text{cv}^\mu, \\ 
&\hat{\bb x}_\text{lasso-cv} = \underset{\bb x}{\arg\min} \: \|\bb y - \bb \Phi \bb x \|_2^2 + \hat{\mu} \|\bb \Psi^T \bb x\|_1.
    \label{eqn:lasso_cv}
\end{align}
Here $\mathtt{CVE}$ is the CV error computation routine, and \textsc{Lasso-Cv} is \textsc{Lasso} with cross-validation.
Also, $\bb y_\text{cv}$ and $\bb \Phi_\text{cv}$ are the held out measurements and the corresponding rows of $\bb \Phi$, and $\bb y_\text{r}$ and $\bb \Phi_\text{r}$ are the remaining measurements or rows used for signal recovery.
Furthermore, $\hat{\bb x}_\text{r}^\mu$ is the signal recovered using the parameter value $\mu$, $\epsilon_\text{cv}^\mu$ is the cross-validation error of $\hat{\bb x}_\text{r}^\mu$, 
$\Gamma$ is the set of possible values of $\mu$ which are tried, $\hat{\mu}$ is the value of $\mu \in \Gamma$ which gives the smallest cross-validation error, and $\hat{\bb x}_\text{lasso-cv}$ is the final estimate of $\bb x^*$ output by \textsc{Lasso-Cv}.

Hence an estimate of $\bb x^*$ could be recovered via \textsc{Lasso} (Eqn.~\ref{eqn:lasso}) or \textsc{Lasso-Cv} (Eqn.~\ref{eqn:lasso_cv}) by putting $\bb \Psi = \boldsymbol{V}_{\text{actual}}$, with recovery guarantees from compressed sensing theory.
A naive method of estimating $\bb x^*$ when $\act{\mathcal{E}}$ is not known is to use the GFT matrix of the nominal graph, $\boldsymbol{V}_{\text{nominal}}$ as the orthonormal basis $\bb \Psi$ in \textsc{Lasso} or \textsc{Lasso-Cv} from Eqn.~\ref{eqn:lasso_cv}.
Thus we have the following estimates:
\begin{align}
     \act{\hat{\bb x}} = \textsc{Lasso-Cv}(\bb y, \bb \Phi, \boldsymbol{V}_{\text{actual}}, \Gamma, m_\text{cv}), \label{eqn:recovery_actual}\\
    \nom{\hat{\bb x}} = \textsc{Lasso-Cv}(\bb y, \bb \Phi, \boldsymbol{V}_{\text{nominal}}, \Gamma, m_\text{cv}).\label{eqn:recovery_nominal}
\end{align}
We refer to the technique of using the GFT basis in \textsc{Lasso} as \textsc{Gft-Lasso}. If the actual graph is used, we call it \textsc{Agft-Lasso}, and if the nominal graph is used, it is called \textsc{Ngft-Lasso}.
If cross-validation is used to determine the value of the parameter $\mu$, then these techniques are termed \textsc{Gft-Lasso-Cv}, \textsc{Agft-Lasso-Cv}, and \textsc{Ngft-Lasso-Cv}, respectively.

Since the set of edges $\nom{\mathcal{E}}$ differs slightly from $\act{\mathcal{E}}$, the GFT matrix of the nominal graph, $\boldsymbol{V}_{\text{nominal}}$ will be a perturbed version of the actual GFT matrix $\boldsymbol{V}_{\text{actual}}$. Indeed, $\boldsymbol{x^*}$ may not even be sparse in $\boldsymbol{V}_{\text{nominal}}$ if the perturbation is significant. 
In the following subsections, we present methods which use the CV error of \textsc{Gft-Lasso-Cv} to select from potential refinements of the nominal graph. Note that even small perturbations to the adjacency matrix can lead to large differences between $\boldsymbol{V}_{\text{actual}}$ and $\boldsymbol{V}_{\text{nominal}}$, and so we cannot exploit any sparsity property of $\|\boldsymbol{V}_{\text{actual}}-\boldsymbol{V}_{\text{nominal}}\|_1$ for deriving $\boldsymbol{V}_{\text{actual}}$.

\subsection{Greedy Edge Selection}
\label{subsec:greedy_edge_selection}

We present the Greedy Edge Selection (\textsc{Ges}) algorithm (Alg.~\ref{alg:greedy_edge_selection}) to solve the CPGR problem.
The main idea is to keep refining the edges of a candidate graph -- initialized with the edges of the nominal graph -- by perturbing one edge at a time, as long as the cross-validation error of the retrieved signal on a held-out set of measurements keeps decreasing.
The hope is that each greedy refinement of the candidate graph brings it closer to the actual graph, such that eventually the actual graph as well as the original graph signal are recovered.
The algorithm performs at most $d_0$ greedy refinement steps.
The edges of the candidate graph after greedy refinement step $t$ are referred to as $\can{\mc{E}}^{(t)}$, with $\can{\mc{E}}^{(0)} =  \nom{\mc E}$.
In the greedy step $t$ of the algorithm, all graphs which can be obtained by adding or removing one edge to or from the edge set of the candidate graph $\can{\mc{E}}^{(t-1)}$ (except those which have already been perturbed in greedy steps $1,\dots,(t-1)$) are considered.
There are ${n \choose 2} - (t-1)$ such graphs.
GFT matrices of each of these graphs are computed via eigendecomposition of their Laplacian matrices.
\textsc{Gft-Lasso-Cv} (Sec.~\ref{subsec:standard_cs_recovery}) is performed to find the smallest CV error and the ideal value of $\mu$ using each GFT matrix.
The edge  $e_{\text{best}}^{(t)}$ which gives the smallest value for CV error is chosen and the candidate graph is updated, provided the CV error decreases by more than a factor $\tau \in (0, 1]$ relative to the CV error for the current candidate graph.
Otherwise, the algorithm stops, and an estimate of the signal is returned by performing \textsc{Lasso} using the GFT matrix and $\mu$ obtained from $\can{\mc E}^{(t-1)}$.

\noindent\textbf{Noise-based stopping criterion:}
From Eqn.~\ref{eqn:cv_error_formula} and Eqn.~\ref{eq:cs_measurement}, we see that the CV error for the ground truth signal $\bb x^*$ is a sum of squares of $m_\text{cv}$ i.i.d.\ Gaussian random variables with mean $0$ and variance $\sigma^2$, and thus has mean equal to $m_\text{cv}\sigma^2$ and standard deviation $\sqrt{2m_\text{cv}}\sigma^2$. 
Hence, before a greedy step begins, the \textsc{Ges} algorithm checks whether the CV error is within a high confidence interval of $m_\text{cv}\sigma^2$ and stops if that is the case, in order to prevent fitting to the measurement noise.

\noindent\textbf{Brute-force algorithm:}
A brute-force version of \textsc{Ges} (referred to as Brute-Force Graph Selection or \textsc{Bfgs}) is also possible, wherein all graphs which are at most $d_0$ perturbations from $\nom{\mc G}$ are considered, signals are recovered using their GFT matrices on $m_\text{r}$ measurements, and the graph which gives the minimum CV error on the remaining $m_\text{cv}$ measurements is chosen for final signal recovery. 
As there are $n \choose 2$ possible edges, the total number of graphs enumerated via brute-force is ${{n \choose 2} \choose 0} + {{n \choose 2} \choose 1} +\dots+ {{n \choose 2} \choose d_0} = O(n^{2d_0})$.

\noindent\textbf{Running time: } The algorithm performs a maximum of $d_0$ greedy steps.
In each greedy step, \textsc{Gft-Lasso}  is performed for a maximum of $n \choose 2$ graphs.
Grid search for $\mu$ is performed over $|\Gamma|$ values.
Hence the total number of \textsc{Gft-Lasso}  optimizations performed is $O(|\Gamma| d_0 n^2)$, which is a significant improvement over the $O(|\Gamma| n^{2d_0})$ \textsc{Gft-Lasso}  optimizations performed by the brute-force algorithm.

\noindent\textbf{Signal recovery accuracy:} 
While the greedy algorithm is not guaranteed to recover the original graph and the signal, we find empirically (Sec.~\ref{subsec:ges_results}) that the mean error in the recovered signal is much lower than that using the nominal graph, and in many cases, the original graph is recovered.
In Theorem~\ref{thm:ges_solution_improvement}, we present conditions under which the solution is guaranteed to improve at any step of the \textsc{Ges} algorithm.

\begin{algorithm}[t]
\textbf{Input:} $\bb y$ : Compressive measurements,
$\bb \Phi$ : measurement matrix,
$\nom{\mathcal{E}}$ : nominal graph edge set,
$ \sigma^2$ : variance of measurement noise,
$\Gamma$ : values of $\mu$ for grid search,
$\tau \in (0, 1]$ : CV error improvement factor,
$m_\text{cv}$ : Number of CV measurements,
$d_0$ : maximum edge perturbations,
$g$ : CV error confidence interval factor

\textbf{Output:} Estimated graph signal $\hat{\bb x}_{\text{greedy}}$
\begin{algorithmic}[1]
\caption{Greedy Edge Selection} \label{alg:greedy_edge_selection}
\State Initialize: $\can{\mathcal{E}}^{(0)} \gets \nom{\mathcal{E}}$,
$\bb V^{(0)} \gets \nom {\bb V}$, 
$\{\epsilon_{\text{cv}}^{(0)}, \mu^{(0)}\} \gets \{\underset{\mu \in \Gamma}{\min, \mathrm{argmin}}\}{\mathtt{CVE}(\bb V^{(0)}}, \mu| \bb y, \bb \Phi,$ $ m_{\text{cv}})$, 
$\mc P^{(0)} = \emptyset $, and $t_{\text{est}} \gets 0$
\For{$t$ in $1,\dots,d_0$}
    \If{$\epsilon_{\text{cv}}^{(t-1)} \leq (m_{\text{cv}}\sigma^2 + g\sqrt{2m_{\text{cv}}}\sigma^2)$
    }
        \State \textbf{break}, to prevent fitting on noise
    \EndIf
    \For{each possible edge $e$ which is not in $\mc P^{(t-1)}$}
        \State Obtain perturbed graph edge set:\\ $\mathcal{E}^{(t)}_e \gets (\can{\E}^{(t-1)} - \{e\}) \cup (\{e\} - \can{\E}^{(t-1)}) $
        \State Compute Laplacian matrix $\bb L_e^{(t)}$ from $\mathcal{E}^{(t)}_e$
        and GFT matrix $\bb V_e^{(t)}$ via eigendecomposition of $\bb L_e^{(t)}$
        \State Compute best CV error and $\mu$ using $\bb V_e^{(t)}$:\\
        $\epsilon_{\text{cv}}^{(e, t)} \gets \underset{\mu \in \Gamma}{\min}{\:\mathtt{CVE}(\bb V_e^{(t)}}, \mu| \bb y, \bb \Phi, m_{\text{cv}})$ \\
        $\mu_e^{(t)} \gets \underset{\mu \in \Gamma}{\mathrm{argmin}}{\:\mathtt{CVE}(\bb V_e^{(t)}}, \mu| \bb y, \bb \Phi, m_{\text{cv}})$
    \EndFor
    \If{$\underset{e}{\min} \: \epsilon_{\text{cv}}^{(e, t)} < \tau \epsilon_{\text{cv}}^{(t-1)}$}
        \State Select edge: $e_{\text{best}}^{(t)} \gets \underset{e}{\arg\min} \: \epsilon_{\text{cv}}^{(e, t)}$
        \State Perform updates:
        $\mc P^{(t)} \gets P^{(t-1)} \cup \{e_{\text{best}}^{(t)}\}$,\\
        $\epsilon_{\text{cv}}^{(t)} \gets \underset{e}{\min} \: \epsilon_{\text{cv}}^{(e, t)}$,
        $\can{\E}^{(t)} \gets \E_{e_{\text{best}}^{(t)}}^{(t)}$,
        ${\bb V}^{(t)} \gets \bb V_{e_{\text{best}}^{(t)}}^{(t)}$, ${\mu}^{(t)} \gets \mu_{e_{\text{best}}^{(t)}}^{(t)}$, and $t_{\text{est}} \gets t$
    \Else
        \State \textbf{break}; insignificant improvement.
    \EndIf
\EndFor
\State Estimate $\hat{\bb x}_{\text{greedy}} $ via \textsc{Lasso} from $\bb y$, $\bb \Phi$, $ \bb V^{(t_{\text{est}})}$, and $\mu^{(t_{\text{est}})}$ using all the measurements:\newline
\hspace*{2em} 
$\hat{\bb x}_{\text{greedy}} \gets \underset{\bb x\in R^n}{\arg\min}\:  \|\bb y - \bb \Phi \bb x\|_2^2 + \mu^{(t_{\text{est}})} \|\bb V^{(t_{\text{est}})T} \bb x\|_1$
\State \Return  $\hat{\bb x}_{\text{greedy}} $
\end{algorithmic}
\end{algorithm}

\subsection{Inferred Linear-Edge Compressive Image Recovery}
\label{subsec:compressive_image_acquisition}
In this section, we use the term `edge' as an element of a graph connecting a pair of vertices, and use the term `image edge' to refer to the entity which separates two regions in an image. The greedy edge selection method presented in Sec.~\ref{subsec:greedy_edge_selection} is a generally applicable technique on any graph.
However, if the graph has some well-known structure to it, it may be possible to select the edges for perturbation in a more structured manner, instead of greedily one at a time.
We present such a method for image reconstruction from compressive measurements taken using a block-based version \cite{kulkarni2016reconnet} of the Rice Single-Pixel Camera \cite{Duarte2008}.
That is, for each $h\times w$ patch $\boldsymbol{x^*}$ of an image, compressive measurements $\bb y$ obtained via a noisy version of $\boldsymbol{\Phi x^*}$ as in Eqn.~\ref{eq:cs_measurement}, are assumed to be available, with $\bb \Phi$ having dimensions $m\times n$ with $n = hw$ and $m \ll n$, and the task is to reconstruct each patch of the image.

\begin{figure}[t]
    \centering
    \begin{subfigure}[b]{0.2\linewidth}
        \includegraphics[width=\linewidth]{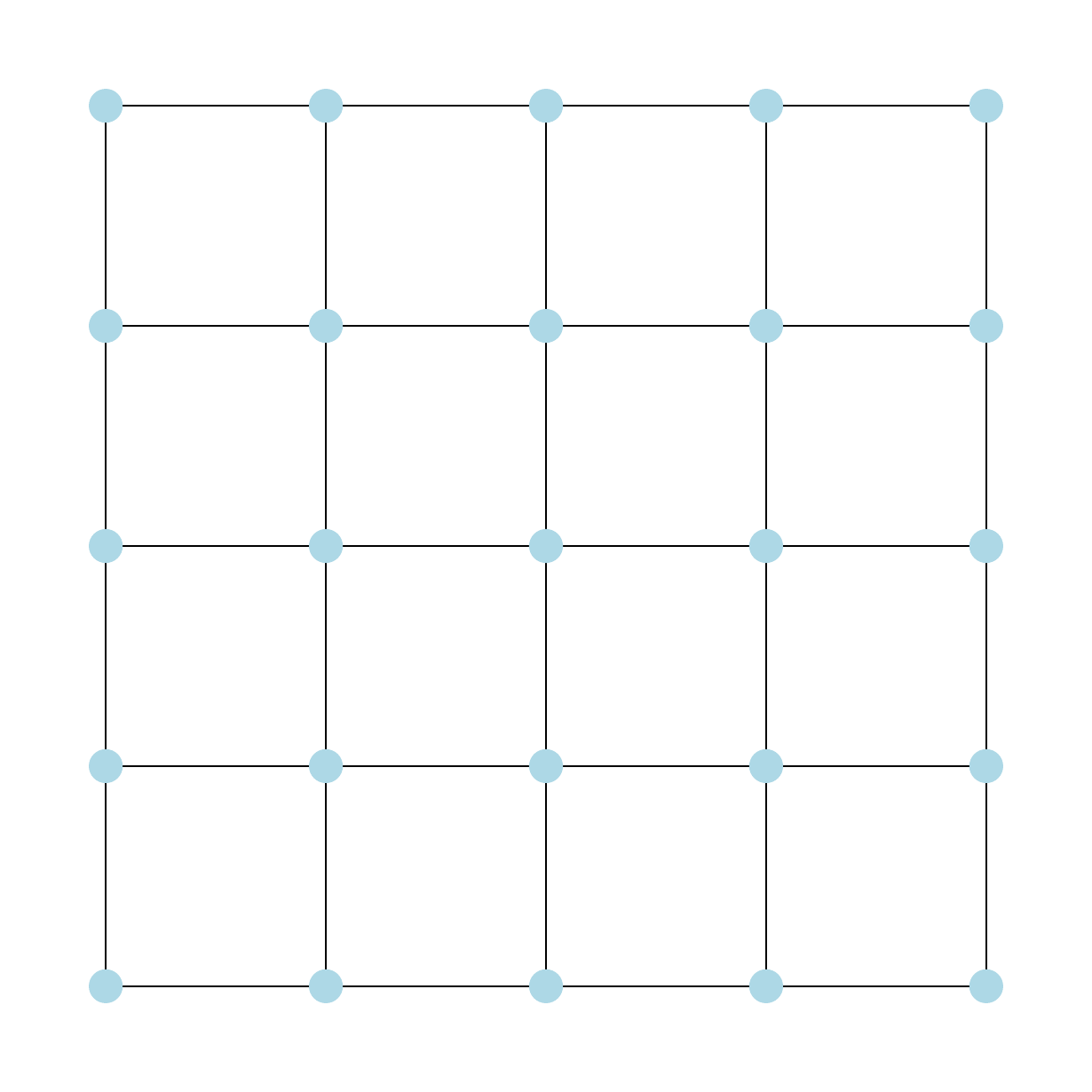}
        \caption{ }
        \label{fig:lattice_graph}
    \end{subfigure}\hfill%
    \begin{subfigure}[b]{0.2\linewidth}
    \includegraphics[width=\linewidth]{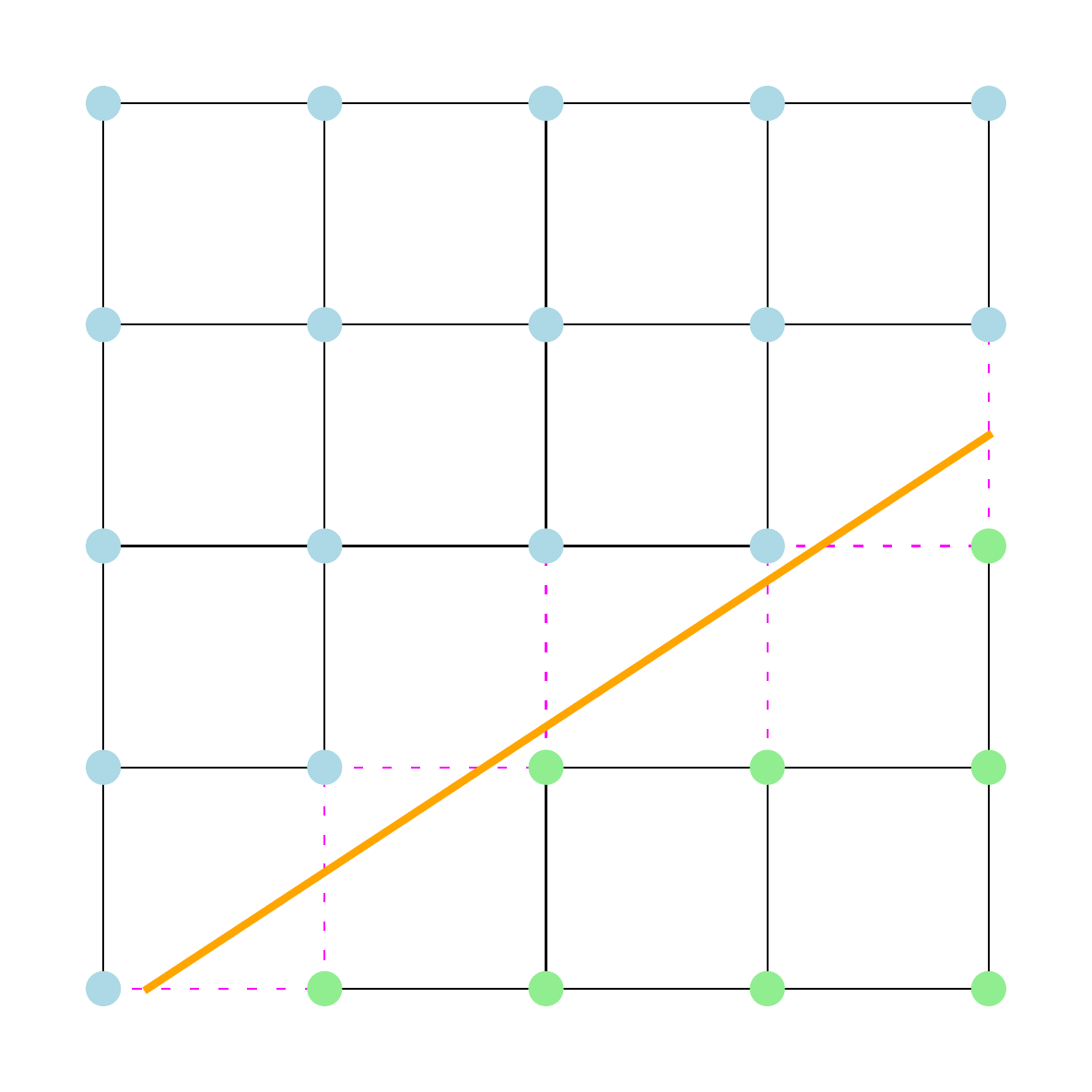} 
        \caption{ }
        \label{fig:lattice_graph_partitioned}
    \end{subfigure}\hfill%
    \begin{subfigure}[b]{0.24\linewidth}
    \includegraphics[width=\linewidth]{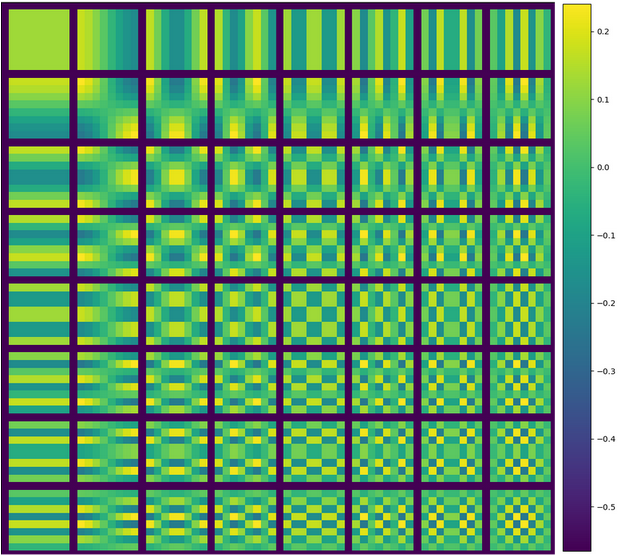} 
        \caption{ }
        \label{fig:dct_basis}
    \end{subfigure}\hfill
    \begin{subfigure}[b]{0.1\linewidth}
    \includegraphics[width=\linewidth]{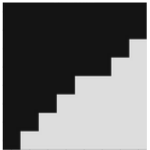} 
        \caption{ }
        \label{fig:patch}
    \end{subfigure}\hfill
    \begin{subfigure}[b]{0.24\linewidth}
    \includegraphics[width=\linewidth]{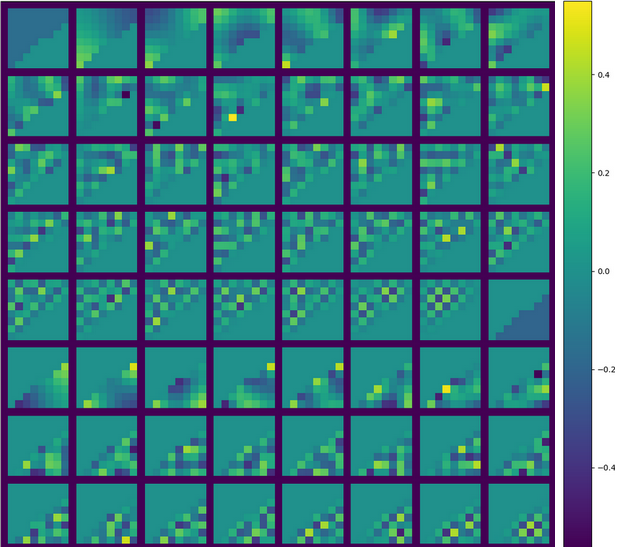} 
        \caption{ }
        \label{fig:seg_aware_basis}
    \end{subfigure}%
    \caption[Segmentation-aware basis vectors of images]{\protect\subref{fig:lattice_graph} A $5\times 5$ 2-D lattice graph whose nodes represent pixels of a $5\times 5$ patch and whose edges represent connections between a pixels and its four neighbors. This forms the nominal graph for the problem of compressive image patch recovery.
    \protect\subref{fig:lattice_graph_partitioned} The lattice graph partitioned by an image edge (orange line). The graph edges going across the image edge are removed (dotted purple lines). Since the image edge is unknown before reconstruction, the image-edge partitioned graph is the (unknown) actual graph for the problem of recovery of an image patch from compressive measurements.
    \protect\subref{fig:dct_basis} All 64 2-D DCT basis vectors of an $8\times 8$ patch.
    \protect\subref{fig:patch} An $8\times 8$ patch with a sharp edge.
    \protect\subref{fig:seg_aware_basis} Segmentation-aware basis vectors for this patch, obtained by computing the eigenvectors of the Laplacian matrix of the graph created by dropping the edges of the $8\times8$ lattice graph whose endpoints lie in different segments of the patch.}
    \label{fig:lattice_graph_combined}
\end{figure}

Any 2-D (two-dimensional) grayscale raster image can be considered to be a graph signal, with each pixel being a node and horizontal and vertical neighbours connected to each other via an edge, with each pixel node mapped to its value in the image.
Such a graph is called a `lattice graph' (Fig.~\ref{fig:lattice_graph}).
The 2-D Discrete Cosine Transform (DCT) basis (Fig.~\ref{fig:dct_basis}) is known to form an eigenbasis of the Laplacian matrix of a 2-D lattice graph {\cite[Proposition 1]{fracastoro2016steerable}}, and thus are a GFT basis for this graph. Piece-wise smooth images, depth-maps, and natural images have a sparse or compressible representation in the DCT basis, because the low-frequency basis vectors in DCT model the correlation between neighbouring pixels well.
This fact is in fact heavily exploited in the well known JPEG standard for image compression.
Thus, it is natural to employ \textsc{Lasso} (Eqn.~\ref{eqn:lasso}) with the DCT basis in order to reconstruct the image patches.

However, neighbouring pixels on either side of a sharp image-edge will not have correlated values, due to which DCT-based recovery will not be accurate near an image-edge.
In such cases, it may be better to use some other transform basis whose vectors maintain no correlation in their values across the image edge.
Since the DCT basis is a GFT basis of the 2-D lattice graph, one possible basis is the GFT basis of the graph obtained by dropping those edges of the 2-D lattice graph which link nodes that are located on two different sides of an image-edge -- see Fig.~\ref{fig:patch} and Fig.~\ref{fig:seg_aware_basis} for an example of such a `segmentation-aware' basis.
We may call this graph the `image-edge partitioned graph' (see Fig.~\ref{fig:lattice_graph_partitioned}) if it is constructed using the true image edges. However, since the image is acquired compressively, it is not straightforward to know the exact location of the image-edges, and hence the image-edge partitioned graph is unknown.
Hence, the problem of decoding of the compressively acquired image using the GFT of the unknown image-edge-partitioned graph while knowing only the 2D lattice graph may be formulated as Problem~\ref{eqn:main_problem}.
The 2D lattice graph is the nominal graph, the image-edge partitioned graph is the actual graph, and the set of pixel values is the unknown graph signal, sparse or compressible in the GFT of the \emph{actual} graph.

\begin{algorithm}[t]
\textbf{Input: }
$\bb y$ : vector of $m$ measurements of an $h\times w$ patch, 
$\bb \Phi$ : $m\times n$ measurement matrix with $n = hw$,
$\bb V_{\text{DCT}}$ : 2-D DCT basis vectors of an $h\times w$ patch,
$\phi$ : set of linear image-edges connecting boundary pixels of an $h\times w$ patch,
$\mc V_{h,w}$ : GFT matrices of the partitioned graphs formed by the image-edges in $\phi$,
$ \sigma^2$ : measurement noise variance,
$\Gamma$ : values of $\mu$ for grid search,
$m_\text{cv}$ : number of CV measurements,
$\tau \in (0, 1]$ : CV error improvement factor,
$g$ : CV error confidence interval factor.

\textbf{Output:} $\hat{P}$ : estimated image patch of size $h\times w$
\begin{algorithmic}[1]
\caption{Inferred Linear-Edge Compressive Image Recovery (\textsc{Ilecir})}
\label{alg:image_reconstruction}
\State Compute CV error and $\mu$ using DCT:\\
$\{\epsilon_{\text{cv}}^{\text{DCT}}, \mu_{\text{DCT}} \} \gets\underset{\mu \in \Gamma}{\{\min, \mathrm{argmin}\}}{\:\mathtt{CVE}(\bb V_{\text{DCT}}}, \mu| \bb y, \bb \Phi, m_{\text{cv}})$

\If{$\epsilon_{\text{cv}}^{\text{DCT}} \leq (m_{\text{cv}}\sigma^2 + g\sqrt{2m_{\text{cv}}}\sigma^2)$ 
i.e. close to noise
}
\State $\bb V_{\text{est}} \gets \bb V_{\text{DCT}}$, 
$\mu_{\text{est}} \gets \mu_{\text{DCT}}$
\Else
    \For{Each linear image edge $i \in \phi$ }
        \State Retrieve corresponding GFT matrix $\bb V_i \in \mc V_{h,w}$
        \State Compute best CV error and $\mu$ using $i$:\\
        $\{\epsilon_{\text{cv}}^{(i)}, \mu_i\} \gets\underset{\mu \in \Gamma}{\{\min, \mathrm{argmin}\}}{\:\mathtt{CVE}(\bb V_i}, \mu| \bb y, \bb \Phi, m_{\text{cv}})$
    \EndFor
    \If{$\underset{i}{\min} \: \epsilon_{\text{cv}}^{(i)} < \tau \epsilon_{\text{cv}}^{\text{DCT}}$}
        \State $\bb V_{\text{est}} \gets \bb V_i$,
        $\mu_{\text{est}} \gets \mu_i$
    \EndIf
\EndIf
\State Estimate the vectorized patch using LASSO: $\hat{\bb x}_\text{ilecir} \gets \underset{x\in R^n}{\arg\min}\:  \|\bb y - \bb \Phi \bb x\|_2^2 + \mu_{\text{est}} \|\bb V_{\text{est}}^T \bb x\|_1$
\State Convert $\hat{\bb x}_\text{ilecir}$ to $h \times w$ patch $\hat{P}$, assuming row-major order
\State \Return $\hat{P}$
\end{algorithmic}
\end{algorithm}

Rather than applying the greedy algorithm from Sec.~\ref{subsec:greedy_edge_selection} to solve this problem, it is better to exploit the inherent structure in image-edges to drop graph edges in a systematic way.
We present our algorithm called Inferred Linear-Edge Compressive Image Recovery (\textsc{Ilecir}) in Algorithm~\ref{alg:image_reconstruction}. Each image patch is assumed to either contain no image-edge or at most a single \emph{linear} image-edge, i.e., a straight line with its endpoints at two boundary pixels of the patch.
Reconstruction of the patch from a reconstruction subset of measurements is performed using the GFTs corresponding to each image-edge partitioned graph as well as DCT, and CV errors on a held-out subset of measurements are computed.
The GFT corresponding to the estimated signal with the lowest cross-validation error is selected, and the final estimate of the patch is reconstructed using all the measurements.
The noise-based stopping criterion from Sec.~\ref{subsec:greedy_edge_selection} is employed here.

\subsection{Recovery Guarantees and Bounds}
\label{subsec:proof_brute_force}
Let $\hat{\bb x}_\mc{G}$ denote the signal recovered using \textsc{Gft-Lasso-Cv}  with the GFT of graph $\mc{G}$.
We make the assumption that each entry of $\bb \Phi$ is independently drawn from a sub-Gaussian distribution of mean 0 and variance 1.
We present the following recovery guarantee for the \textsc{Bfgs} algorithm from Sec.~\ref{subsec:greedy_edge_selection}:
\begin{theorem}[Brute-Force Algorithm Recovery Guarantee]
\label{thm:brute_force_guarantee}
If in the \textsc{Bfgs} algorithm in Sec.~\ref{subsec:greedy_edge_selection}, the number of CV measurements $m_\text{cv}$ obeys
    \begin{align}
    \label{eqn:mcv}
        m_{\text{cv}} \: \geq \: 4\Big(1+ \frac{2c}{(c-1)^2}\Big) \Big\{ \ln |\Gamma| + \ln{(d_0 + 1)} \nonumber\\ + 2d_0 \ln n + \ln {\frac{1}{\delta}}\Big\}
    \end{align}
    for arbitrary constants $c \in (1, \infty)$ and $\delta \in (0, 1)$, then its recovery error is bounded as
    \begin{align}
    \label{eqn:error_bound}
        \|\hat{\bb{x}}_{\text{bf}} - \bb x^* \|_2^2 \: < \: c \|\hat{\bb{x}}_{\text{actual}} - \bb x^*\|_2^2 + (c-1)\sigma^2
    \end{align}
    with probability more than $1-\delta$. As a consequence, if 
    \begin{align}
        \label{eqn:G}
        \|\hat{\bb{x}}_{\mc{G}} - \bb x^* \|_2^2 \: \geq \: c 
        \|\hat{\bb{x}}_{\text{actual}} - \bb x^*\|_2^2 + (c-1)\sigma^2,
    \end{align}
    for all graphs $\mc G \neq \mc G_{\text{actual}}$ which are upto $d_0$ edge perturbations from $\mc G_\text{nominal}$, then with probability more than $1-\delta$, $\hat{\bb x}_{\text{bf}} = \hat{\bb x}_{\text{actual}}$, and the actual graph is recovered.
\end{theorem}
The proof is provided in the supplemental material (Sec.~\ref*{sec:appendix_proof}).
It is based on a theorem from \cite{Zhang2014} regarding how well the CV error predicts the recovery error for compressed sensing with Gaussian random matrices\footnote{The theorems in \cite{Zhang2014} can however be easily extended to handle any sub-Gaussian matrix by using an appropriate formula for the variance of the particular sub-Gaussian distribution used in Eqn.~50 in \cite{zhang2016cross}, which will reflect accordingly in the bounds in Theorem~\ref{thm:brute_force_guarantee}.}.
We use this theorem and the union bound to lower bound the probability that the CV error of the signal recovered using \textsc{Gft-Lasso-Cv} with the actual graph is lower than the CV errors for all signals having a recovery error satisfying the condition in Eqn.~\ref{eqn:G}.

The result shows that there is a tradeoff between the number of CV measurements used and the quality of recovery, encapsulated by the parameter $c$.
If $c$ is close to $1$, the recovery error is close to or equal to that achieved using \textsc{Gft-Lasso-Cv}  on the actual graph.
However in such a case, $m_\text{cv}$ will be large.
On the other hand, a large value of $c$ will allow for a smaller value of $m_\text{cv}$ -- however in that case, the recovered signal may have error higher than if the actual graph was known.
It also illustrates that if the recovery error using the nominal graph is close to that using the actual graph or if the noise level is too high, then it is harder to distinguish between the two.
This may be case when perturbing the actual graph did not significantly perturb the eigenvectors on which $\bb x^*$ had support.
Dependence of the bound on $d_0$ in Eqn.~\ref{eqn:mcv} shows that the method works well if the nominal graph and the actual graph are only a few perturbations away from each other.
In particular, if a nominal graph were not known, then the brute-force algorithm must enumerate all graphs, making $d_0 = {n\choose 2}$.
In this case, $m_\text{cv} = \Omega(n^2\ln n)$, and we are no longer in the regime of CS.
The dependence of the bound in Eqn.~\ref{eqn:mcv} on $|\Gamma|$ seems to be an artifact of our proof technique.
Our proof does not exploit the fact that there may be many `bad' graphs in the search space, for which the recovery error (and the CV error) will be high regardless of the value of $\mu \in \Gamma$ used.
Instead, the proof proceeds by assuming that the CV errors for a `bad' graph with different values of $\mu$ are not correlated, leading to an overestimation of the bound for $m_\text{cv}$.
Perhaps some other proof technique might be used which removes or weakens the dependence of the bound on $|\Gamma|$.

\noindent\textbf{Remark on Graph Recovery:}
While the brute-force method is guaranteed to recover the signal to good accuracy as per Thm. ~\ref{thm:brute_force_guarantee}, we do not know if it always recovers the actual graph, even in the absence of measurement noise.
For example, there might be cases wherein a graph signal has a sparser or equally-sparse representation in the GFT basis of a graph which is not the actual graph, which might get chosen by the algorithm.
In particular, it is known that the eigenvectors of the Laplacian matrix of a graph and its complement graph\footnote{The complement of a graph is a graph in which every edge present in the original graph is absent in the complement graph and every edge which is absent in the original graph is present in the complement graph.} are the same.
The authors do not know if there exist two graphs which are a small number of perturbations from each other and share some eigenvectors or have eigenvectors which are the linear combination of a small number of the eigenvectors of the other graph.
Interestingly, the signal recovery bound in Eqn.~\ref{eqn:error_bound} does not depend on the structure of the graph -- it only depends on $m_{\text{cv}}$ and $\sigma$.

Analogous to Thm.~\ref{thm:brute_force_guarantee}, we present two theorems for solution improvement using the greedy \textsc{Ges} method (Algorithm~\ref{alg:greedy_edge_selection}) and \textsc{Ilecir} (Algorithm~\ref{alg:image_reconstruction}).
Let $\hat{\bb x}^{(t)}$ be the estimate by the greedy edge selection algorithm after $t$ steps, and let $\hat{\bb x}_{\text{best}}^{(t)}$ be the estimate with the lowest recovery error amongst the signals recovered at step $t$.
\begin{theorem}[Greedy Edge Selection Solution Improvement Guarantee]
\label{thm:ges_solution_improvement}
If in the greedy edge selection algorithm (Algorithm~\ref{alg:greedy_edge_selection}) with $\tau = 1$ the number of CV measurements $m_\text{cv}$ obeys
    \begin{align}
        m_{\text{cv}} \: \geq \: 4\Big(1+ \frac{2c}{(c-1)^2}\Big) \Big\{ \ln |\Gamma| + \ln \frac{n(n+1)}{2} + \ln {\frac{1}{\delta}}\Big\}
    \end{align}
    for arbitrary constants $c \in (1, \infty)$ and $\delta \in (0, 1)$, then the recovery error after step $t$ is bounded as
    \begin{align}
        \|\hat{\bb{x}}^{(t)} - \bb x^* \|_2^2 \: < \: c \|\hat{\bb{x}}_{\text{best}}^{(t)} - \bb x^*\|_2^2 + (c-1)\sigma^2
    \end{align}
    with probability more than $1-\delta$.
    As a consequence, if
    \begin{align}
        \|\hat{\bb{x}}^{(t-1)} - \bb x^* \|_2^2 \: \geq \: c \|\hat{\bb{x}}_{\text{best}}^{(t)} - \bb x^*\|_2^2 + (c-1)\sigma^2,
    \end{align}
    then the solution is guaranteed to improve at step $t$ with probability more than $1-\delta$.
\end{theorem}
Let $\hat{\bb x}_\text{best-edge}$ denote the (vectorized) patch estimate with the lowest recovery error amongst all the patch estimates in the \textsc{Ilecir} algorithm for a single patch. Note that $\hat{\bb x}_\text{best-edge}$ is unobservable. 
\begin{theorem}[\textsc{Ilecir} Solution Improvement Guarantee]
\label{thm:ilecir_solution_improvement}
        If in the \textsc{Ilecir} algorithm (Algorithm~\ref{alg:greedy_edge_selection}) with $\tau = 1$ the number of CV measurements $m_\text{cv}$ obeys
    \begin{align}
        m_{\text{cv}} \: \geq \: 4\Big(1+ \frac{2c}{(c-1)^2}\Big) \Big\{ \ln |\Gamma| + \ln({|\phi| + 1}) + \ln {\frac{1}{\delta}}\Big\}
    \end{align}
    for arbitrary constants $c \in (1, \infty)$ and $\delta \in (0, 1)$ and where $\phi$ is the set of all possible linear image edges in an $h\times w$ patch with endpoints at the boundary, then the recovery error is bounded as
    \begin{align}
        \|\hat{\bb{x}}_\text{ilecir} - \bb x^* \|_2^2 \: < \: c \|\hat{\bb{x}}_{\text{best-edge}} - \bb x^*\|_2^2 + (c-1)\sigma^2
    \end{align}
    with probability more than $1-\delta$.
    As a consequence, if
    \begin{align}
        \|\hat{\bb{x}}_{\text{dct}} - \bb x^* \|_2^2 \: \geq \: c \|\hat{\bb{x}}_{\text{best-edge}} - \bb x^*\|_2^2 + (c-1)\sigma^2,
    \end{align}
    then the solution is guaranteed to improve over \textsc{Dct-Lasso-Cv} with probability more than $1-\delta$.
\end{theorem}
We omit the proofs for Thms.~\ref{thm:ges_solution_improvement} and \ref{thm:ilecir_solution_improvement} since they are very similar to that for Thm.~\ref{thm:brute_force_guarantee}.
The main difference is in the lower bound for $m_\text{cv}$, which is due to the different number of perturbed graphs considered in each algorithm.

\subsection{Alternatives to Eigendecomposition for GFT basis computation}
\label{subsec:ges_ceci_barbarossa}
The \textsc{Ges} algorithm requires computing the eigen-decomposition (step 16 in Alg.~\ref{alg:greedy_edge_selection}) of the Laplacian matrix $\boldsymbol{L_e}^{(t)}$ of the graph obtained by perturbing the candidate edge set $\can{\mc E}^{(t-1)}$ with edge $e$, for each possible edge $e$, at each time step $t$.
This step may take a long time for large graphs.
It may be made more efficient by using an approximate formula given in \cite{barbarossa2020}, to obtain the eigenvectors of $\boldsymbol{L_e}^{(t)}$ from the eigenvectors of $\nom{\boldsymbol{L}}$, and the set of perturbations $\mc{P}^{(t-1)} \cup \{e\}$, instead of performing eigendecomposition of $\boldsymbol{L_e}^{(t)}$.
If an original Laplacian matrix $\boldsymbol{L}$ is perturbed by a set of edges $\mc P$ to obtain a perturbed Laplacian matrix $\boldsymbol{\Tilde L}$, and their eigenvectors are $\bb v_1\dots\bb v_n$ and $\Tilde{\bb v}_1\dots \Tilde{\bb v}_n$ respectively, then from \cite[Eqn. 7 and 11]{barbarossa2020} we have the approximation:
\begin{align}
    \label{eqn:eigvec_perturbation_formula}
    \Tilde{\bb v}_k \simeq \bb v_k + \sum_{\{i, j\} \in \mc P} \sigma_{i, j} (\bb v_k(i) - \bb v_k(j)) \sum_{\substack{l=2 \\ l \neq k}}^n \frac{\bb v_l(i) - \bb v_l(j)}{\lambda_k - \lambda_l} \bb v_l,
\end{align}
where $\sigma_{i, j} = 1$ for edge addition, and $\sigma_{i, j} = -1$ for edge deletion.
The approximation for $\Tilde{\bb v}_k$ is valid only under the condition that
\begin{align}
\label{eqn:gsp_barbarossa_eigengap_condition}
    \lambda_{k-1} - \lambda_{k}\ll \sum_{\{i, j\} \in \mc P} \sigma_{i, j} (\bb v_k(i) - \bb v_k(j))^2 \ll \lambda_{k+1} - \lambda_k.
\end{align}
We discuss some intuition behind this approximation and the condition under which it is valid.
First, note that each new edge $\{i, j\}$ in $\mc P$ introduces a rank-one perturbation of $\boldsymbol{L}$ with the matrix $\sigma_{i,j} \bb a_{i,j}\bb a_{i,j}^T$, where $\bb a_{i,j}$ is a vector with $\bb a_{i,j}(i) = 1, \bb a_{i,j} (j) = -1$, and the remaining entries of $\bb a_{i,j}$ are equal to zero.
That is, $\boldsymbol{\Tilde{L}} = \boldsymbol{L} + \boldsymbol{\Delta L}$, where the perturbing matrix $\boldsymbol{\Delta L} = \sum\limits_{\{i,j\} \in \mc P} \sigma_{i,j} \bb a_{i,j}\bb a_{i,j}^T$.
Note that $\bb a_{i,j}^T \bb v_k = (\bb v_k(i) - \bb v_k(j))$ and hence $ \bb v_k^T  \boldsymbol{\Delta L} \bb v_k = \sum\limits_{\{i, j\} \in \mc P} \sigma_{i, j} \bb v_k^T \bb a_{i,j} \bb a_{i,j}^T \bb v_k = \sum\limits_{\{i, j\} \in \mc P} \sigma_{i, j} (\bb v_k(i) - \bb v_k(j))^2$.
It is easily verified that if $\bb a_{i,j}^T \bb v_k = 0 $ (i.e. $(\bb v_k(i) - \bb v_k(j)) = 0$) for all the perturbing edges $\{i, j\} \in \mc P$, then $\bb v_k$ is also an eigenvector of  $\boldsymbol{\Tilde{L}}$.
Recall that the normalized eigenvectors of a matrix $\boldsymbol{L}$ are the critical points of the Rayleigh quotient $\bb x^T \boldsymbol{L} \bb x$ on the unit sphere $\bb x^T \bb x = 1$, with the eigenvalue being the value of the Rayleigh quotient at the corresponding eigenvector.
Eqn.~\ref{eqn:gsp_barbarossa_eigengap_condition} means that the perturbation of the Rayleigh quotient at $\bb v_k$ must be much smaller than the difference in values of the Rayleigh quotient at the closest critical points.
In Eqn.~\ref{eqn:eigvec_perturbation_formula}, the perturbation of $\bb v_k$ by edge $\{i, j\}$ is negligible if $|\bb a_{i,j}^T \bb v_k|$ is small.
Similarly, if $|\bb a_{i,j}^T \bb v_l|$ is small, then the perturbation of other eigenvectors in the direction of $\bb v_l$ due to the edge $\{i, j\}$ is small. Finally, a small eigen-gap $\lambda_k - \lambda_{k-1}$ or $\lambda_{k+1} - \lambda_k$ means that the Rayleigh quotient is relatively flat between the two critical points, and hence the critical point may be changed easily via a perturbation. We evaluate the suitability of this approximation to our method in  Sec.~\ref{subsec:ges_results}.

\section{Empirical Evaluation}
\label{sec:gsp_results}
We perform an empirical evaluation of our \textsc{Ges} and \textsc{Ilecir} methods, the details of which are discussed in Sec.~\ref{subsec:ges_results} and \ref{subsec:ilecir_exp} respectively.

\subsection{Greedy Edge Selection on synthetic graphs}
\label{subsec:ges_results}
\begin{figure}[t]
    \centering
    \includegraphics[width=0.49\linewidth]{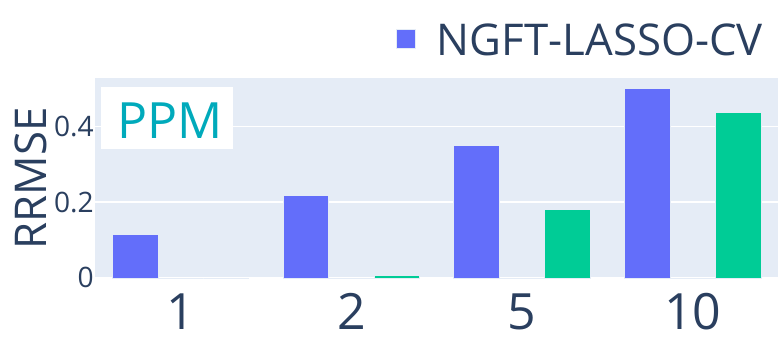}\hfill%
    \includegraphics[width=0.49\linewidth]{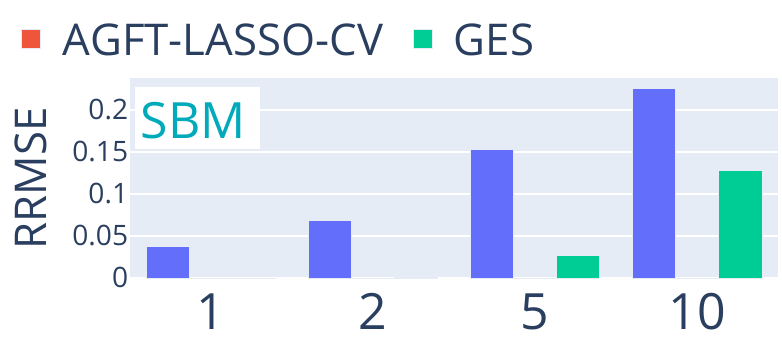}\\%
    \includegraphics[width=0.49\linewidth]{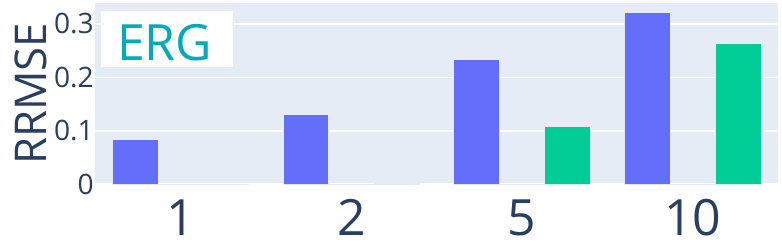}\hfill%
    \includegraphics[width=0.49\linewidth]{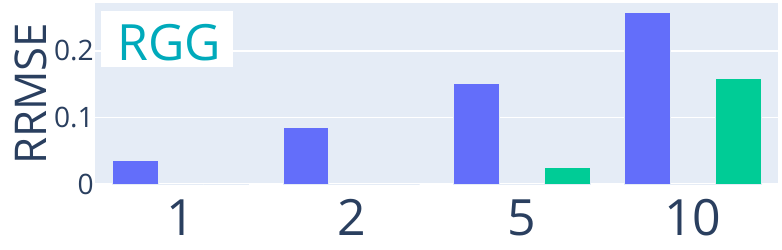}\\%
    \includegraphics[width=0.49\linewidth]{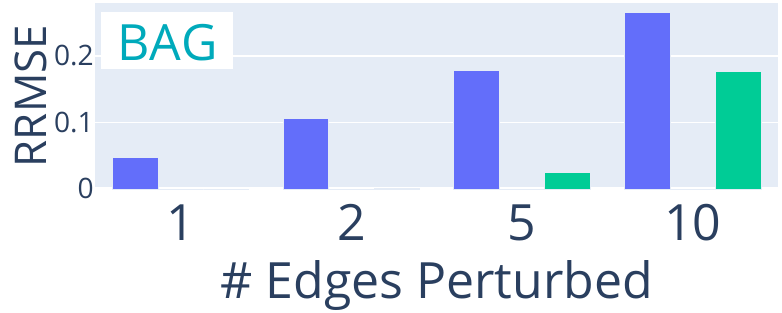}\hfill%
    \includegraphics[width=0.49\linewidth]{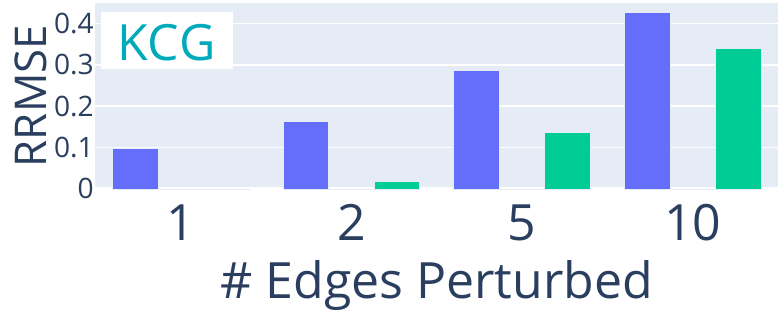}
    \caption[RRMSE of signal recovered via Greedy Edge Selection (\textsc{Ges})]{RRMSE of signal recovered from noiseless measurements ($m = 50, n = 100$ except for KCG where $m = 17, n = 34$) via Greedy Edge Selection (\textsc{Ges}), \textsc{Lasso} with the nominal graph (\textsc{Ngft-Lasso-Cv}), and \textsc{Lasso} with the actual graph (\textsc{Agft-Lasso-Cv}), for various number of perturbed edges.
    Some bars are not visible due to the value being close to zero.
    }
    \label{fig:ges_rmse}
\end{figure}

\begin{figure}[t]
    \centering
    \includegraphics[width=0.49\linewidth]{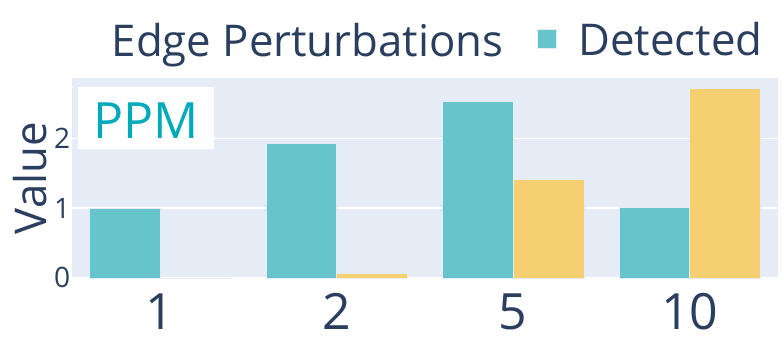}\hfill%
    \includegraphics[width=0.49\linewidth]{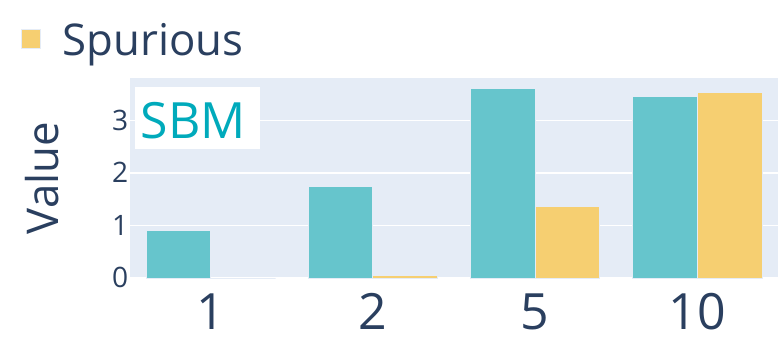}\\%
    \includegraphics[width=0.49\linewidth]{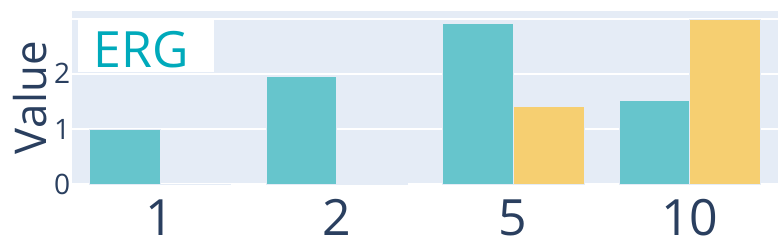}\hfill%
    \includegraphics[width=0.49\linewidth]{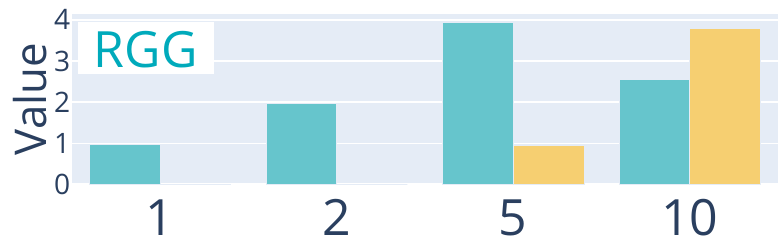}\\%
    \includegraphics[width=0.49\linewidth]{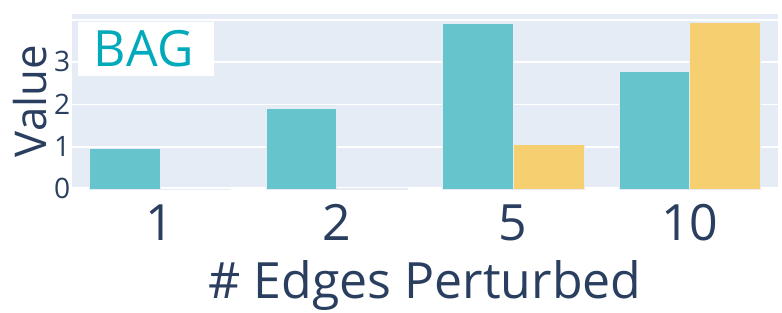}\hfill%
    \includegraphics[width=0.49\linewidth]{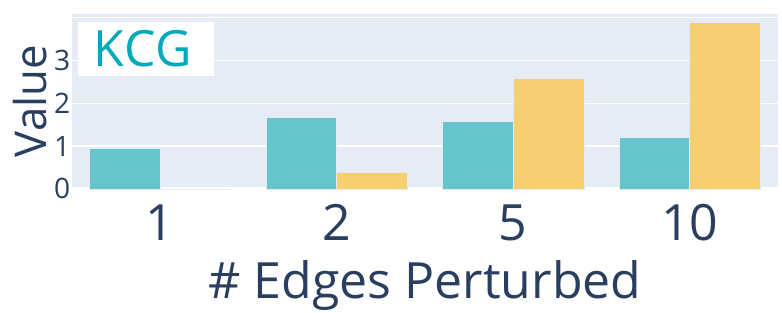}
    \caption[\textsc{Ges} edge perturbation recovery performance]{The average number of edge perturbations successfully detected and the average number of spurious edge perturbations reported by \textsc{Ges}, for various number of perturbed edges on different graph models ($m = 50, n = 100$ except for KCG where $m = 17, n = 34$).}
    \label{fig:ges_edges_detected}
\end{figure}

\noindent\textbf{Experiment Setup:}
{
We test \textsc{Ges} on the following graphs commonly used in the network science literature:  Planted Partition Model (PPM), Stochastic Block Model (SBM), Erd\H{o}s–R\'{e}nyi (ER) Graph, Random Geometric Graph (RGG), Barabasi-Albert (BA) Graph and Karate Club graph -- all are random graph models, other than the Karate Club graph, which is a two-community social network having $n=34$ nodes. More details regarding the random graph models can be found in the supplemental material in Sec.~\ref*{sec:details_RGM}. 
The actual graphs used in our experiments are instantiated with $n=100$ nodes for the random graph models.
The PPMs have $l = 5$ communities, each of size $r=20$, with $p=0.9$ and $q=0.01$ (intra-cluster and inter-cluster edge probabilities, respectively), with the expected number of edges being $895$.
The SBMs have $l=5$ communities with sizes $\{r_1,\dots,r_5\} = \{5, 10, 20, 25, 40\}$, and the probabilities $p,q$ are set to be the same as the PPM. The ER Graphs have $p = \frac{895}{4950}$.
The RGGs have $r = 0.27$.
The BA graphs have $r=10$.
The graph parameters are chosen so that the expected number of edges of the graphs are roughly equal, except the SBM, for which the edge probabilities are matched with the PPM.

Nominal graphs are generated by perturbing $d \in \{1, 2, 5, 10\}$ edges of the actual graph.
The set of edges to be perturbed -- known as the perturbation set $\omega$, with $|\omega| = d$ -- is chosen to be a subset of a prior set $\Omega$ of $100$ ``potentially faulty edges'' decided at the time of data generation. For each graph, all $n \choose 2$ possible edges are assigned to different categories, and an approximately equal number of edges from each of these categories are sampled to form $\Omega$.
The edge categories are based on presence/absence of an edge in the actual graph, whether the edge is inter-cluster or intra-cluster for community-structured graphs, and whether the endpoints of the edge are both low-degree or both high-degree or one low-degree and one high-degree node for the BA graph, making a total of $2$ categories for ER graphs and RGGs, $4$ for PPMs, SBMs and the Karate Club graph, and $6$ for the BA graph.

We generate $10$ instances of each graph type described above, with $10$ sparse-spectrum graph signals for each graph, making it a total of $N = 100$ sparse-spectrum signals for each graph type.
The signals are linear combinations of $s = 5$ ($s=2$ for Karate Club graph) eigenvectors (chosen randomly for each signal) of the Laplacian matrices of the graphs.
We use a measurement matrix (of size $50\times 100$ or $17\times 34$) whose entries are i.i.d.\ $\mc N(0, 1)$.
Furthermore, i.i.d.\ $\mc N(0, \sigma^2)$ noise is added to each measurement, with $\sigma = \beta \frac{\|\bb \Phi \bb x^*\|_1}{m}$, with $\beta \in \{0, 0.01, 0.02, 0.05\}$.
Corresponding to each signal, a nominal graph is generated as described above.

The algorithm \textsc{Ges} is compared with the baseline algorithm \textsc{NGft-Lasso-Cv} (Eqn.~\ref{eqn:recovery_nominal}) and the ideal algorithm \textsc{AGft-Lasso-CV} (Eqn.~\ref{eqn:recovery_actual}) using the RRMSE (Relative Root Mean-Square Error) metric, defined as 
$\mathtt{RRMSE}(\hat{\bb x}) = \frac{\|\hat{\bb x} - \bb x^*\|_2}{\|\bb x^*\|_2}$,
where $\hat{\bb x}$ is the estimated signal, and $\bb x^*$ is the ground-truth signal.
The set $\Gamma := \{10^{-3+\frac{6}{19}t} : t \in \{0, 1, 2,\dots, 19\}\}$ (i.e. 20 values between $0.001$ and $1000$, evenly spaced in logarithm) for all algorithms.
For \textsc{Ges}, the loss improvement factor is $\tau = 0.99$, and the max iterations $d_0 = 2d$, for each $d\in \{1, 2, 5, 10\}$ (in practice, the algorithm stops much earlier than the max number of iterations due the CV error increasing in an iteration or being close to the noise variance, which is also evidenced by the fact that the total number of edge perturbations reported by the algorithm (Fig.~\ref{fig:ges_edges_detected}) is in general close to $d$ and much smaller than $d_0$.)

The \textsc{Ges} algorithm is run with the following practical modifications:
(1) grid search over $\Gamma$ to estimate the \textsc{Lasso} regularization parameter $\mu$ is performed only for the candidate graph at the end of a step and is re-used for the perturbed graphs in the next step in order to speed up the experiments;
(2) $K$-fold (or multi-fold) cross-validation (with $K=5$) is used instead of single-fold CV to improve accuracy since the total number of available measurements is small -- in $K$-fold cross-validation, the CV error is computed $K$ times, each time with a different subset of the $m$ measurements, each subset of size $\lfloor \frac{m}{K} \rfloor$ (with reconstruction done on the remaining $m - \lfloor \frac{m}{K} \rfloor$ measurements), and the sum of the $K$ CV errors is returned;
and (3) the prior set of possible faulty edges $\Omega$ (of size $|\Omega| = 100$ as specified earlier) is provided to the \textsc{Ges} so that only these edge perturbations are tried, which speeds up the experiments.
The latter is justified since in real-world applications, such prior information may be available.
For example, in a contact tracing application used in epidemiology, two people may have come in contact if their phones use the same WiFi access points, even though Bluetooth-based contact tracing may suggest otherwise (e.g. by being located in the same office room at a far-enough distance).

}
\noindent\textbf{Results:}
\label{subsubsec:ges_results}
The RRMSE of the Greedy Edge Selection (\textsc{Ges}) algorithm for recovery of sparse-spectrum signals from noiseless measurements ($\sigma = \beta=0$) is presented in Fig.~\ref{fig:ges_rmse}, for each graph model mentioned earlier, with $d = \{1, 2, 5, 10\}$ edges perturbed to create the nominal graph.
We see that for each graph type, \textsc{Ges} outperforms the baseline method of recovery using the nominal graph, i.e. \textsc{Ngft-Lasso}.
As expected, the RRMSE increases as the number of perturbations are increased.
\begin{figure}[t]
    \centering
    \vspace{-1mm}
    \includegraphics[width=0.8\linewidth]{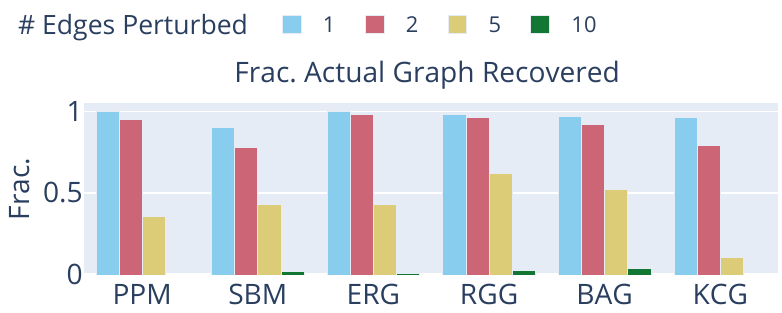}%
    \caption[Fraction of cases in which actual graph was recovered by \textsc{Ges}]{Fraction of cases in which actual graph was recovered by \textsc{Ges}, for various number of perturbed edges, for $m = 50, n = 100$ except for KCG where $m = 17, n = 34$.}
    \vspace{-3mm}
    \label{fig:ges_graph_recovery}
\end{figure} 
For one or two edge perturbations, the RRMSE is close to zero or very small, whereas for upto five perturbations, the RRMSE is at most half of that obtained using \textsc{Ngft-Lasso}, and often much lower than that.
Even for ten edge perturbations, the RRMSE of \textsc{Ges} is significantly lower than \textsc{Ngft-Lasso}. 
Recovery using the actual graph gives an RRMSE close to zero, and hence the bars for it are not visible in Fig.~\ref{fig:ges_rmse}.

Fig.~\ref{fig:ges_graph_recovery} presents the fraction of cases in which \textsc{Ges} successfully recovers (i.e. recovers with \emph{zero} error) the actual graph from the nominal graph. We see that for most graph types, the actual graph is recovered in close to $100\%$ cases when up to two edges are perturbed.
Despite being a greedy algorithm, the actual graph gets recovered in a significant fraction of cases even when five edges perturbations were performed. In some cases the actual graph is recoverable via \textsc{Ges} even if ten edge perturbations were induced.
Fig.~\ref{fig:ges_edges_detected} shows the average number of edge perturbations \emph{correctly} detected by \textsc{Ges}, and the average number of spurious edge perturbations reported by it.
Close to $100\%$ of edge perturbations are correctly detected for upto two edge perturbations made to the actual graph, and around $50\%$ to $80\%$ for most graphs when five edge perturbations are made.
The number of spurious edge perturbations reported by \textsc{Ges} is close to zero for upto two edge perturbations, and small for five edge perturbations.
With ten edge perturbations, the number of edge perturbations correctly detected drops and the number of spurious edge perturbations reported increases.
But it is worth noting that the problem of edge perturbation recovery from \emph{compressive measurements} of a graph signal is quite challenging due to the eigenvectors of the nominal graph being significantly perturbed for even a small number of edge perturbations.
For comparison, the work in \cite{barbarossa2020} considers recovery of upto four edge perturbations from a known, \emph{uncompressed} graph signal with Gaussian priors. For analysis of edge recovery by category (intra- or inter-cluster), see Sec.~\ref*{sec:recovery_analysis_edges} of the supplemental material. 

\begin{figure}[t]
    \centering
    \vspace{-0.5mm}
    \includegraphics[width=0.4\linewidth]{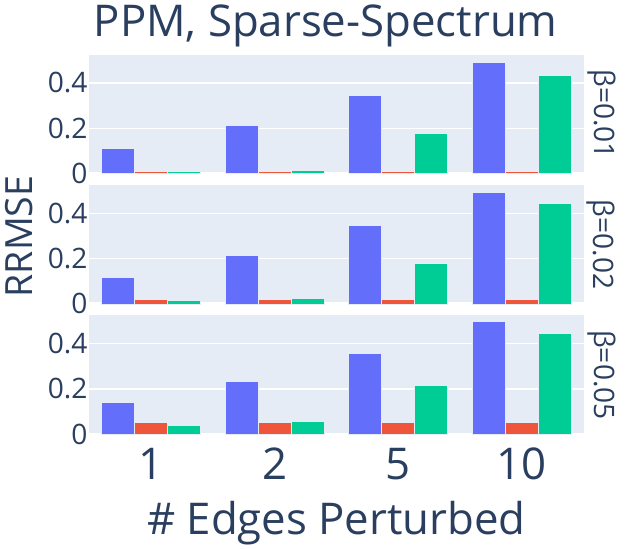}%
    \includegraphics[width=0.6\linewidth]{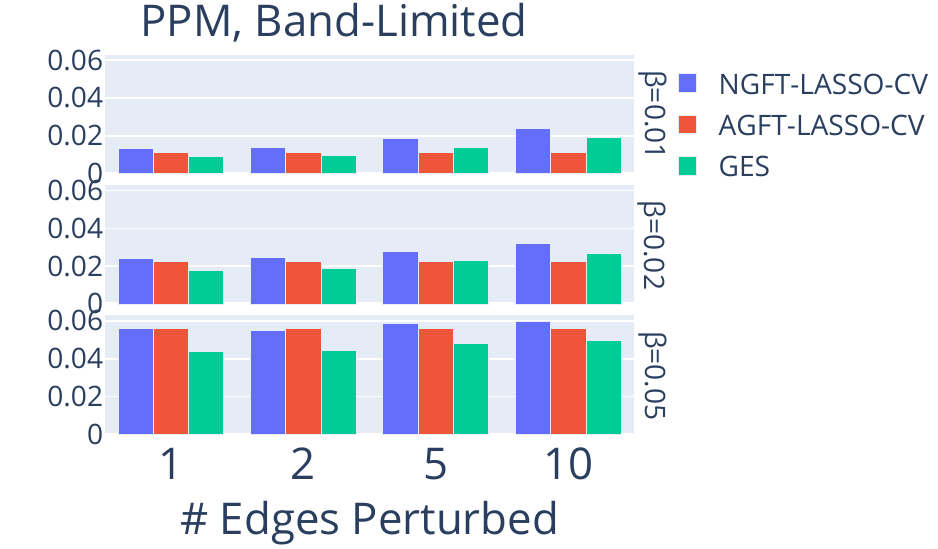}
    \caption{Performance of \textsc{Ges} with measurement noise of different levels given by $\beta \in \{0.01,0.02,0.05\}$ ($m = 50, n = 100$).}
    \vspace{-5mm}
    \label{fig:ges_noisy}
\end{figure}
The performance of \textsc{Ges} with noisy measurements is shown in Fig.~\ref{fig:ges_noisy}, for the PPM.
We find that the \textsc{Ges} algorithm performs well even in the presence of noise.
Its RRMSE is significantly less than the RRMSE of \textsc{NGft-Lasso-Cv} for upto $5$ edge perturbations, and is comparable to that of \textsc{AGft-Lasso-Cv} when the number of edge perturbations is $1$ or $2$.
Occasionally, we find that \textsc{Ges} is able to do slightly better than \textsc{Agft-Lasso-Cv}.
We suspect that this is because only $20$ different values were tried during grid search for the \textsc{Lasso} regularization parameter $\mu$, due to which \textsc{Ges} could find a better graph than the actual graph for the given values of $\mu$ for the purpose of signal recovery. A more fine-grained grid search would lead to the expected behaviour of \textsc{AGft-Lasso-Cv} performing better than \textsc{Ges}.

\begin{figure}[t]
    \centering
    \includegraphics[width=0.36\linewidth]{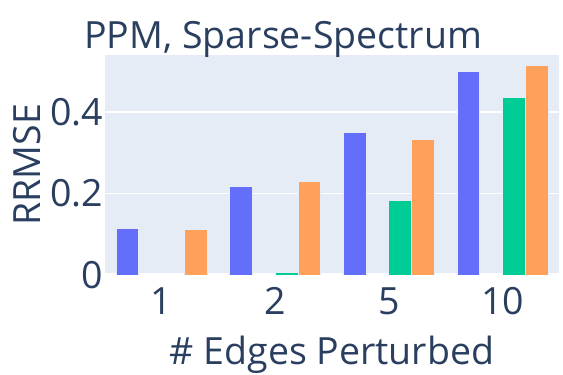}\hfill%
    \includegraphics[width=0.60\linewidth]{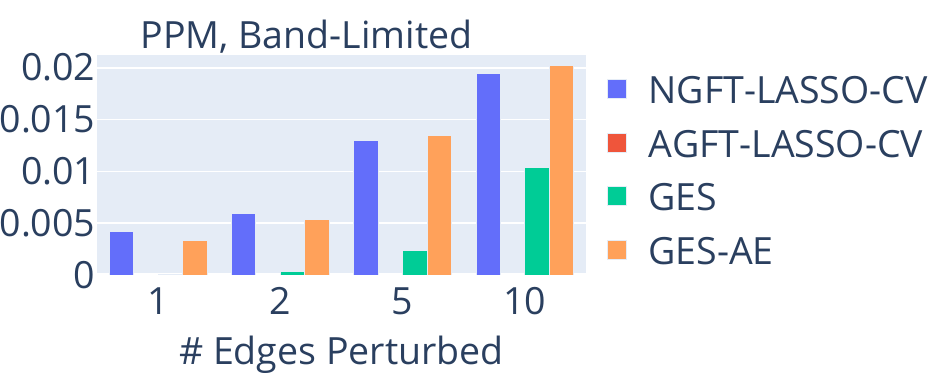}
    \caption[Performance of \textsc{Ges} with approximated eigenvectors (\textsc{Ges-Ae})]{Performance of Greedy Edge Selection with approximated eigenvectors (\textsc{Ges-Ae}) compared to that of \textsc{Ges}, \textsc{Agft-Lasso-Cv}, and \textsc{Ngft-Lasso-Cv} for sparse-spectrum (left) and bandlimited (right) signals, all with $m = 50, n = 100$. 
    In some cases where the error for \textsc{Ges}, \textsc{Agft-Lasso-Cv} is close to 0, the error bars are not visible.
    }
    \vspace{-3mm}
    \label{fig:ges_ceci_barbarossa}
\end{figure}

\noindent\textbf{Using closed-form approximations for eigenvector perturbation:}
Fig.~\ref{fig:ges_ceci_barbarossa} shows the RRMSE obtained using the \textsc{Ges-Ae} algorithm, in which the eigenvector perturbation approximation \cite{Ceci2020total} (Eqn.~\ref{eqn:eigvec_perturbation_formula}) is used in place of eigendecomposition in \textsc{Ges} (see Sec.~\ref{subsec:ges_ceci_barbarossa}),
for compressive recovery of sparse-spectrum and band-limited signals on PPMs with a few edge perturbations.
We find that for both sparse-spectrum and band-limited signals, the RRMSE of \textsc{Ges-Ae} is significantly higher than that for \textsc{Ges}, across a range of edge perturbations.
Moreover, the RRMSE of \textsc{Ges-Ae} is close to the RRMSE obtained with the nominal graph, and sometimes slightly worse. Hence we do not find the eigenvector perturbation approximation (a first order approximation) to be useful when applied to greedy edge selection with cross-validation.
Perhaps a better approximation with higher order terms in it might be more useful. A major reason for the performance of \textsc{Ges-Ae} could be that the condition under which the approximation is applicable -- given by Eqn.~\ref{eqn:gsp_barbarossa_eigengap_condition} -- gets violated often. We deem the condition to be violated for an edge $\{i, j\}$ and the $k^\text{th}$ eigenvector $\bb v_k$ if $(\bb v_k(i) - \bb v_k(j))^2 > 4 (\lambda_{k+1} - \lambda_k))$  for edge addition or $(\bb v_k(i) - \bb v_k(j))^2 > 4 (\lambda_k - \lambda_{k-1})$ in case of edge deletion. In our experiments, the fraction of times this condition was violated for at least one eigenvector out of the first five (i.e. the band-limited eigenvectors) for a PPM graph with $n=100, l=5, p = 0.9, q = 0.01$ was $87.7\%$, if an inter-cluster edge is perturbed.
Similarly, we found that the condition in Eqn.~\ref{eqn:gsp_barbarossa_eigengap_condition} was violated $74.7\%$ of the times for at least one eigenvector out of five for a random choice of five eigenvectors and any kind of edge.
While this condition does not get violated for intra-cluster edges in the band-limited case, the amount of perturbation in the low-frequency eigenvectors by such edges is very small, such that the RRMSE for these cases is close to zero even for \textsc{NGft-Lasso-Cv}, and hence these cases do not contribute much to the average RRMSE.

\subsection{Inferred Linear-Edge Compressive Image Recovery}
\label{subsec:ilecir_exp}

\noindent\textbf{Experiment Setup:}
For evaluation of \textsc{Ilecir}, we simulate the patch-wise compressive acquisition of $10$ images each from the following datasets:
(1) A synthetic dataset we created, containing piece-wise smooth images in the form of a union of regions individually containing intensity values expressed by polynomials of $x,y$; (2) The Berkeley segmentation dataset\footnote{\url{https://www2.eecs.berkeley.edu/Research/Projects/CS/vision/bsds/}}, a set of natural images used in image segmentation tasks; (3) The Tom and Jerry dataset\footnote{\url{https://www.kaggle.com/datasets/balabaskar/tom-and-jerry-image-classification}}, a dataset of still frames from the popular Tom and Jerry cartoon; and (4) The NYU Depth dataset\footnote{\url{https://cs.nyu.edu/~silberman/datasets/nyu_depth_v2.html}}, a dataset of depth-map images of indoor scenes.
While ground-truth segmentation is available for the synthetic dataset, the available segmentations for the other datasets are either human-labelled or machine-generated. In each dataset, the segmentation information is used to construct the actual graph, by dropping the edges of the lattice graph which connect pixels in two different segments of a patch. 

Patch-wise compressive image acquisition is simulated for an image in the following manner.
The image is divided into non-overlapping patches of size $8\times 8$.
Each patch is arranged in row-major order to produce a vector of $n = 64$ dimensions, which forms the signal $\bb x^*$.
Measurements are generated using Eqn.~\ref{eq:cs_measurement}, with a $m \times n$ measurement matrix whose entries are i.i.d.\ $\mc N(0, 1)$, with $m \in \{20, 30, 40\}$.
Furthermore, i.i.d.\ $\mc N(0, \sigma^2)$ noise is added to each measurement, with $\sigma = \beta \frac{\|\bb \Phi \bb x^*\|_1}{m}$, with $\beta \in \{0, 0.01, 0.02, 0.05\}$.
These measurements are then decoded using the \textsc{Ilecir} algorithm, and the $8\times 8$ patch is estimated.
For gray-scale images, the estimated values are clipped to be between $0$ and $255$ and rounded to the nearest integer.
For depth maps, negative values are set to $0$.
The recovered patches are stitched together to reconstruct the image.
The results of \textsc{Ilecir} are compared to those with DCT-based recovery (\textsc{Dct-Lasso-Cv}), and using the GFT of the ground-truth segmentation (\textsc{SegGFT-Lasso-Cv}) in terms of the RRMSE and Structural Similarity Index Measure (SSIM)\footnote{\url{https://en.wikipedia.org/wiki/Structural_similarity}} of the recovered images relative to the ground truth.
For the algorithms, $\Gamma \triangleq \{10^{-3+\frac{6}{19}t} : t \in \{0, 1, 2,\dots, 19\}\}$, $K=5$ folds are used for cross-validation, and the loss improvement factor $\tau$ is set to 0.99. To speed up \textsc{Ilecir}, grid search over $\Gamma$ is performed only for DCT-based recovery, and the same value of $\mu$ is re-used for recovery with the image-edge partitioned graphs.
However, after an image-edge is chosen, $\mu$ is re-computed using CV to yield the final patch estimate.

\begin{figure}[t]
\setlength{\gridcolwidth}{.47\linewidth}
\settoheight{\gridrowheight}{\includegraphics[width=\gridcolwidth]{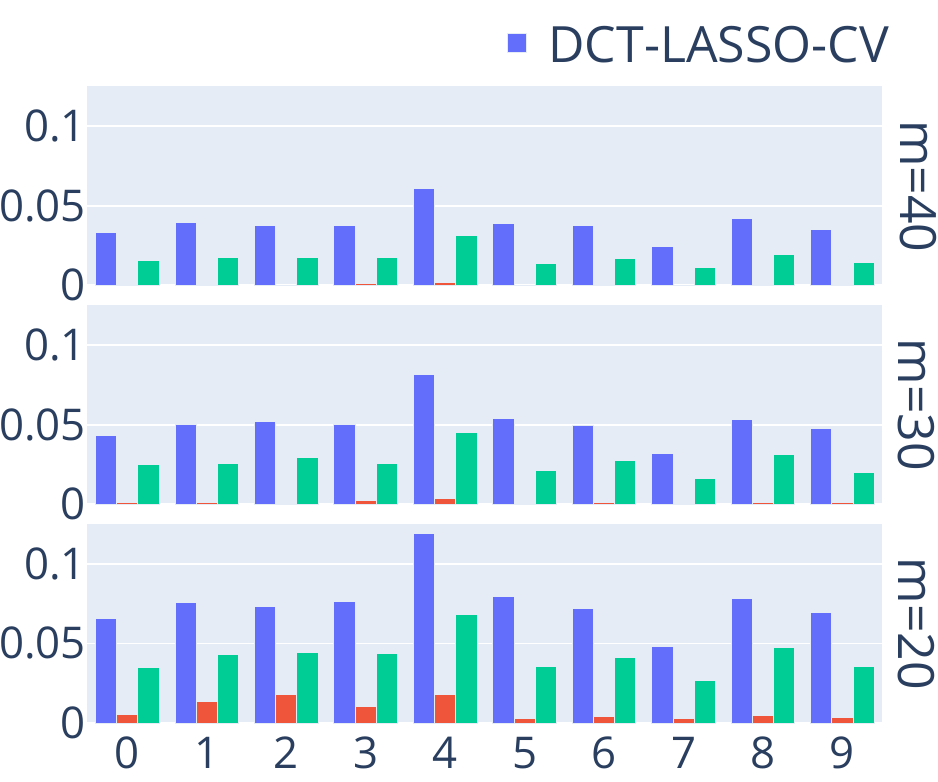}}%
    \centering
    \hspace{\baselineskip}
    \columnname{\textbf{RRMSE}}\hfill%
    \columnname{\textbf{SSIM}}\\
    \vspace{1mm}
    \rowname{\scriptsize Synthetic}
    \begin{subfigure}[b]{\gridcolwidth}
        \includegraphics[width=\linewidth]{figs/ilecia/synthetic_bar_chart_rmse.pdf}
    \end{subfigure}\hfill%
    \begin{subfigure}[b]{\gridcolwidth}
        \includegraphics[width=\linewidth]{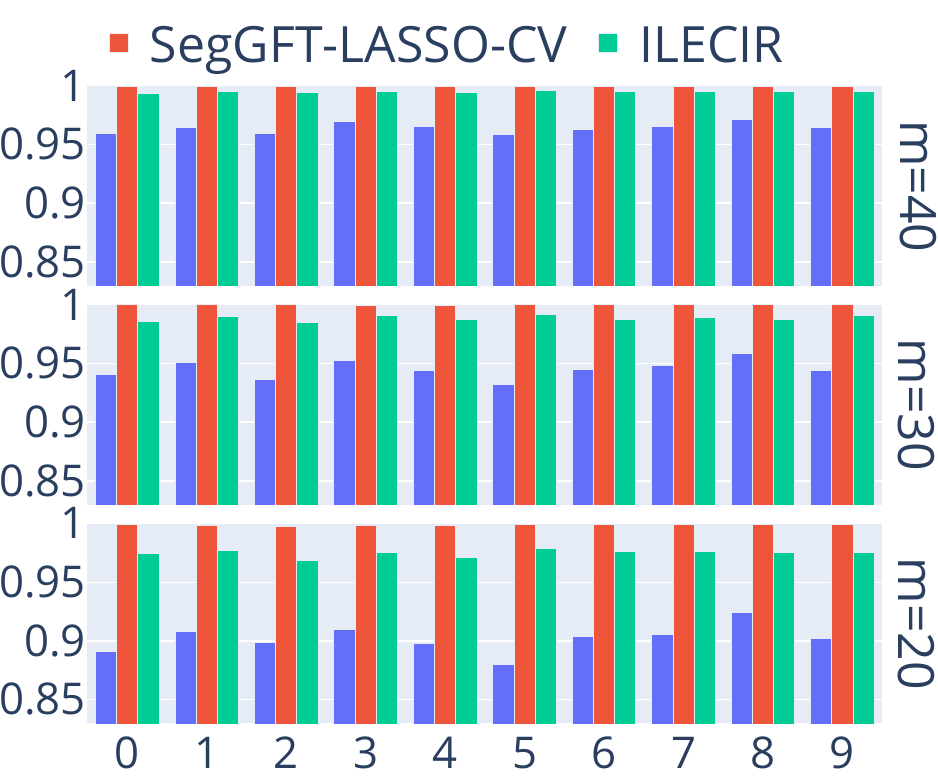}
    \end{subfigure}\\
    \settoheight{\gridrowheight}{
        \includegraphics[width=\gridcolwidth]{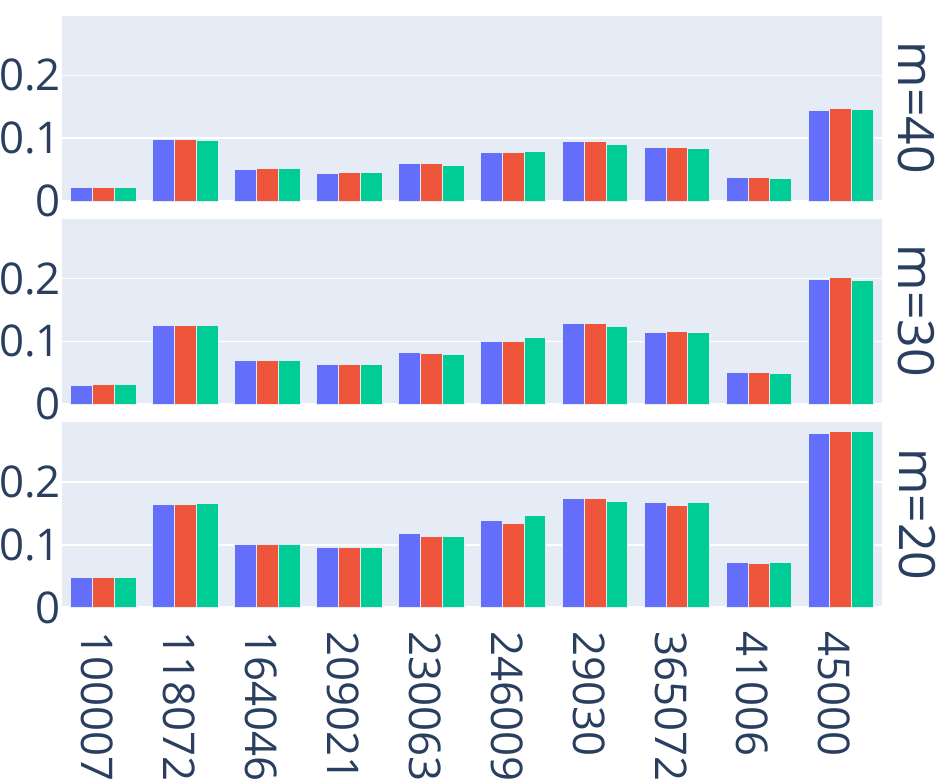}
    }
    \rowname{\scriptsize Berkeley}
    \begin{subfigure}[b]{\gridcolwidth}
        \includegraphics[width=\linewidth]{figs/ilecia/berkeley_bar_chart_rmse.pdf}
    \end{subfigure}\hfill%
    \begin{subfigure}[b]{\gridcolwidth}
        \includegraphics[width=\linewidth]{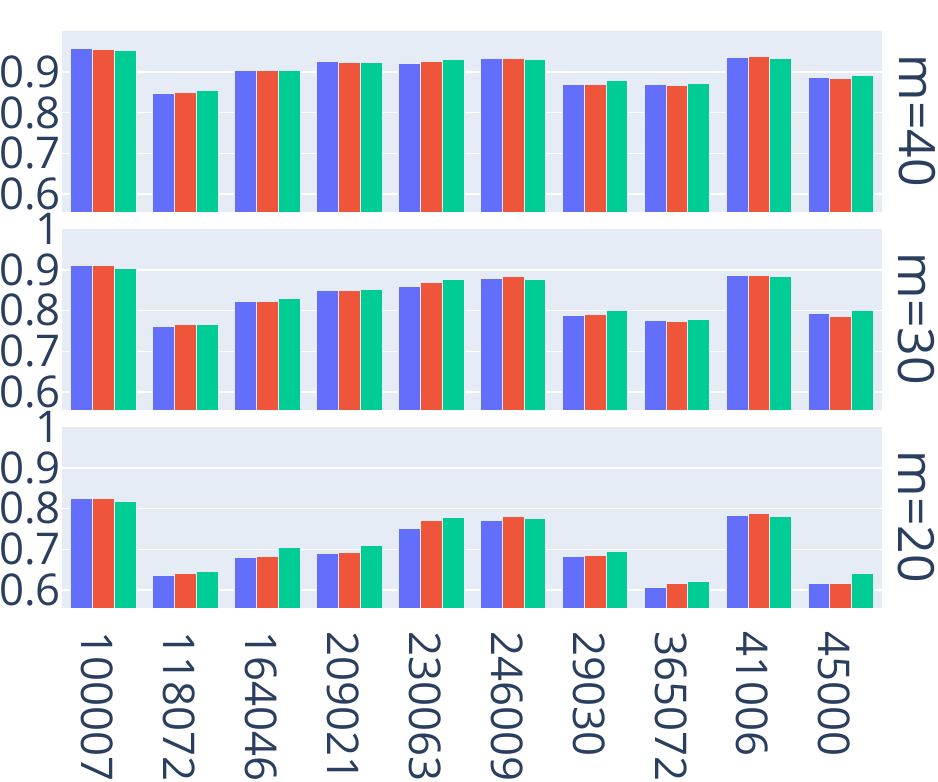}
    \end{subfigure}\\
    \settoheight{\gridrowheight}{
        \includegraphics[width=\gridcolwidth]{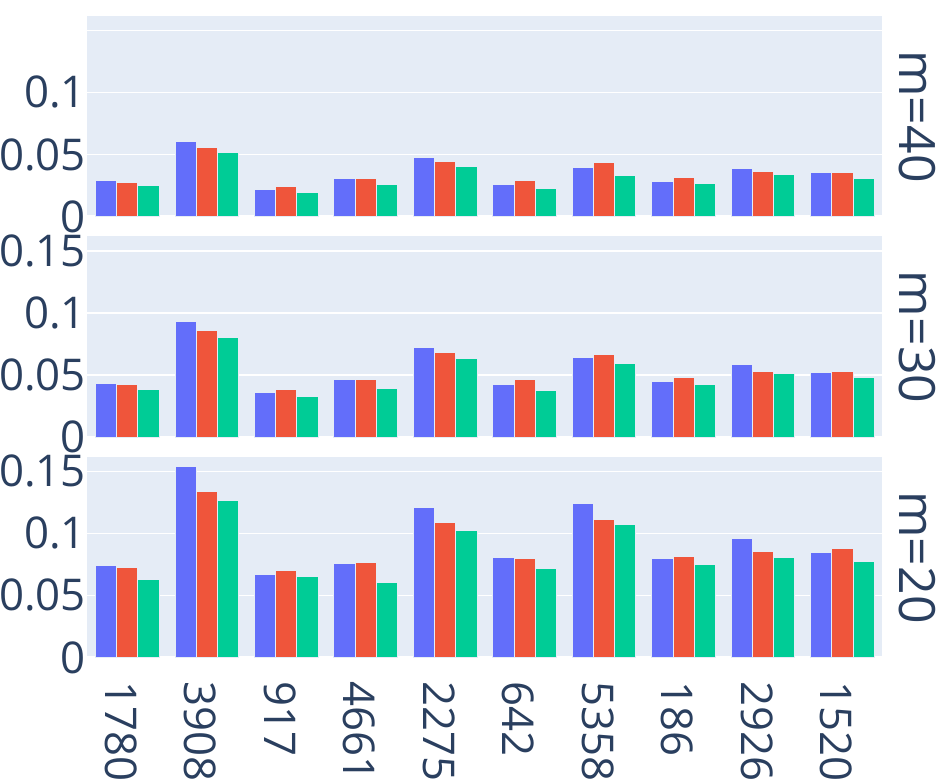}
    }
    \rowname{\scriptsize Tom and Jerry}
    \includegraphics[width=\gridcolwidth]{figs/ilecia/tom_and_jerry_bar_chart_rmse.pdf}\hfill%
    \includegraphics[width=\gridcolwidth]{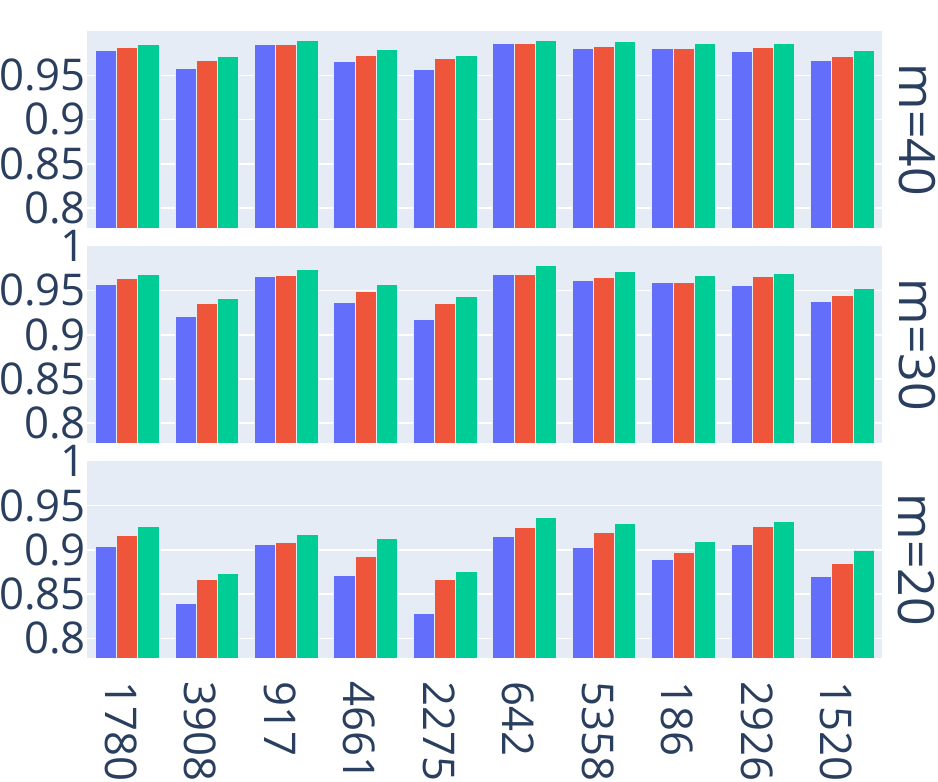}\\
    \settoheight{\gridrowheight}{
        \includegraphics[width=\gridcolwidth]{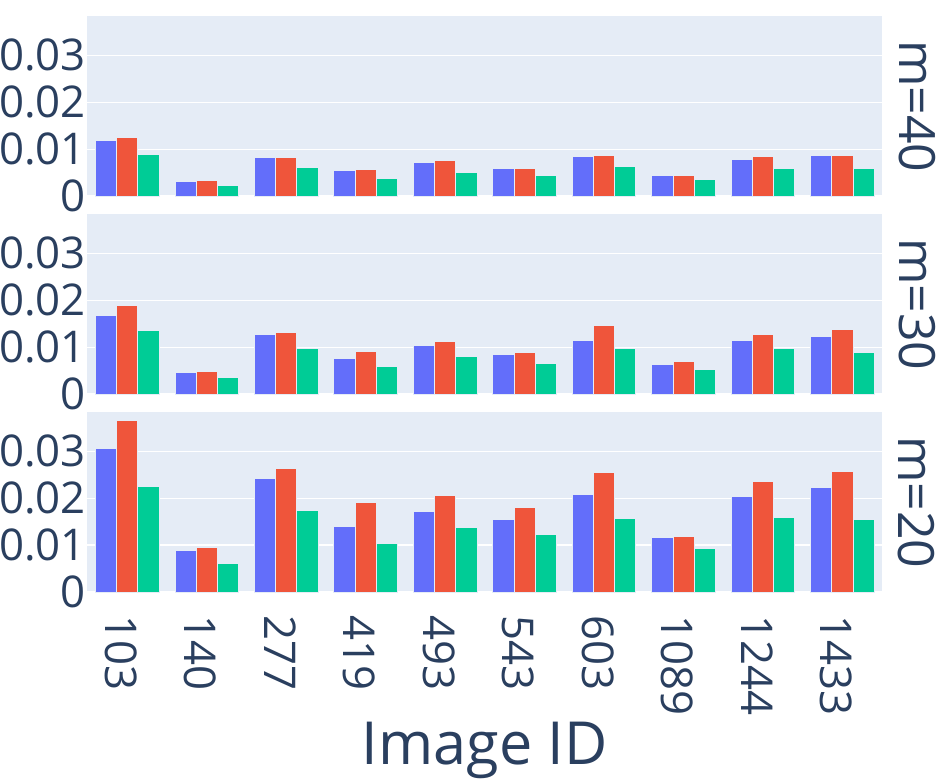}
    }
    \rowname{\scriptsize NYU Depth}
    \includegraphics[width=\gridcolwidth]{figs/ilecia/nyu_depth_bar_chart_rmse.pdf}\hfill%
    \includegraphics[width=\gridcolwidth]{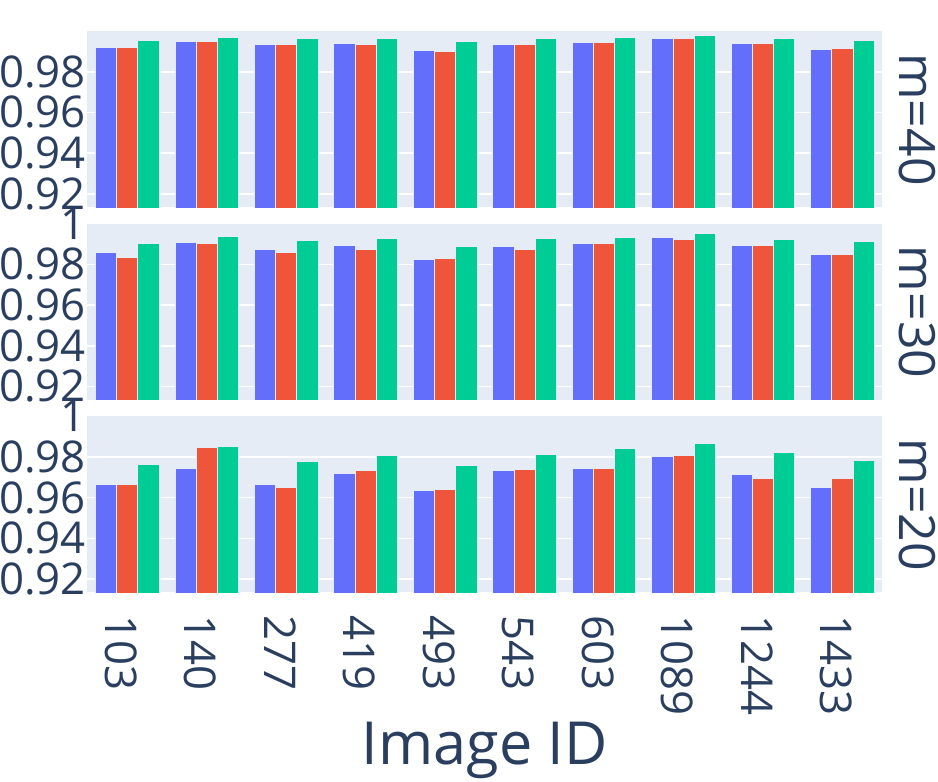}\\
    \caption[RRMSE and SSIM for compressive image reconstruction with $m \in  \{20,30,40\}$ using \textsc{Ilecir}]{RRMSE and SSIM using \textsc{Ilecir}, \textsc{Dct-Lasso-Cv} and \textsc{SegGft-Lasso-Cv} (Segmentation-aware recovery) in the noiseless regime, with $n = 64$ for $8 \times 8$ patches.}
    \vspace{-5mm}
    \label{fig:ilecir_rmse_ssim}
\end{figure}

\noindent\textbf{Quantitative comparison in the noiseless setting:} 
Fig.~\ref{fig:ilecir_rmse_ssim} shows the RRMSE and SSIM of images recovered using \textsc{Ilecir}, \textsc{Dct-Lasso-Cv} and \textsc{SegGft-Lasso-Cv} from $m\in\{20, 30, 40\}$ linear measurements, for 10 images each from the aforementioned datasets.
We note that \textsc{Ilecir} substantially improves over \textsc{Dct-Lasso-Cv} in terms of the RRMSE and SSIM of the recovered images on the Synthetic and the NYU Depth datasets.
It also shows some improvement for the images in the Tom and Jerry dataset. For Tom and Jerry and the NYU Depth datasets, the improvement is most pronounced when $m=20$ linear measurements are used. Such improvement is expected since these are piece-wise smooth images -- while the smooth regions are accurately recovered by \textsc{Dct-Lasso-Cv}, the patches containing sharp edges are not recovered very accurately. \textsc{Ilecir} performs better on such patches, while maintaining the accuracy afforded by \textsc{Dct-Lasso-Cv} on smooth patches. The \textsc{SegGft-Lasso-Cv} method has near-perfect recovery for the Synthetic dataset. This substantiates the rationale behind using the GFT basis of the 2-D lattice graph partitioned according to the segments of a patch for recovery. Since \textsc{Ilecir} only models linear image edges and does not model more than one image edge in a patch, recovery using it is not perfect. Notably, \textsc{Ilecir} often performs better than the \textsc{SegGft-Lasso-Cv} method for the Tom and Jerry and NYU Depth Datasets.
This means that it is able to perform edge detection slightly better than the algorithm used to perform edge detection for the images from the Tom and Jerry dataset.
\begin{figure}[t]
\centering
\vspace{-1mm}
\begingroup
\captionsetup{justification=raggedright}
\begin{subfigure}[b]{0.16\linewidth}
        \centering
        \includegraphics[trim={5mm 0mm 8mm 14mm}, clip, width=\linewidth]{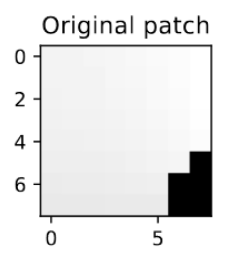}
        \label{subfig:nyu_orig_patch}
\end{subfigure}\hfill%
\begin{subfigure}[b]{0.16\linewidth}
        \centering
        \includegraphics[trim={5mm 0mm 8mm 15mm}, clip, width=\linewidth]{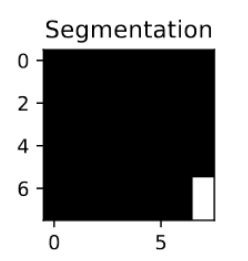}
        \label{subfig:nyu_orig_seg}
\end{subfigure}\hfill%
\begin{subfigure}[b]{0.16\linewidth}
        \centering
        \includegraphics[trim={11mm 0mm 15mm 11mm}, clip, width=\linewidth]{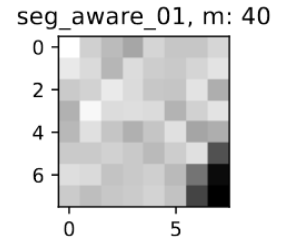}
        \label{subfig:nyu_seggft_patch}
\end{subfigure}\hfill%
\begin{subfigure}[b]{0.16\linewidth}
        \centering
        \includegraphics[trim={0mm 0mm 3mm 0mm}, clip, width=\linewidth]{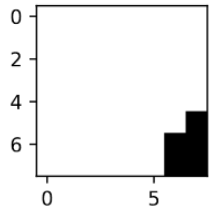}
        \label{subfig:nyu_ilecir_seg}
\end{subfigure}\hfill%
\begin{subfigure}[b]{0.16\linewidth}
        \centering
        \includegraphics[trim={0mm 0mm 10mm 0mm}, clip, width=\linewidth]{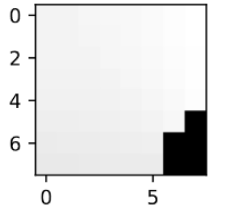}
        \label{subfig:nyu_ilecir_patch}
\end{subfigure}\hfill%
\begin{subfigure}[b]{0.16\linewidth}
        \centering
        \includegraphics[trim={4mm 0mm 5mm 10mm}, clip, width=\linewidth]{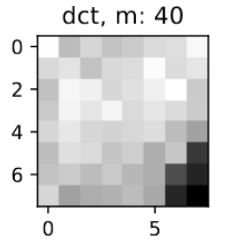}
        \label{subfig:nyu_dct_patch}
\end{subfigure}\hfill%
\endgroup
\vspace{-5mm}
    \caption
    {Left to right: An $8\times 8$ patch from an image in the NYU depth dataset (leftmost), its \emph{incorrectly} labelled segmentation (second from left), signal recovery by \textsc{SegGFT-Lasso-Cv} (poor quality), recovery of correction segmentation by
    \textsc{Ilecir}, recovery of patch by \textsc{Ilecir}, patch recovered via \textsc{Dct-Lasso-Cv}. Note: the two regions of the segmentation maps are colored arbitrarily.
    }
    \vspace{-5mm}
    \label{fig:nyu_segs}
\end{figure}
For the NYU Depth Dataset, we found that the provided segmentation maps were not properly aligned with the depth map images (see Fig.~\ref{fig:nyu_segs} for an example of an incorrectly labelled segmentation map of a patch from this dataset).
Due to this, the \textsc{SegGft-Lasso-Cv} method performs \emph{worse} than \textsc{Dct-Lasso-Cv} for the NYU Depth dataset.
This demonstrates a strength of the \textsc{Ilecir} method -- since the underlying segmentation is automatically inferred, it circumvents possible errors in the segmentation map if it were available.

RRMSE and SSIM of images recovered using \textsc{Ilecir} are higher than those recovered using \textsc{Dct-Lasso-Cv} for some images of the Berkeley dataset.
Since natural images contain vast regions of detailed texture and not just gradient of intensity, it is hard to improve upon the baseline \textsc{Dct-Lasso-Cv} method by modelling just a single linear edge.
This is also substantiated by the performance of the \textsc{SegGft-Lasso-Cv} method, which is similar to \textsc{Dct-Lasso-Cv} and \textsc{Ilecir}.
However, a visual inspection of the images recovered by \textsc{Ilecir}, \textsc{SegGft-Lasso-Cv} and \textsc{Dct-Lasso-Cv} would reveal that \textsc{Ilecir} and \textsc{SegGft-Lasso-Cv} perform better along the edges in these images (see later in this section).

\begin{figure}[t]
\setlength{\gridcolwidth}{.23\linewidth}
\settoheight{\gridrowheight}{\includegraphics[width=\gridcolwidth]{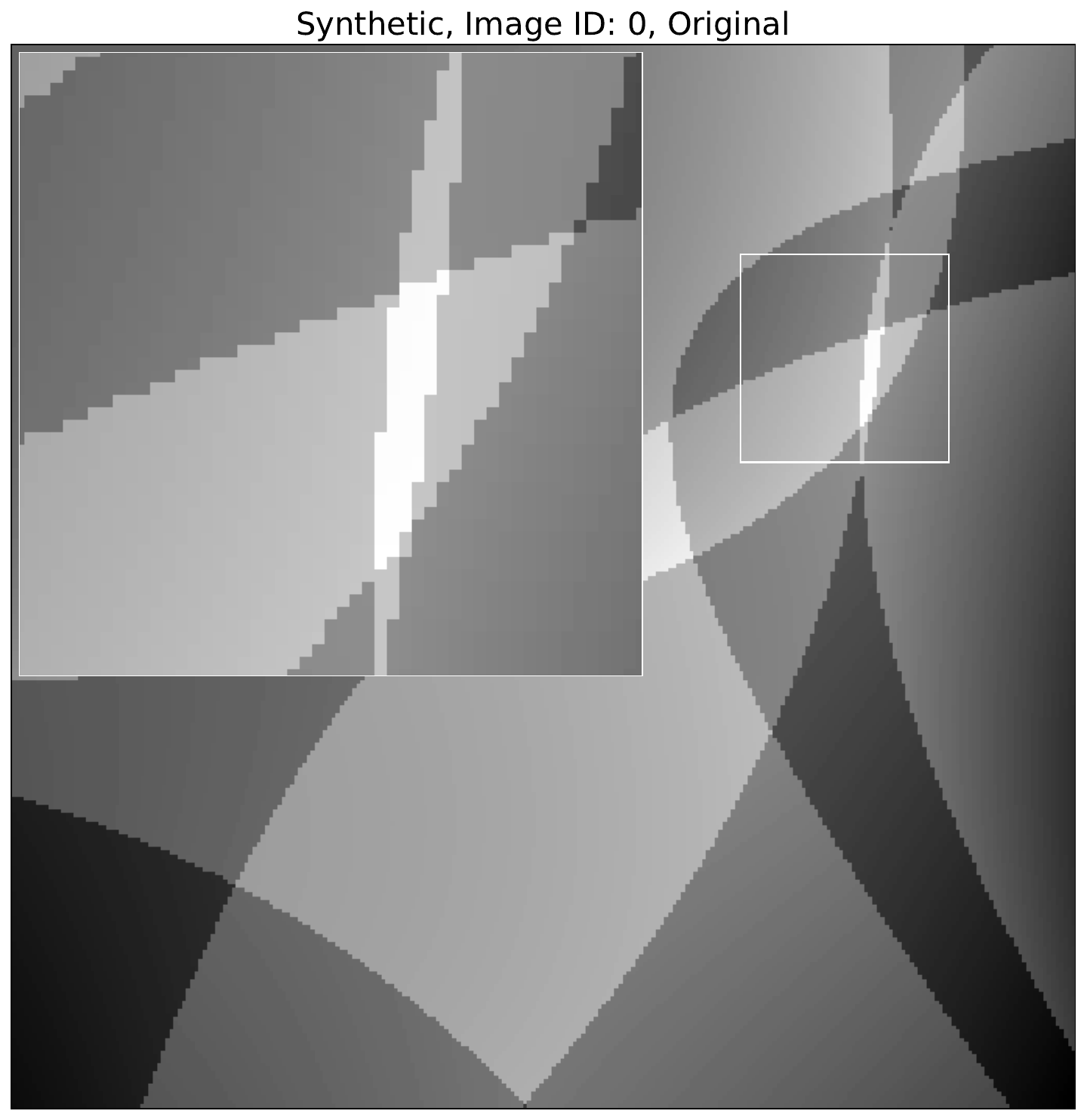}}%
    \centering
    \hspace{\baselineskip}
    \columnname{Original}\hfil%
    \columnname{\textsc{SegGFT}}\hfil%
    \columnname{\textsc{Dct}}\hfil%
    \columnname{\textsc{Ilecir}}\\
    \rowname{Synthetic}
    \begin{subfigure}[b]{\gridcolwidth}
        \includegraphics[trim={0 0 0 10mm},clip,width=\linewidth]{figs/zoomed/synthetic_noise_0/000/image_000_Original_m_40_zoomed_inset.pdf}
    \end{subfigure}\hfil%
    \begin{subfigure}[b]{\gridcolwidth}
        \includegraphics[trim={0 0 0 10mm},clip,width=\linewidth]{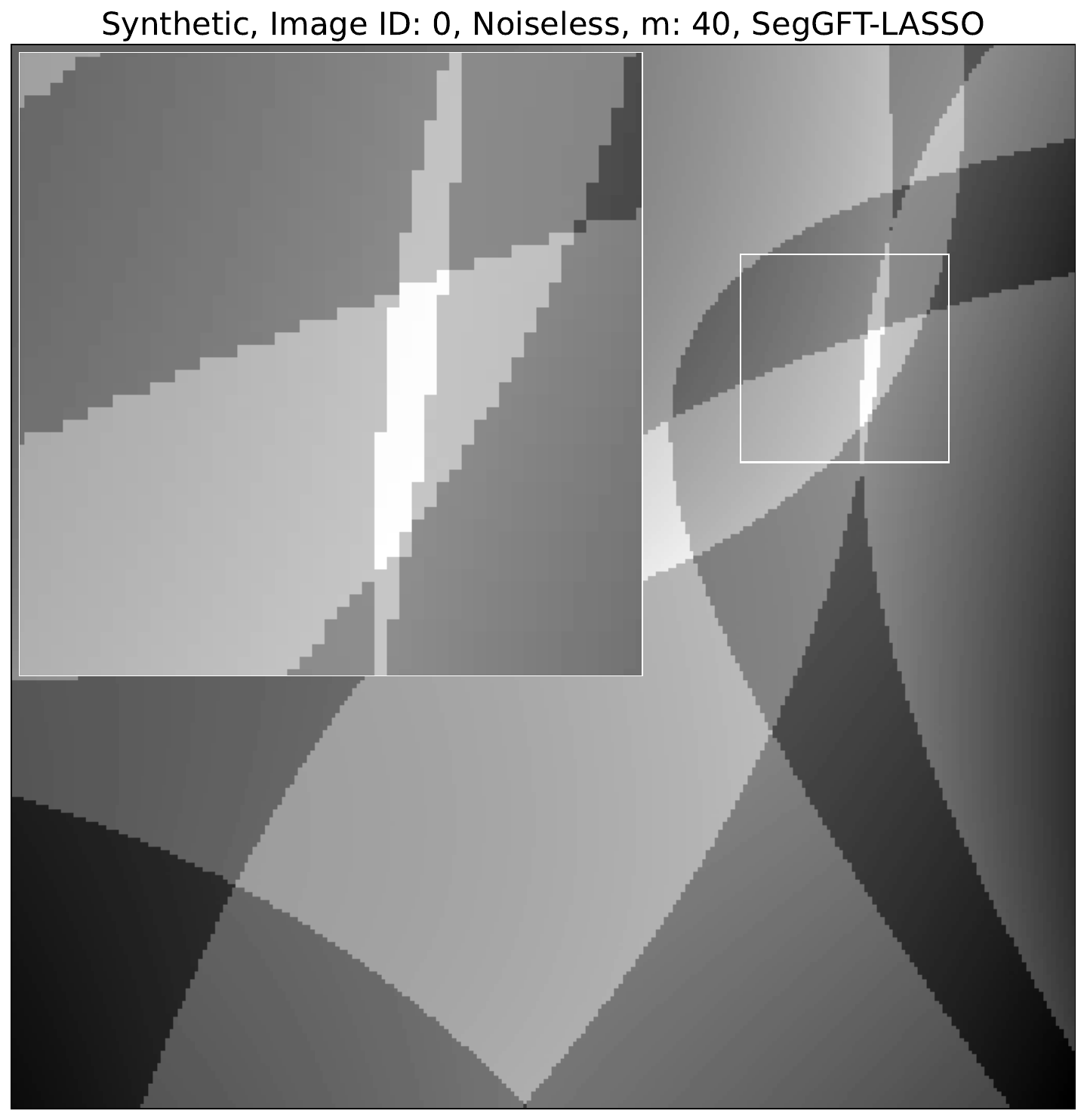}
    \end{subfigure}\hfil%
    \begin{subfigure}[b]{\gridcolwidth}
        \includegraphics[trim={0 0 0 10mm},clip,width=\linewidth]{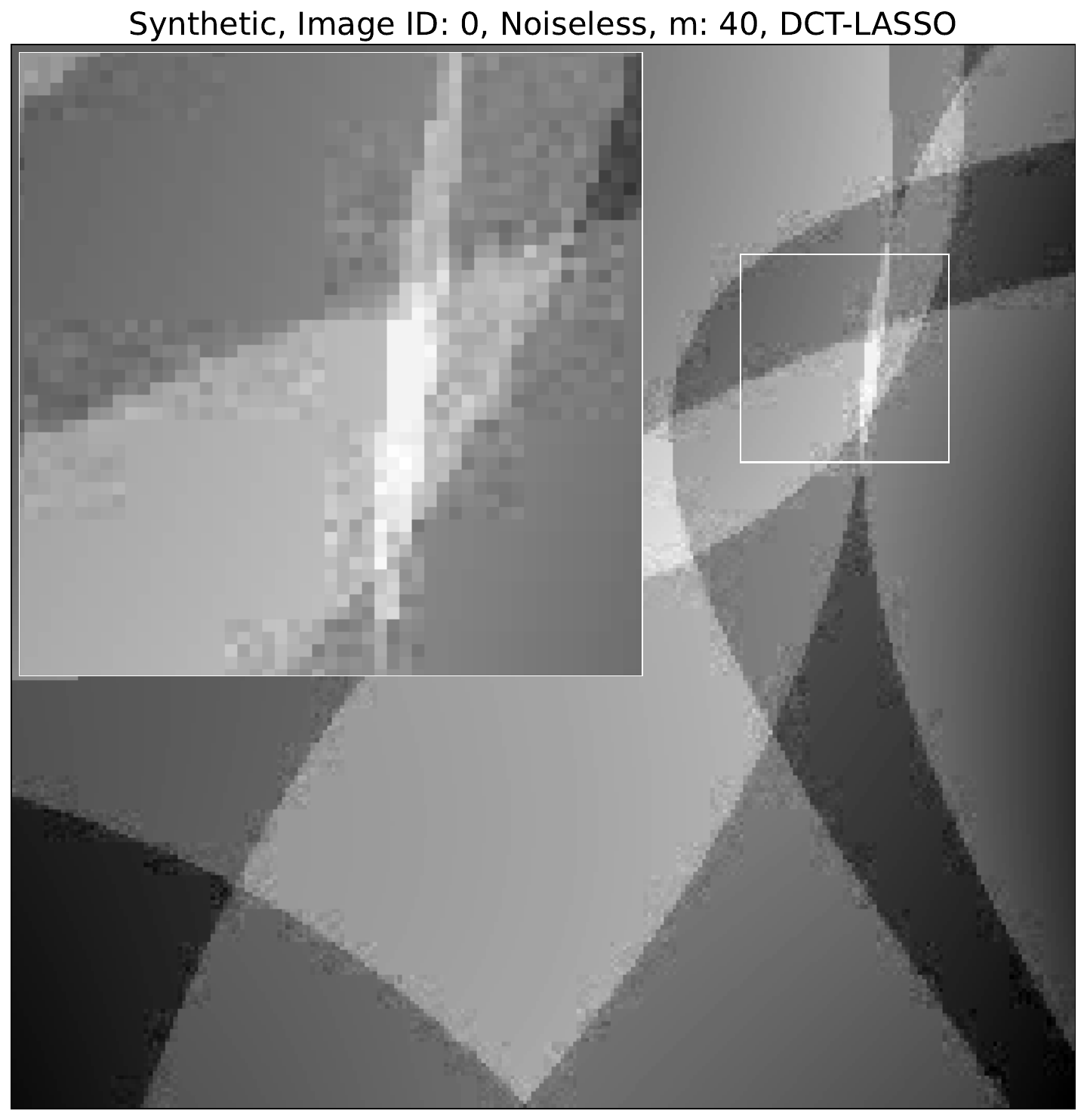}
    \end{subfigure}\hfil%
    \begin{subfigure}[b]{\gridcolwidth}
        \includegraphics[trim={0 0 0 10mm},clip,width=\linewidth]{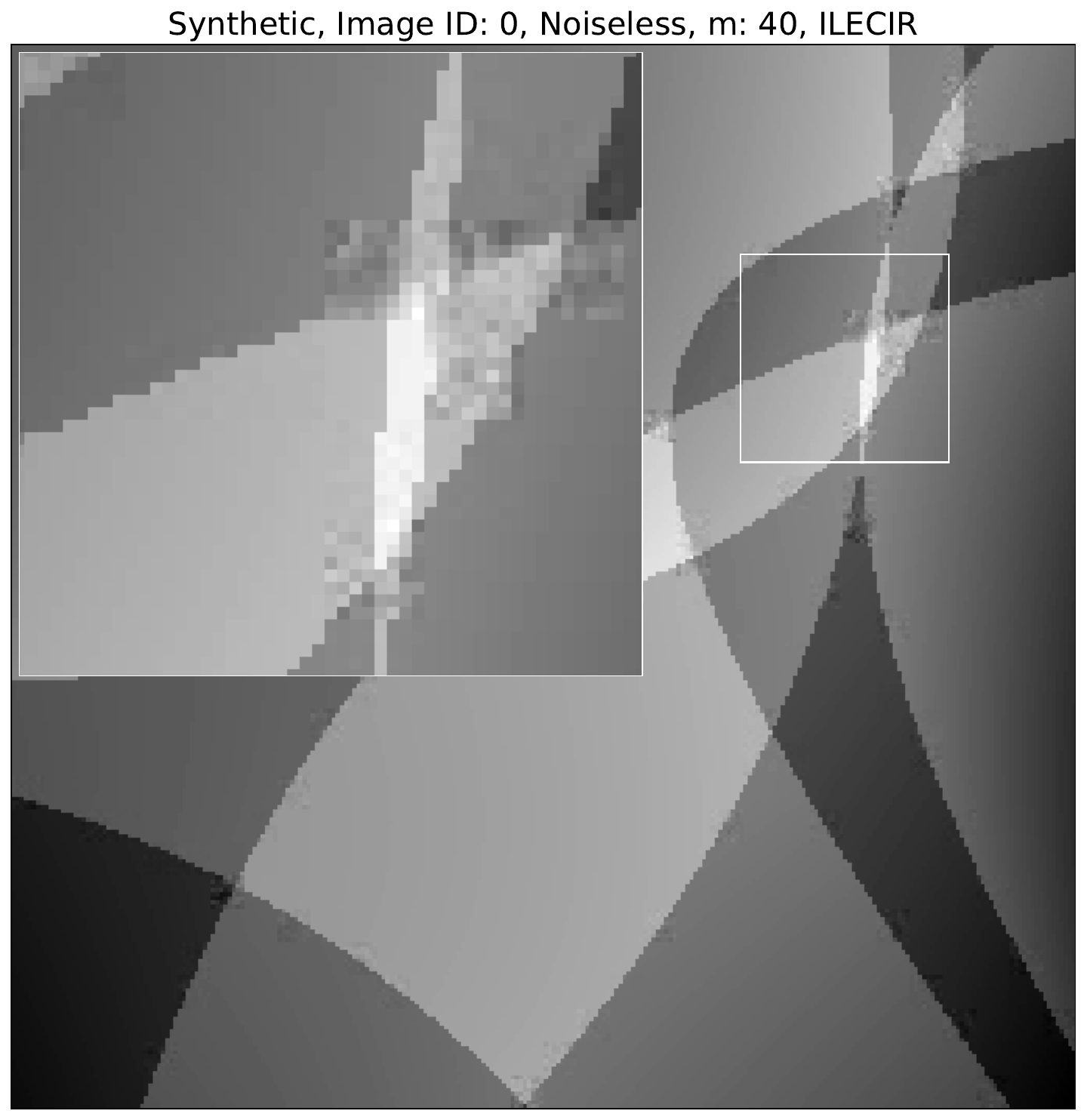}
    \end{subfigure}\\
    \settoheight{\gridrowheight}{
        \includegraphics[trim={0 0 0 10mm},clip,width=\gridcolwidth]{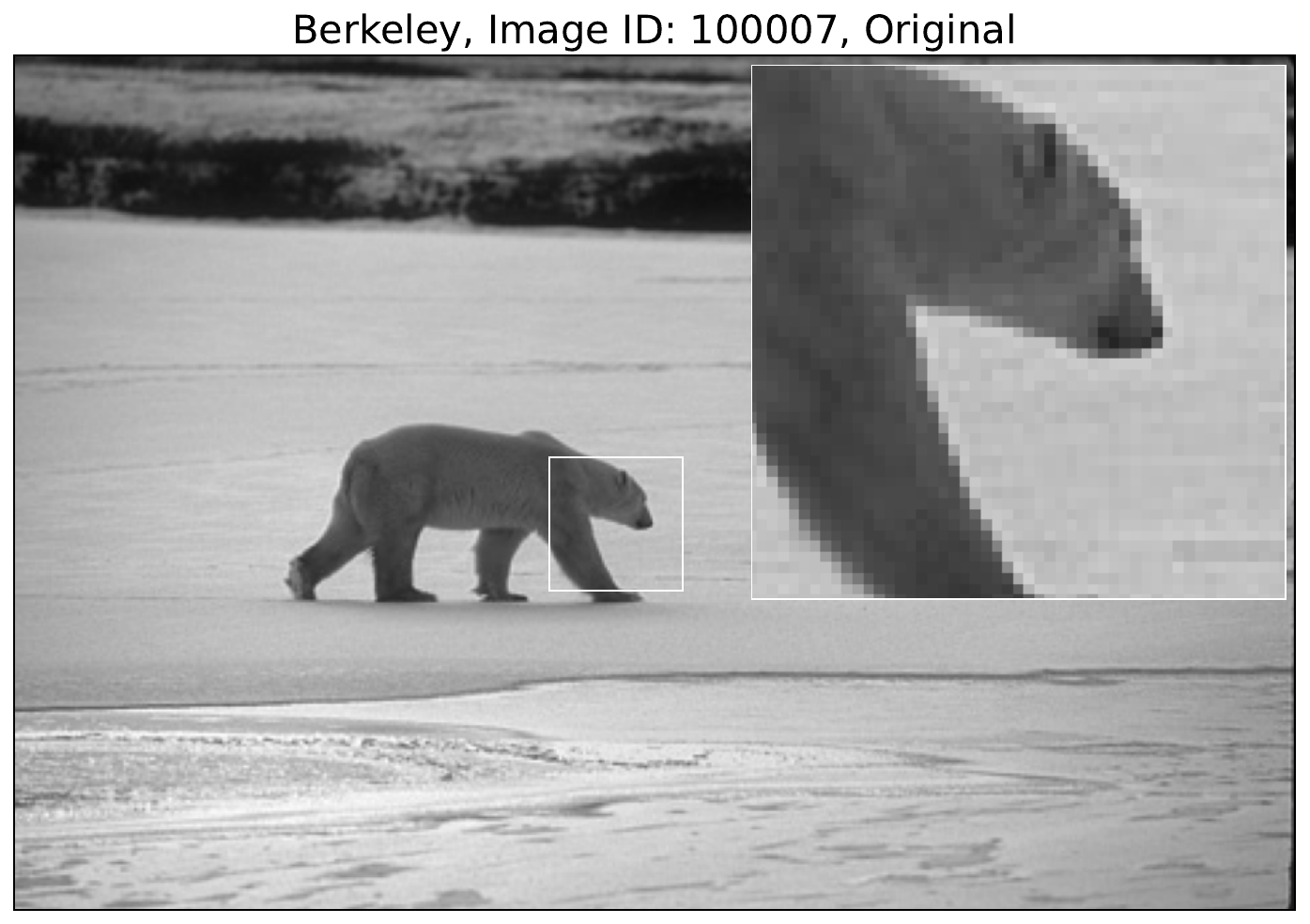}
    }
    \vspace{1.5mm}
    \rowname{Berkeley}
    \begin{subfigure}[b]{\gridcolwidth}
        \includegraphics[trim={0 0 0 10mm},clip,width=\linewidth]{figs/zoomed/berkeley_noise_0/000/image_000_Original_m_40_zoomed_inset.pdf}
    \end{subfigure}\hfil%
    \begin{subfigure}[b]{\gridcolwidth}
        \includegraphics[trim={0 0 0 10mm},clip,width=\linewidth]{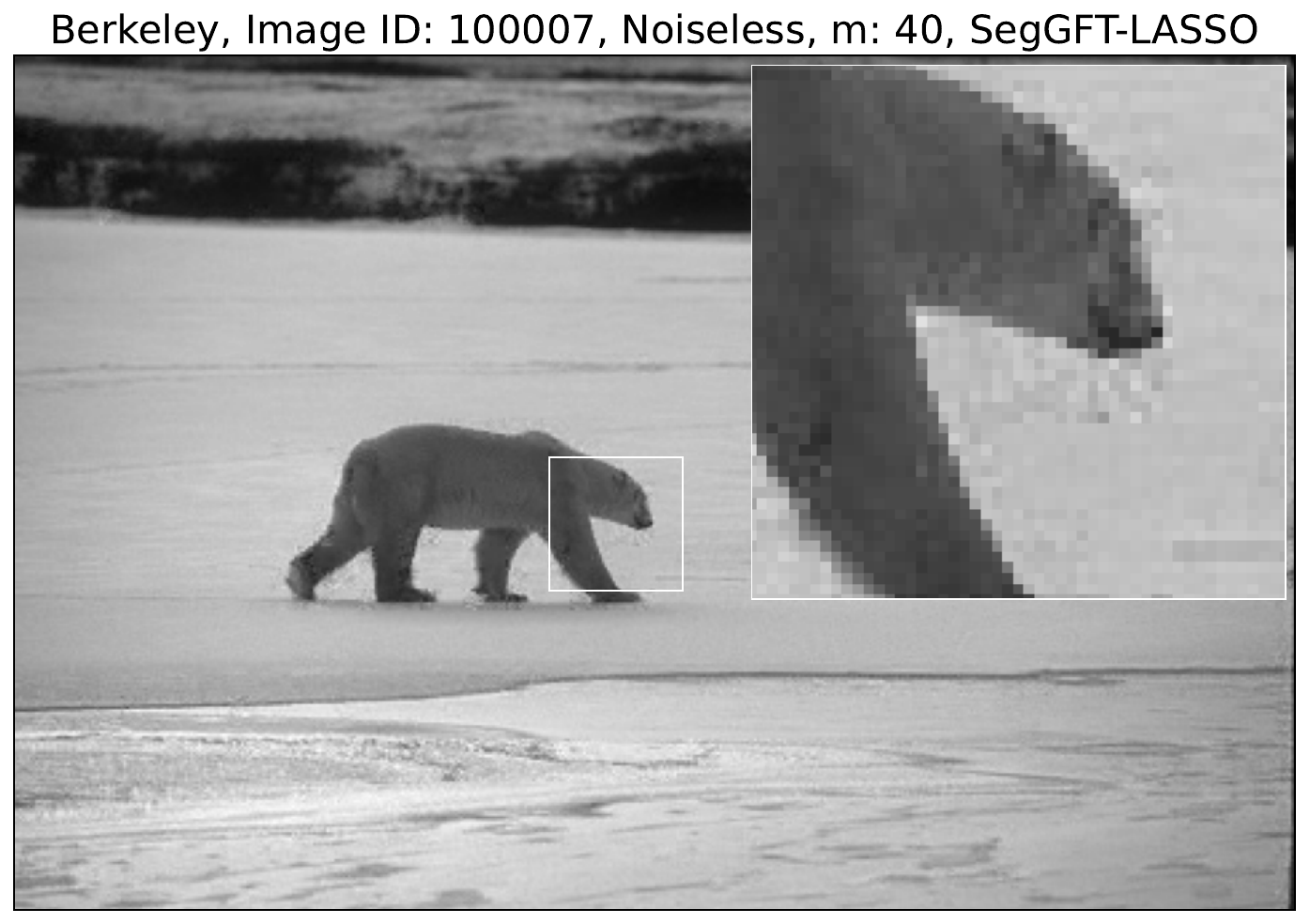}
    \end{subfigure}\hfil%
    \begin{subfigure}[b]{\gridcolwidth}
        \includegraphics[trim={0 0 0 10mm},clip,width=\linewidth]{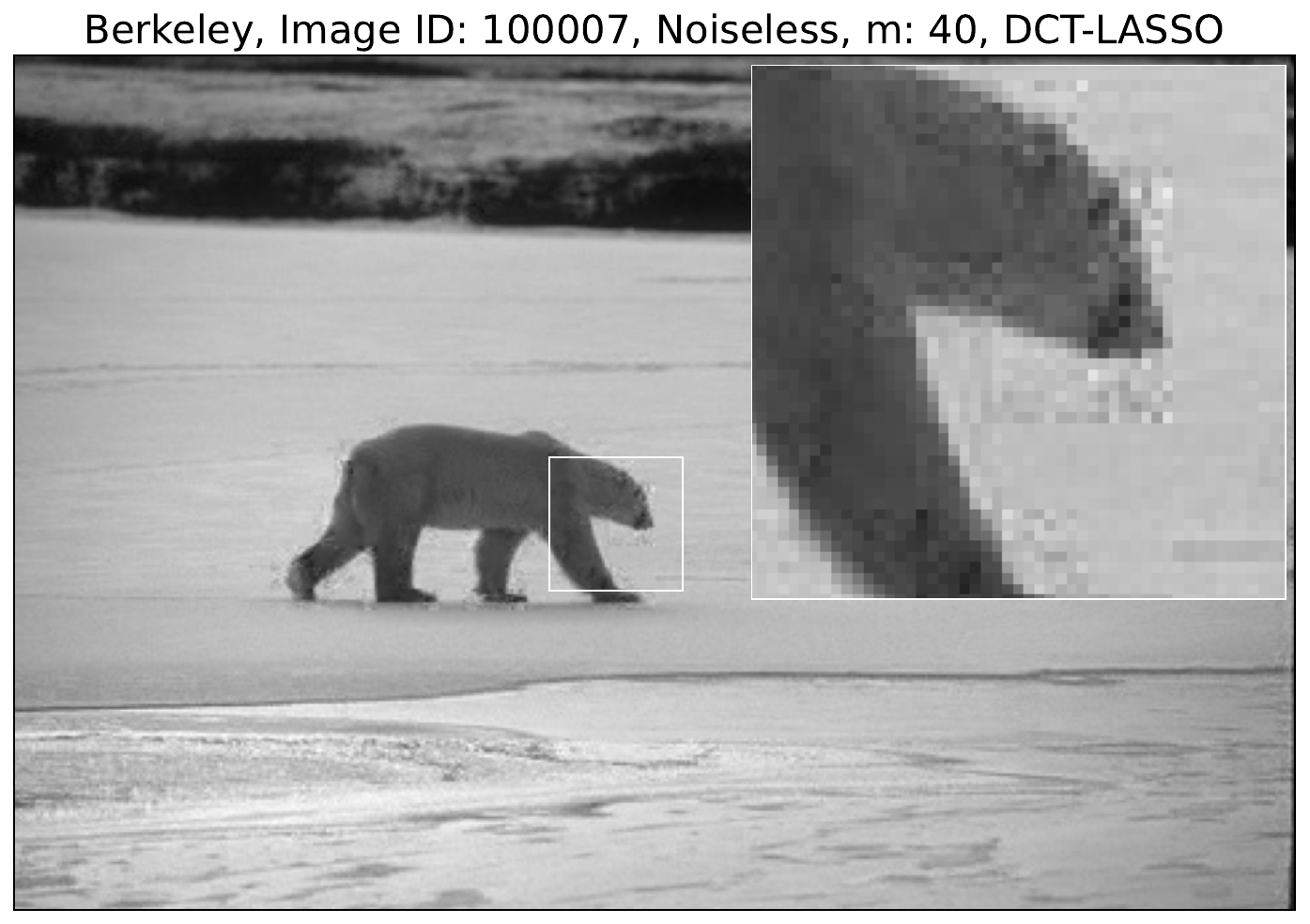}
    \end{subfigure}\hfil%
    \begin{subfigure}[b]{\gridcolwidth}
        \includegraphics[trim={0 0 0 10mm},clip,width=\linewidth]{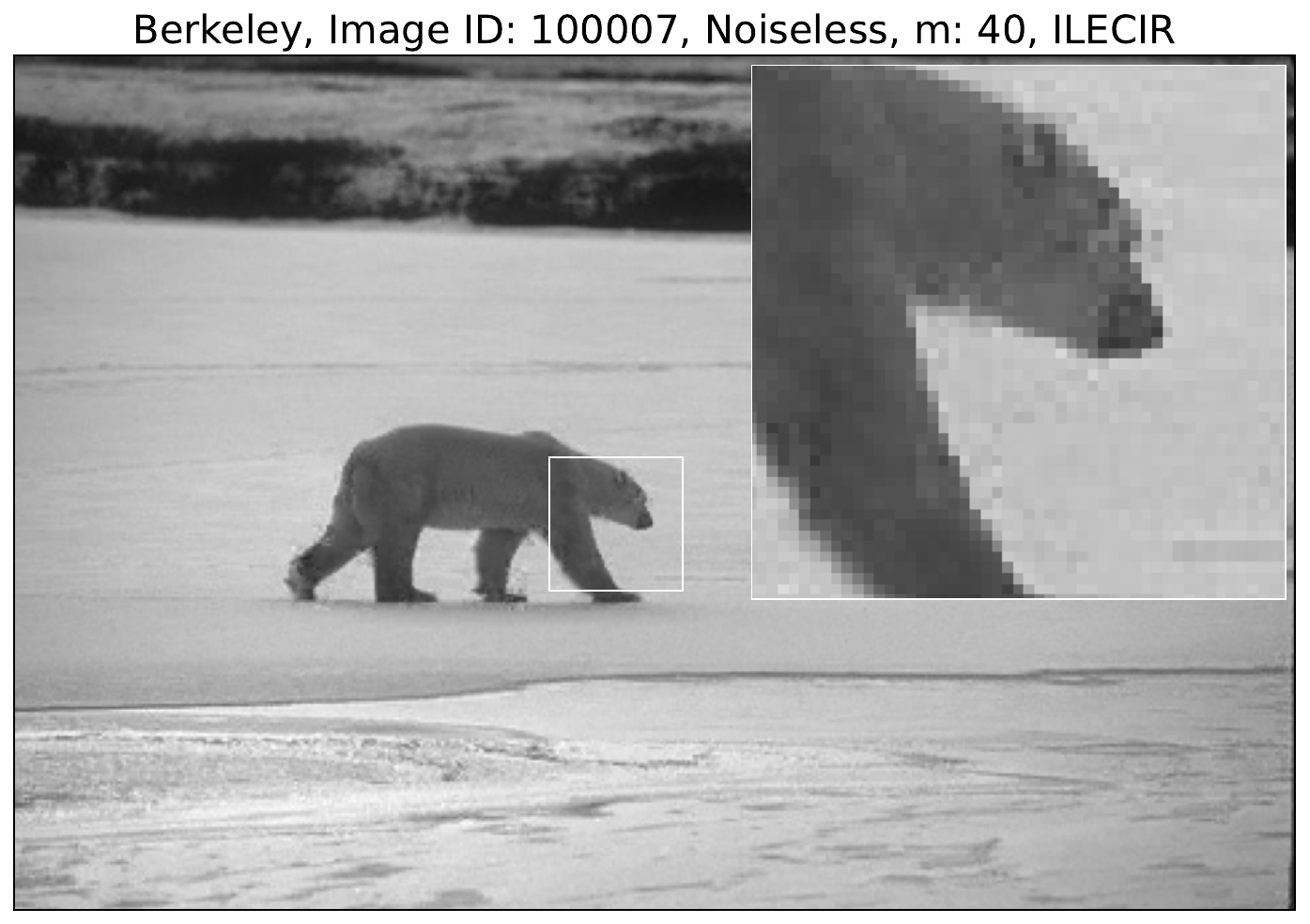}
    \end{subfigure}\\%
    \settoheight{\gridrowheight}{
        \includegraphics[trim={0 0 0 10mm},clip,width=\gridcolwidth]{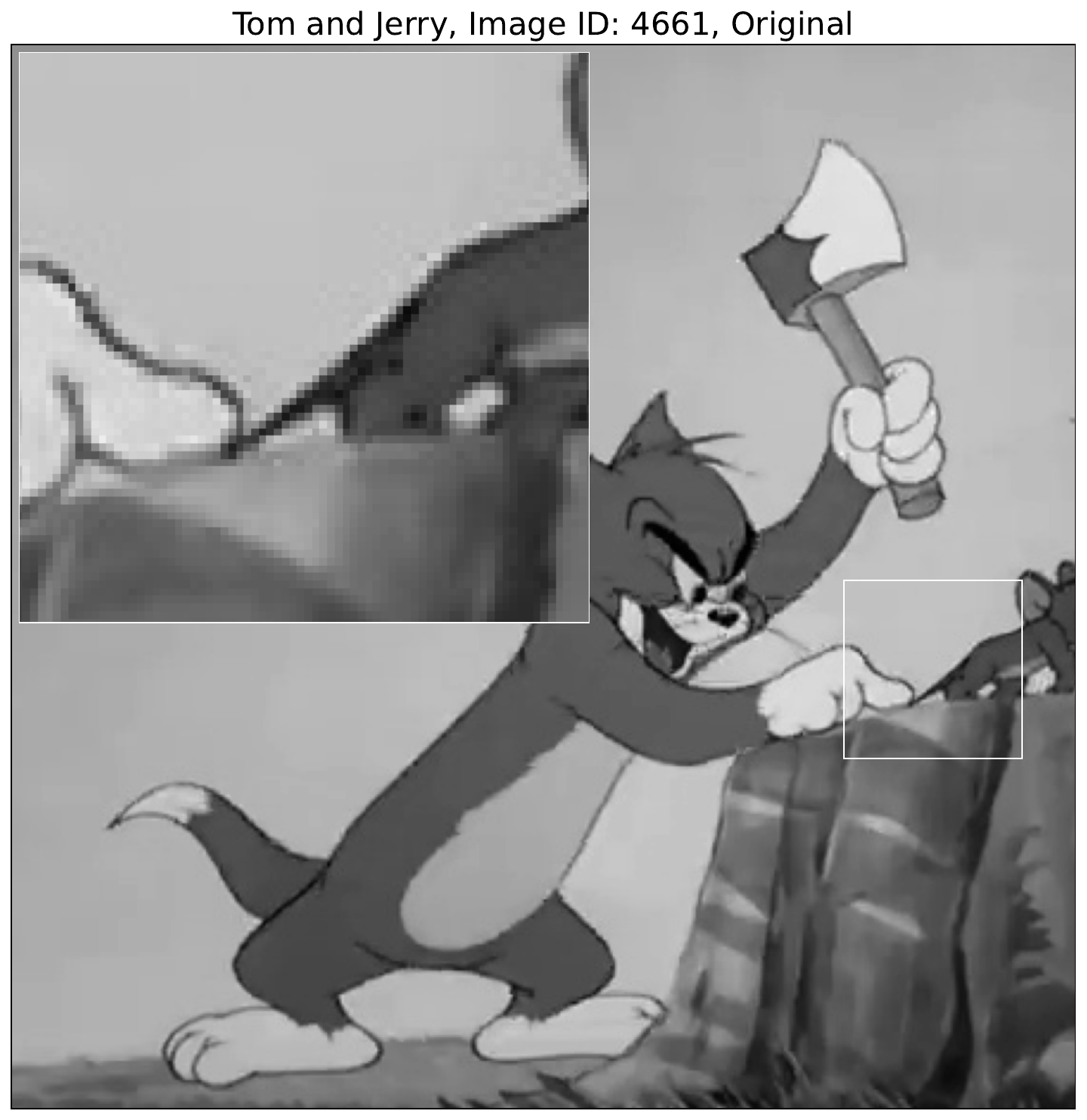}
    }
    \vspace{1.5mm}
    \rowname{Tom and Jerry}
    \includegraphics[trim={0 0 0 10mm},clip,width=\gridcolwidth]{figs/zoomed/tom_and_jerry_noise_0/003/image_003_Original_m_40_zoomed_inset.pdf}\hfil%
    \includegraphics[trim={0 0 0 10mm},clip,width=\gridcolwidth]{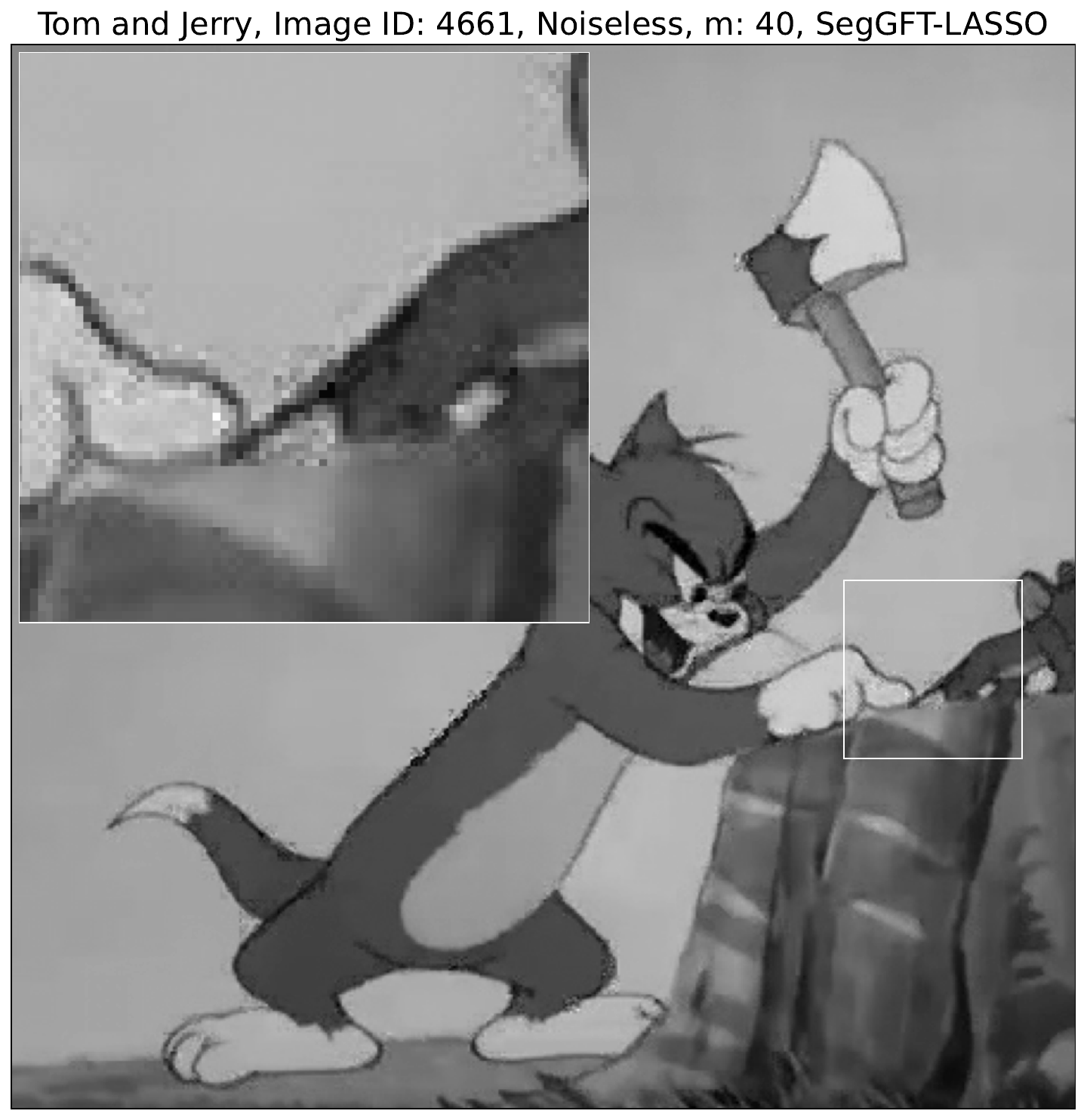}\hfil%
    \includegraphics[trim={0 0 0 10mm},clip,width=\gridcolwidth]{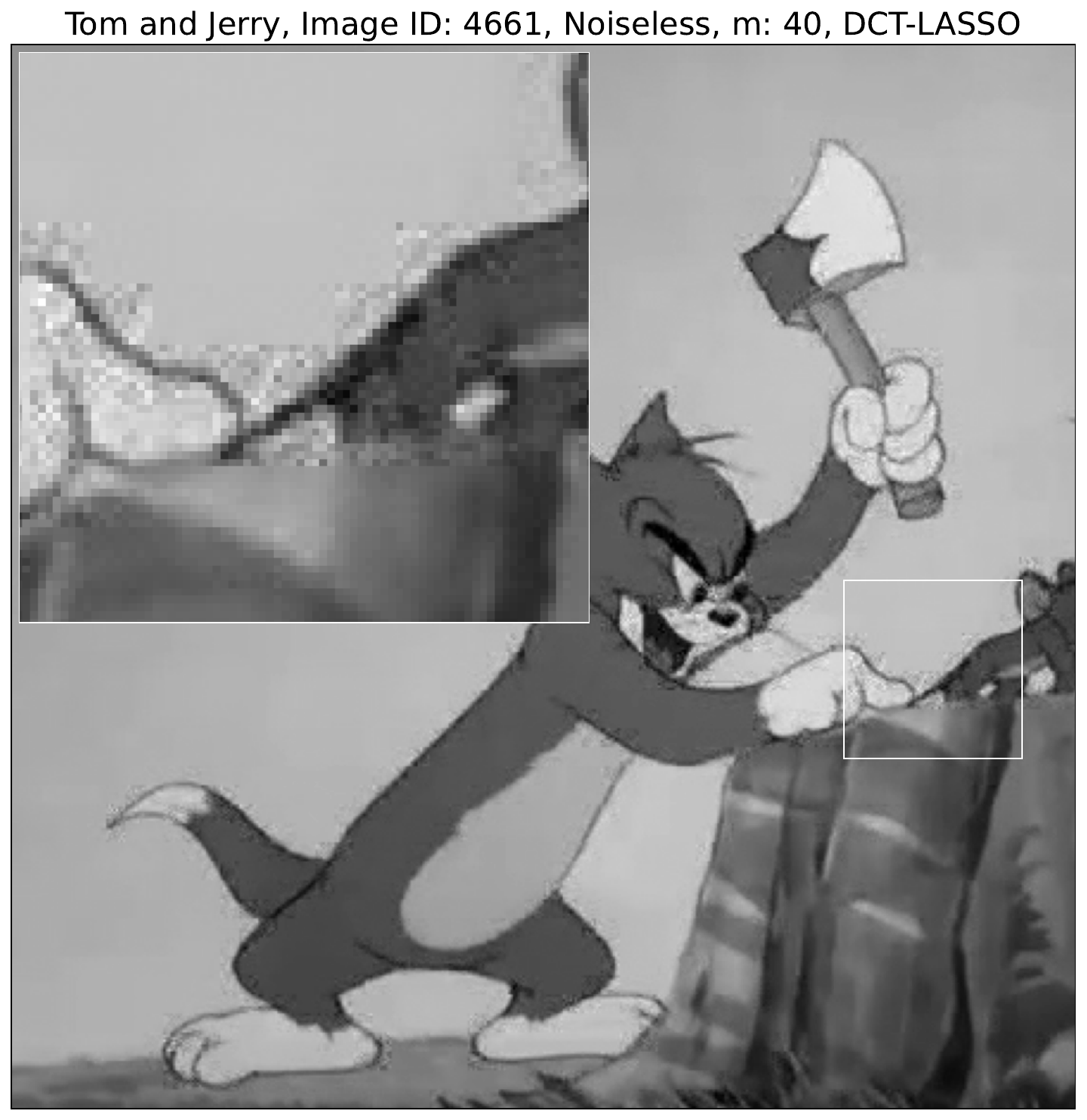}\hfil%
    \includegraphics[trim={0 0 0 10mm},clip,width=\gridcolwidth]{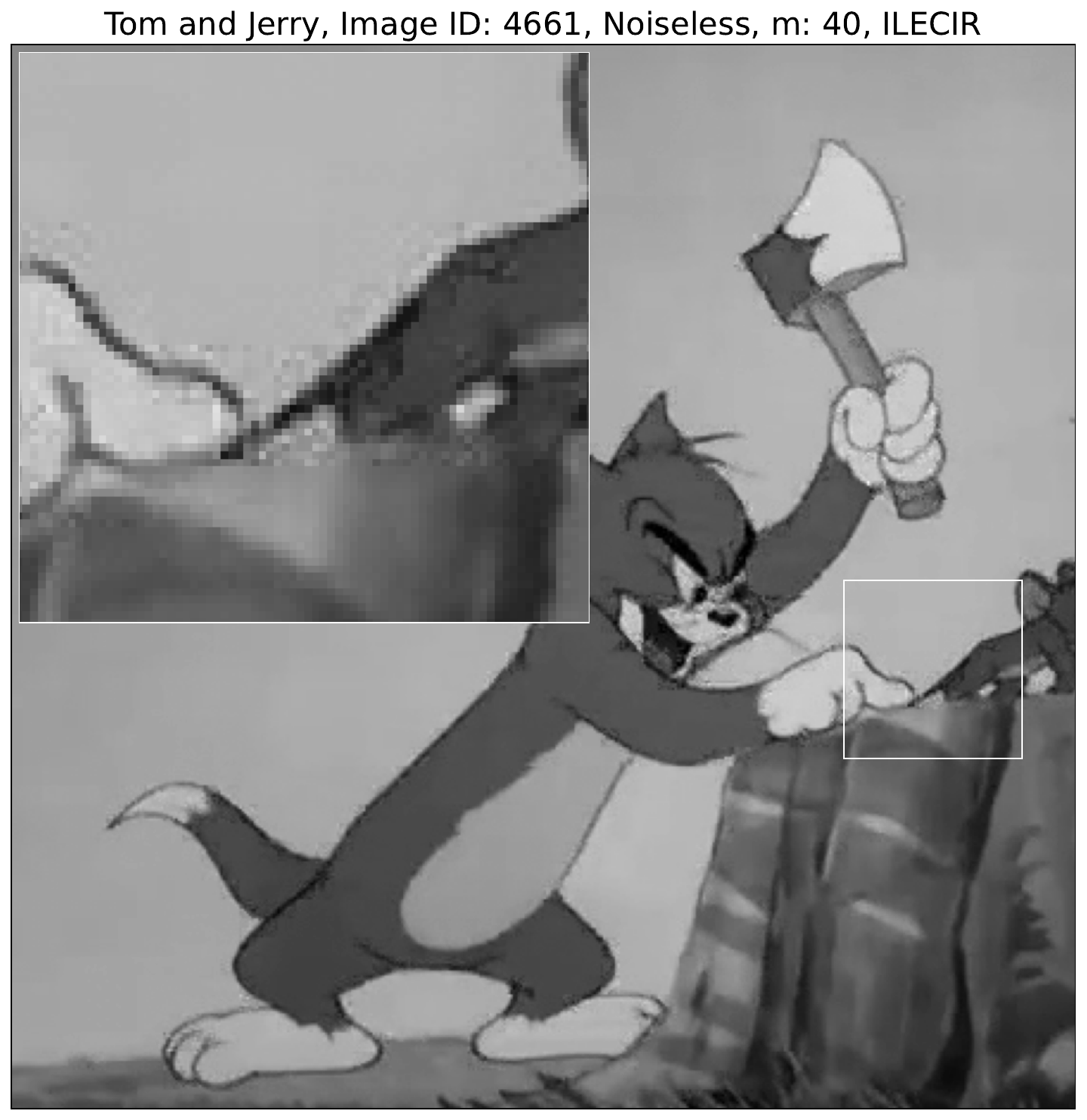}\\
    \settoheight{\gridrowheight}{
        \includegraphics[trim={0 0 0 10mm},clip,width=\gridcolwidth]{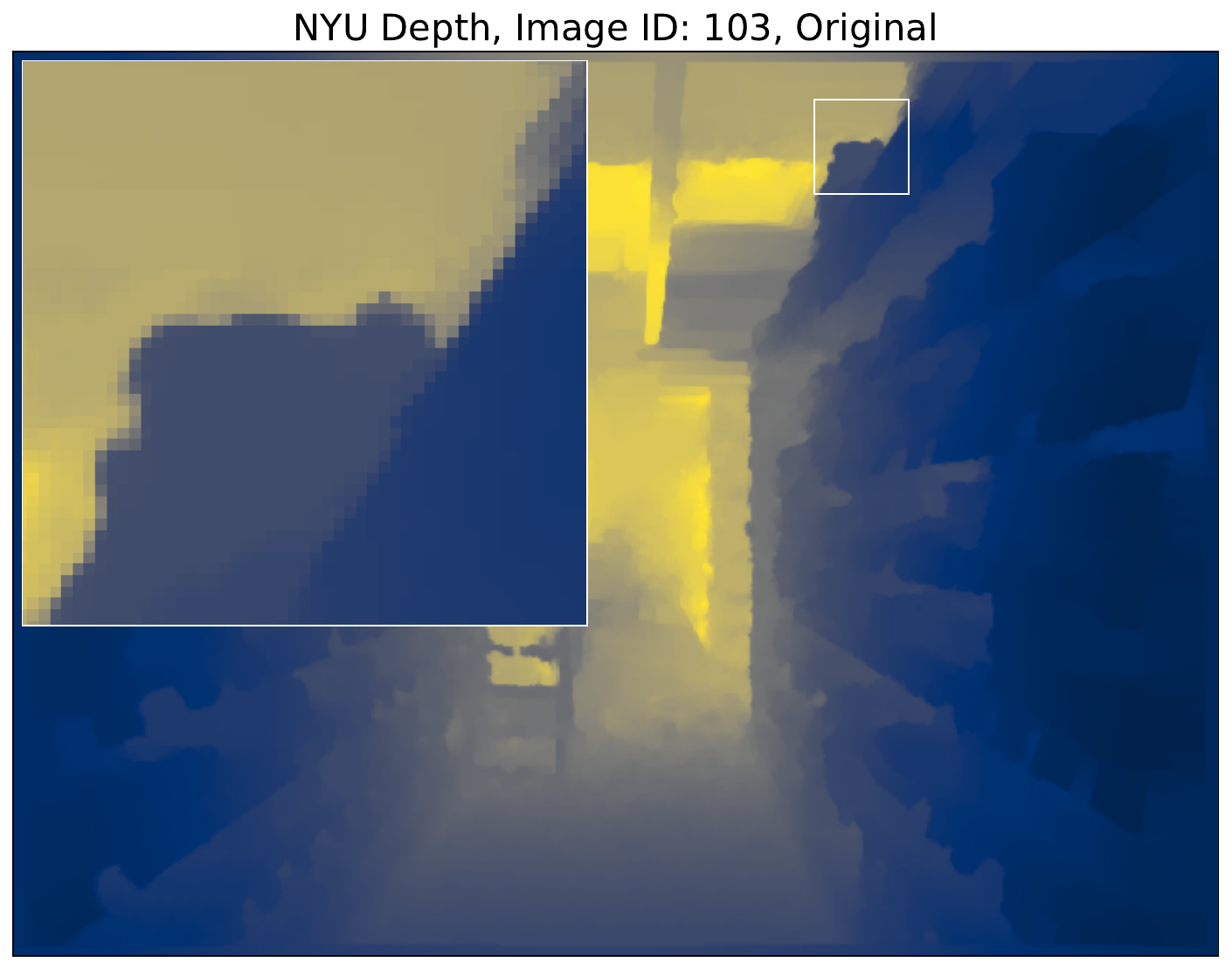}
    }
    \vspace{1.5mm}
    \rowname{NYU Depth}
    \includegraphics[trim={0 0 0 10mm},clip,width=\gridcolwidth]{figs/zoomed/nyu_depth_noise_0/103/image_103_Original_m_40_zoomed_inset.pdf}\hfil%
    \includegraphics[trim={0 0 0 10mm},clip,width=\gridcolwidth]{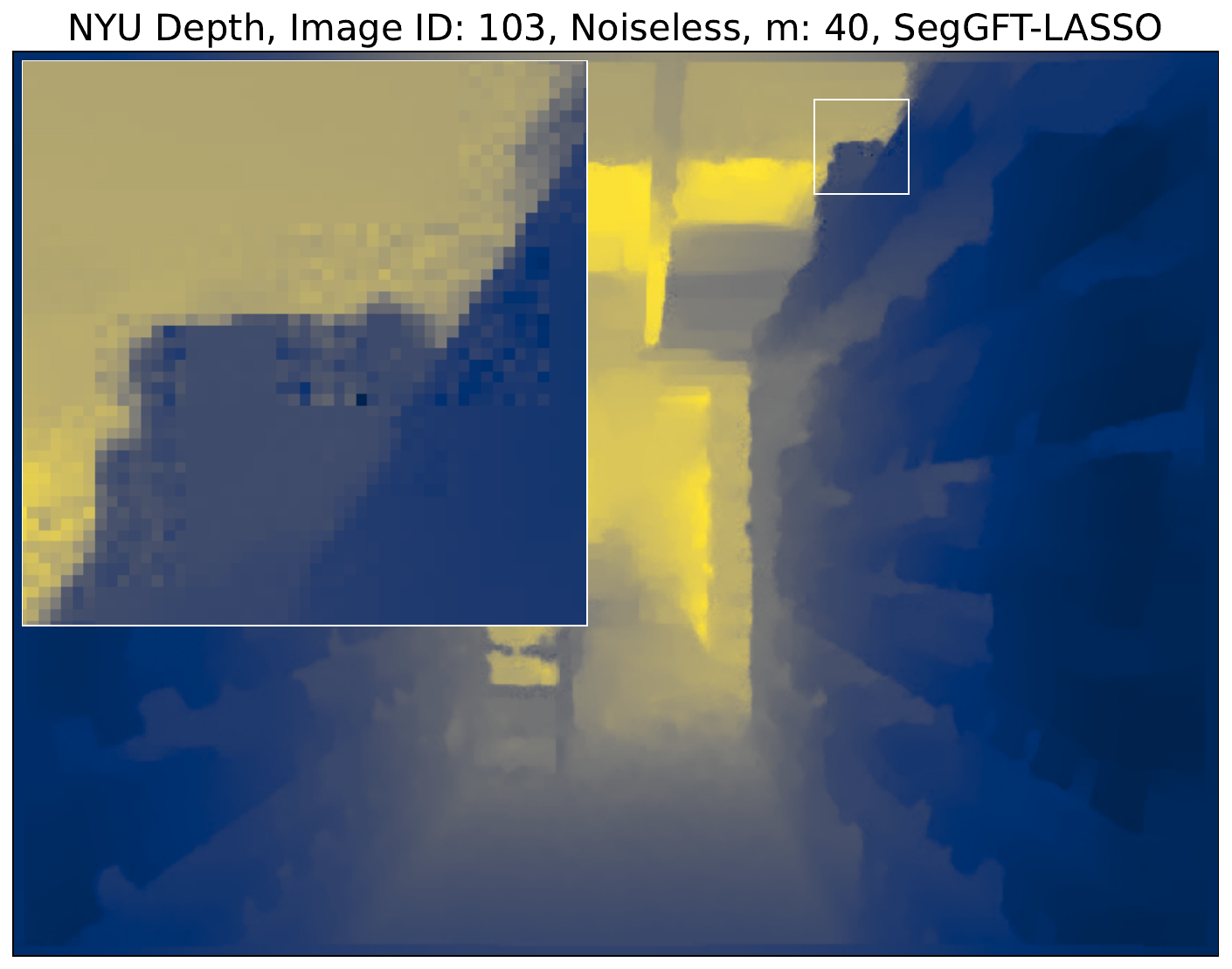}\hfil%
    \includegraphics[trim={0 0 0 10mm},clip,width=\gridcolwidth]{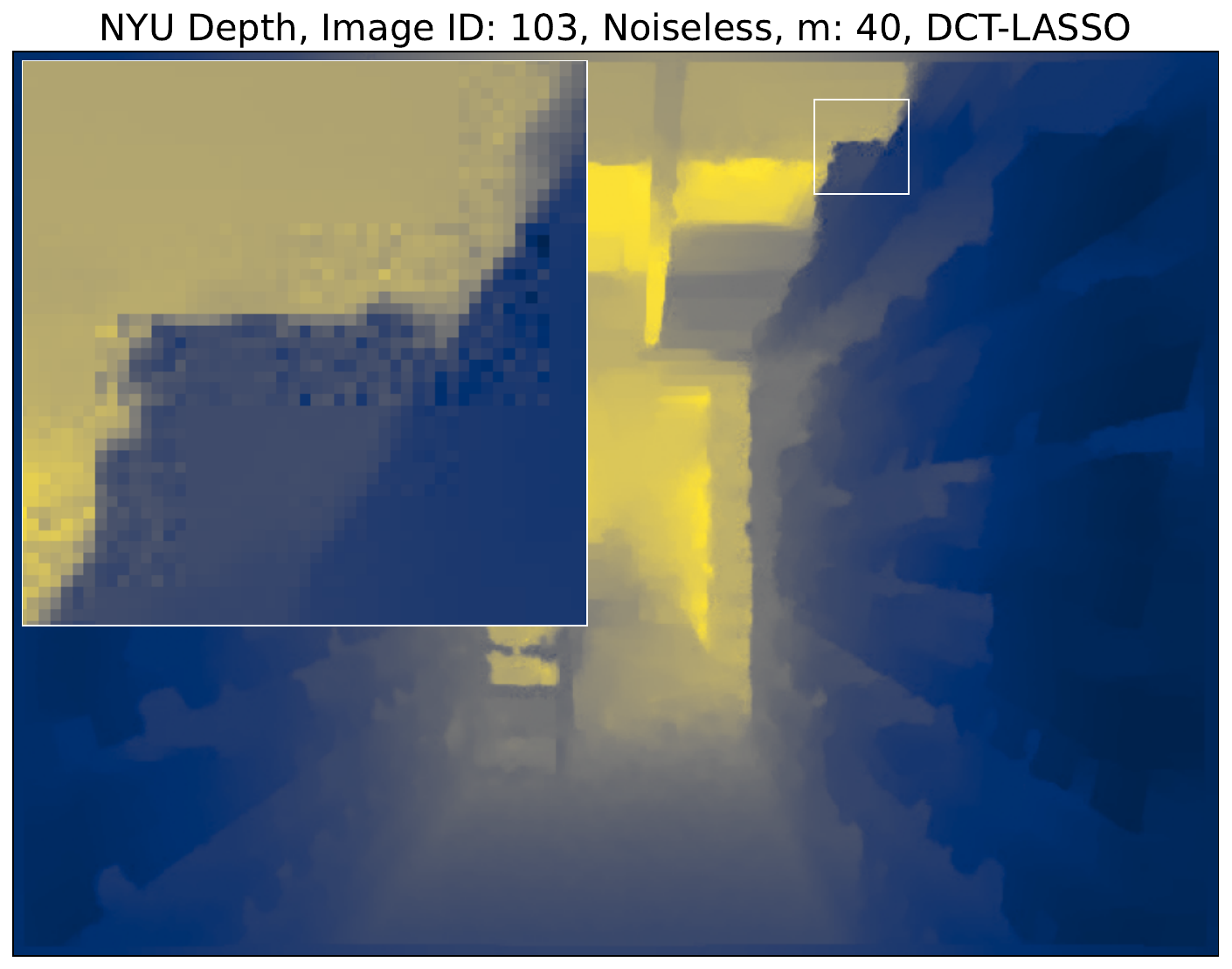}\hfil%
    \includegraphics[trim={0 0 0 10mm},clip,width=\gridcolwidth]{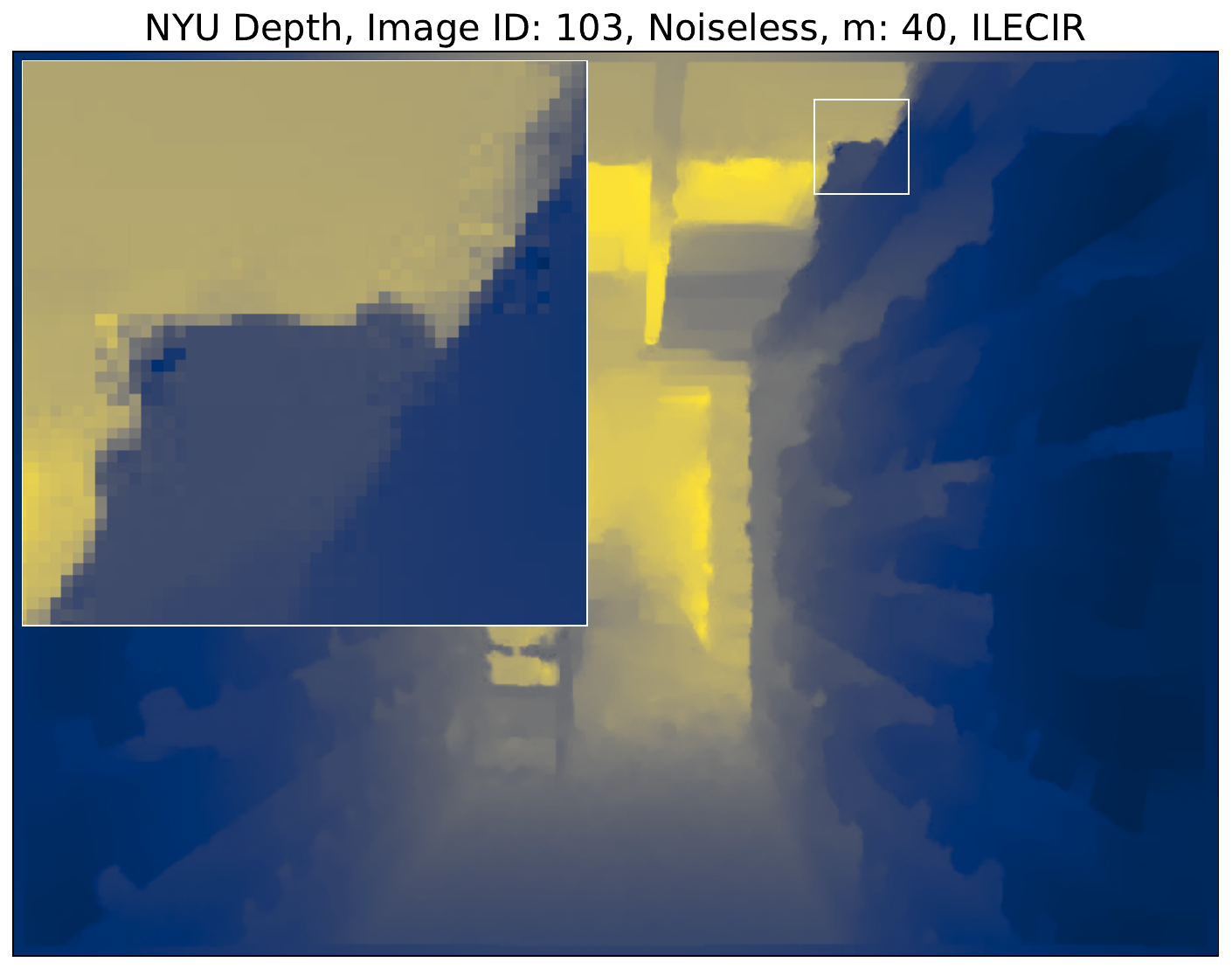}
    \caption[Image recovery comparison for \textsc{Ilecir}]{Zoomed view of images from the Synthetic (top row), Berkeley (second row), Tom and Jerry (third row), and NYU Depth (bottom row) datasets, recovered patch-wise from $m = 40$ compressive measurements per $8\times 8$ patch via various algorithms in the noiseless setting.
    The columns denote, from left to right:
    Col 1 -- the original image,
    Col 2 -- image recovered via \textsc{SegGft-Lasso-Cv},
    Col 3 -- image recovered via \textsc{Dct-Lasso-Cv},
    Col 4 -- image recovered via \textsc{Ilecir}.
    }\vspace{-5mm}
    \label{fig:all_zoomed_noiseless}
\end{figure}

\noindent\textbf{Qualitative Comparison in the noiseless setting:} The results of image recovery via \textsc{SegGft-Lasso-Cv}, \textsc{Dct-Lasso-Cv}, and \textsc{Ilecir} for one image of each dataset are shown in
Fig.~\ref{fig:all_zoomed_noiseless},
along with the original image in each case.
A careful inspection of the images reveals that in each case, the images recovered by \textsc{Ilecir} have higher fidelity than the corresponding image recovered by \textsc{Dct-Lasso-Cv} along sharp edges in the image.
The quality of the recovered images remains the same in regions where there are no sharp edges.
In Fig.~\ref{fig:all_zoomed_noiseless} (Row 1), for the image from the Synthetic dataset, we clearly see that \textsc{Ilecir} has good recovery in regions where there is only a single edge which is close to linear.
In contrast, \textsc{Dct-Lasso-Cv} produces many artifacts along the edges.
The image recovered by \textsc{Ilecir} also possesses some artifacts along the edges, but much fewer than the one recovered by \textsc{Dct-Lasso-Cv}. These artifacts may appear for \textsc{Ilecir} whenever the ground-truth image edge in the patch does not exactly match any of the image edges that \textsc{Ilecir} tests for.
\textsc{Ilecir} does not do well around corners, due to the presence of more than one image edge in the patch.
However, a careful inspection reveals that for some patches, the recovery quality for \textsc{Ilecir} seems to be better than \textsc{Dct-Lasso-Cv} even in the presence of two edges, and there does not appear to be a case where \textsc{Ilecir} has worse recovery quality than \textsc{Dct-Lasso-Cv}.
This is due to the high-probability RMSE improvement guarantee of \textsc{Ilecir} (due to cross-validation) as established by Thm.~\ref{thm:ges_solution_improvement}.

In Fig.~\ref{fig:all_zoomed_noiseless} (Row 2), there are much fewer artifacts around the face of the polar bear in the image recovered by \textsc{Ilecir} as compared to the one recovered by \textsc{Dct-Lasso-Cv}.
In the same region, we also see that while there are fewer artifacts for \textsc{SegGft-Lasso-Cv} compared to \textsc{Dct-Lasso-Cv}, the recovery by \textsc{Ilecir} is much better than for \textsc{SegGft-Lasso-Cv}.
This once again highlights the strength of \textsc{Ilecir} compared to using human-labelled segmentation (which may contain errors) for recovery and is not available during compressive recovery. In Fig.~\ref{fig:all_zoomed_noiseless} (Row 3), there are much fewer artifacts for \textsc{Ilecir} around the ears and the head of Tom (the cat), compared to for \textsc{Dct-Lasso-Cv}.
However, there are many artifacts on the face of Tom for both \textsc{Ilecir} and \textsc{Dct-Lasso-Cv}.
This highlights the inability of \textsc{Ilecir} to improve upon \textsc{Dct-Lasso-Cv} in regions containing many edges.
In Fig.~\ref{fig:all_zoomed_noiseless} (Row 4), we observe the same pattern -- \textsc{Ilecir} performs better than \textsc{Dct-Lasso-Cv} and even \textsc{SegGft-Lasso-Cv} around the edges in the depth map. The reconstruction by each algorithm on an individual patch of an image from the NYU Depth dataset is also shown in Fig.~\ref{fig:nyu_segs}, which clearly demonstrates the recovery of the ground-truth segmentation and the original patch using \textsc{Ilecir}.

\begin{figure}[htbp]
    \centering
\includegraphics[width=0.23\linewidth]{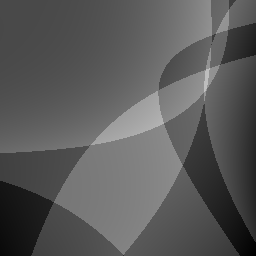}\hfil%
\includegraphics[width=0.23\linewidth]{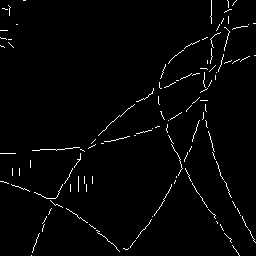}\hfil%
\includegraphics[width=0.23\linewidth]{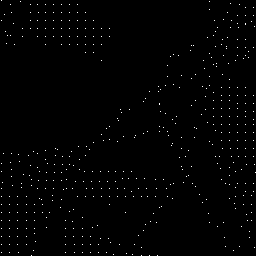}\hfil%
\includegraphics[width=0.23\linewidth]{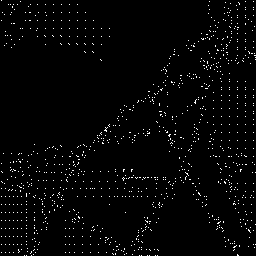}
    \caption[Edgemap recovery comparison between \textsc{Ilecir} and \textsc{Ges}]{
    A synthetic image (col 1), and its edgemap as recovered patch-wise from $m=30$ compressive measurements per $8\times 8$ patch via the \textsc{Ilecir} (col 2), \textsc{Ges-1} (col 3), and \textsc{Ges-10} (col 4) algorithms.
    }
    \label{fig:edgemap}
\end{figure}
\noindent\textbf{Edgemap Recovered by \textsc{Ilecir}:}
Since the \textsc{Ilecir} algorithm discovers the best linear image edge for each patch, an edge-map of the original image may be recovered by stitching together the edge-maps of each patch.
Fig~\ref{fig:edgemap} shows an image from the synthetic dataset (col 1), and its edge-map (col 2) recovered by the \textsc{Ilecir} algorithm.
The original edges are correctly detected for most patches.

\begin{figure}[htbp]
    \centering
    \raisebox{-0.5\height}{\includegraphics[width=0.37\linewidth]{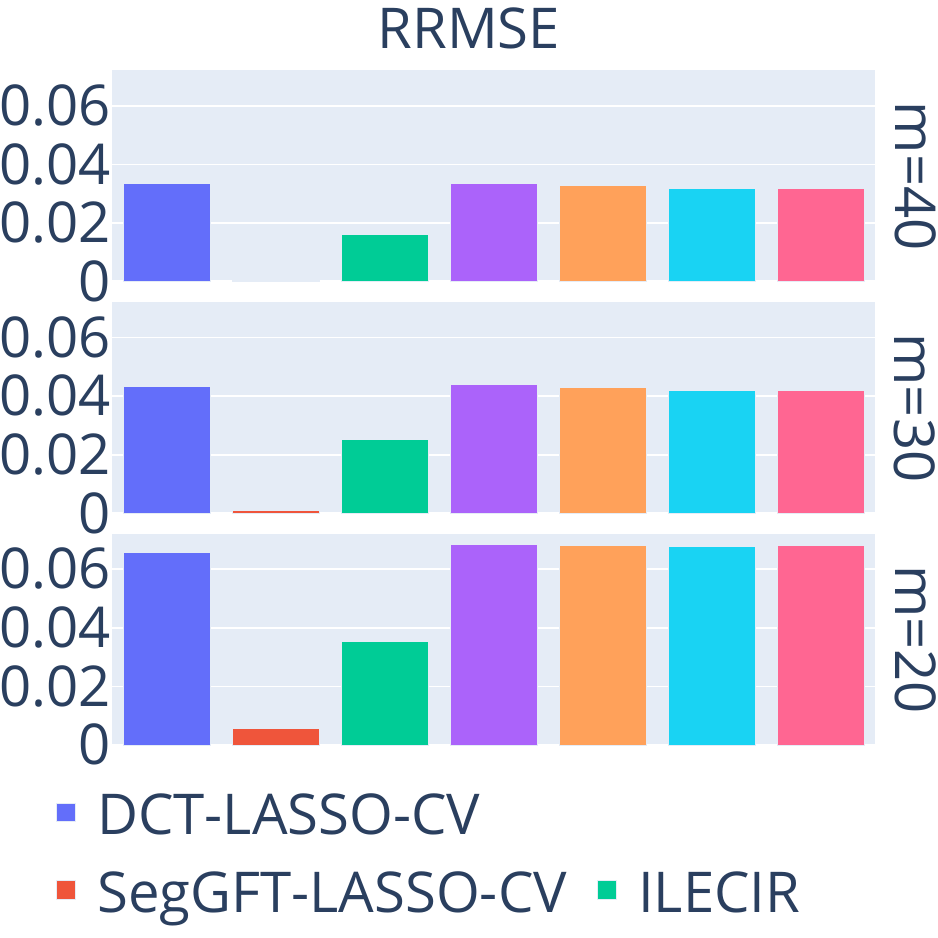}}\hfill%
    \raisebox{-0.5\height}{\includegraphics[width=0.37\linewidth]{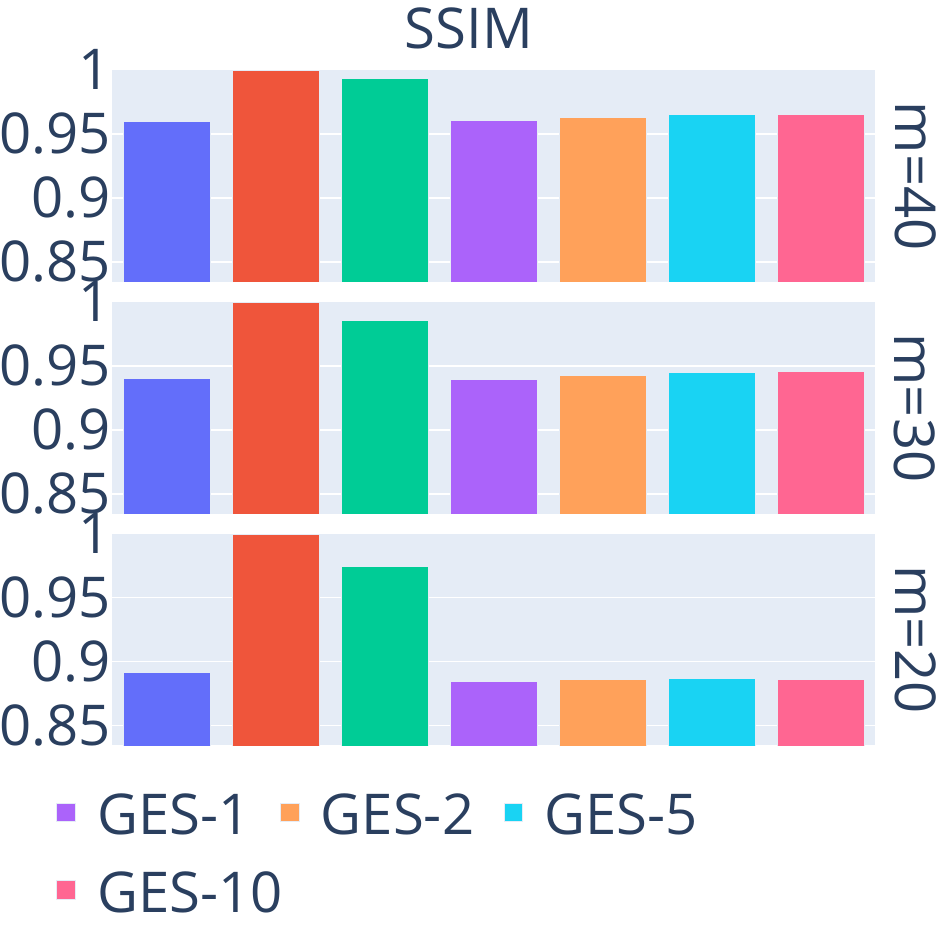}}\hfill%
    \raisebox{-0.5\height}{\includegraphics[trim={0 0 0 10mm},clip,width=0.23\linewidth]{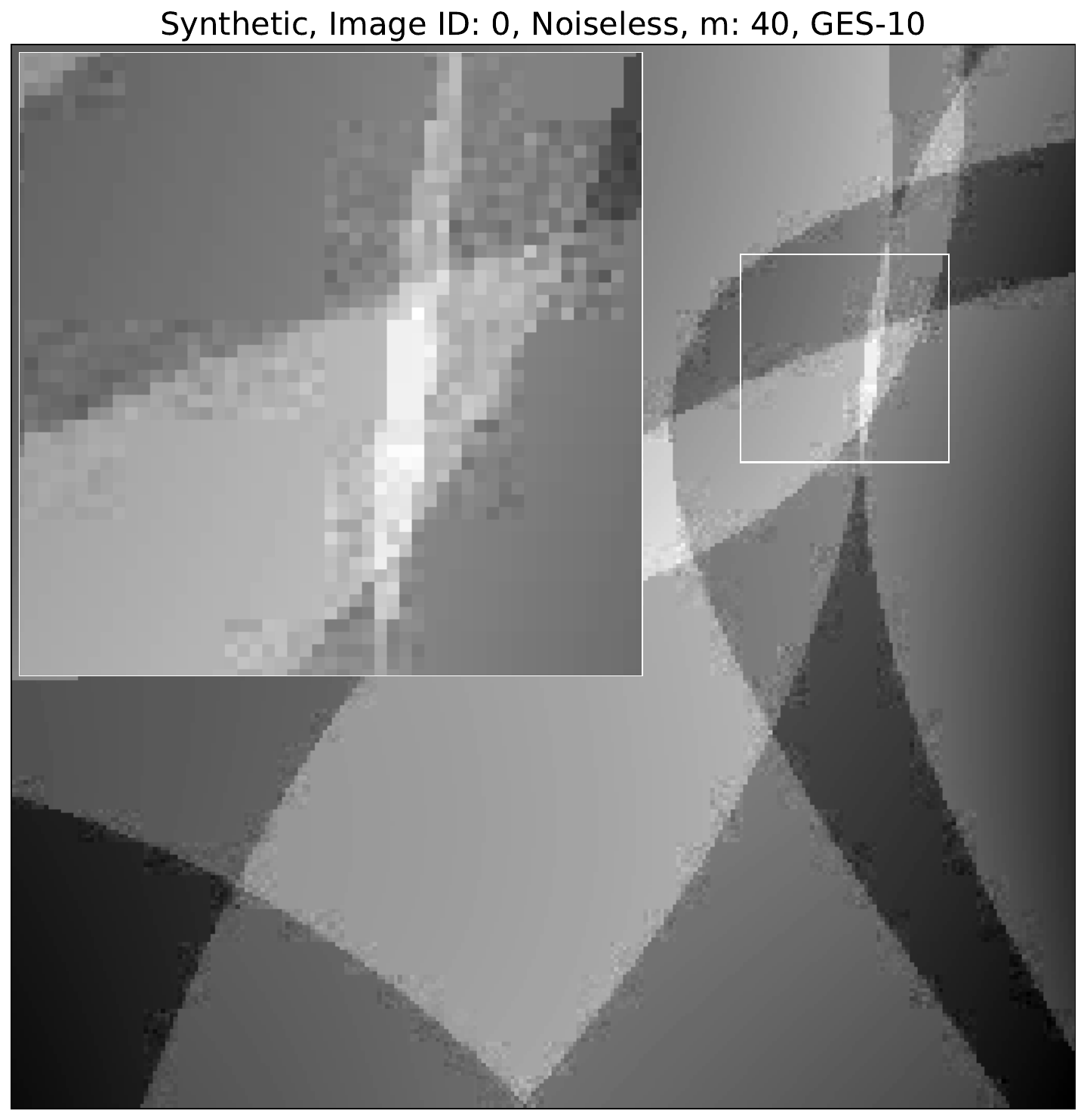}}
    \caption[Performance of \textsc{Ges} on compressive image recovery]{Performance of Greedy Edge Selection (\textsc{Ges}) on Image ID 0 of the Synthetic dataset in the noiseless regime (recovery done patch-wise on $8 \time 8$ patches).
    RRMSE (left) and SSIM (middle) with perturbation budget 1, 2, 5, and 10 (\textsc{Ges-1}, \textsc{GES-2}, \textsc{Ges-5}, \textsc{Ges-10}).
    Plots show comparison with DCT-based recovery (\textsc{Dct-Lasso-Cv}), Segmentation-aware recovery (\textsc{SegGft-Lasso-Cv}) and \textsc{Ilecir}.
    Right: The image recovered using \textsc{Ges-10} (compare with the \textsc{Ilecir} result in Fig.~\ref{fig:all_zoomed_noiseless}).
    }
    \label{fig:ges_rmse_ssim_image_acquisition}
\end{figure}
\noindent\textbf{Performance of Greedy Edge Selection on Compressive Image Recovery:}
\label{subsubsec:ges_image_acquisition_results}
We test the effectiveness of \textsc{Ges} on the problem of patch-wise compressive image recovery on an image from the Synthetic dataset.
Fig.~\ref{fig:ges_rmse_ssim_image_acquisition} shows the RRMSE and SSIM of the recovered images when \textsc{Ges} is used for recovery, with a maximum of $d_0 \in \{1, 2, 5, 10\}$ edges of the 2-D lattice graph 
of a patch being allowed to be perturbed by the algorithm.
We see that using \textsc{Ges} (especially when $d_0 = 10$ graph edges are allowed to be perturbed) is marginally better than using \textsc{Dct-Lasso-Cv} -- however, it is much better to use \textsc{Ilecir}.
This is evident in Fig.~\ref{fig:ges_rmse_ssim_image_acquisition}, as well -- a careful inspection reveals that the image recovered by \textsc{Ges}-10 ($d_0 = 10$) has slightly fewer artifacts than the one recovered by \textsc{Dct-Lasso-Cv}, but recovery via \textsc{Ilecir} (see Fig.~\ref{fig:all_zoomed_noiseless}) is much better.
Fig.~\ref{fig:edgemap}
shows the edgemap recovered by various algorithms, including \textsc{Ges}-5 and \textsc{Ges}-10 (cols 3 and 4).
An edgemap for \textsc{Ges} is recovered in the following manner -- whenever the \textsc{Ges} algorithm drops a (2-D lattice graph) edge between two pixels, the right pixel or the bottom pixel of the graph edge is marked with white color (denoting a pixel belonging to an image edge), and remaining pixels are marked with black color (denoting background pixels).
We see that the edgemap recovered by \textsc{Ges} follows the image edges very coarsely.
It consists of many isolated dots, instead of groups of pixels connected as an edge.
In comparison, the edgemap recovered by \textsc{Ilecir} consists of linear edges in each patch.
The \textsc{Ilecir} algorithm takes advantage of the structure in perturbations of the graph edges, and is able to outperform \textsc{Ges}, which perturbs graph edges in an unstructured manner.

\begin{figure}[t]
    \centering
    \includegraphics[width=0.49\linewidth]{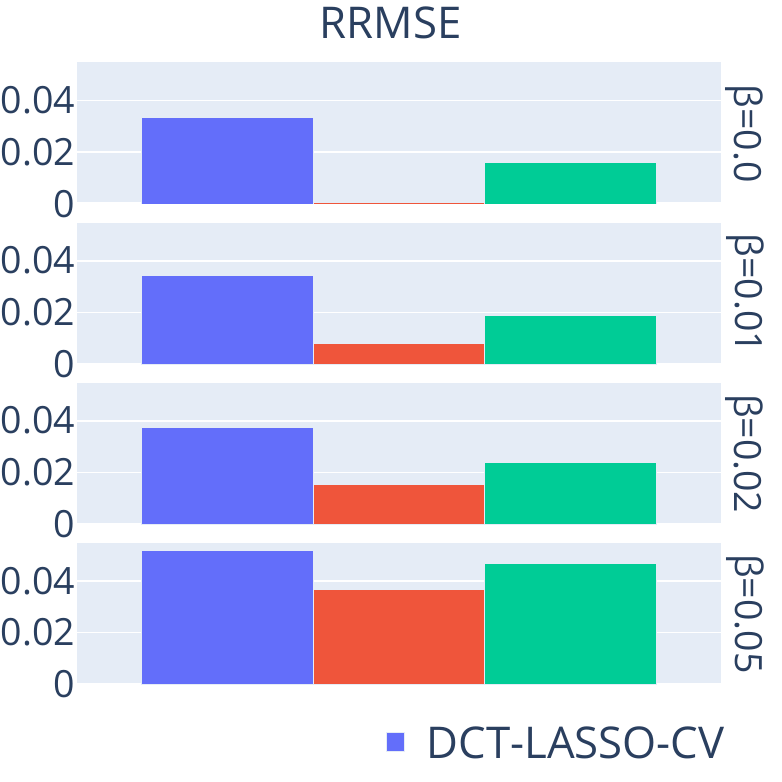}
    \includegraphics[width=0.49\linewidth]{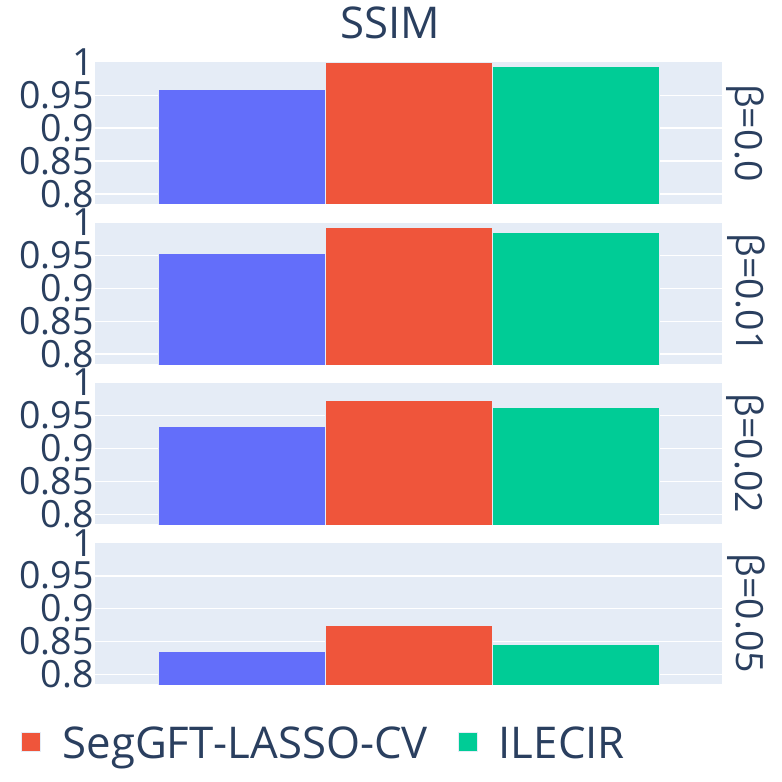}
    \caption[RRMSE and SSIM of \textsc{Ilecir} for various noise levels]{
    RRMSE and SSIM of \textsc{Ilecir}, \textsc{Dct-Lasso-Cv} and \textsc{SegGft-Lasso-Cv}  for Image ID 0 of the Synthetic dataset for different noise levels ($\beta \in \{0, 0.01, 0.02, 0.05\}, m = 40, n = 64$).
    }
    \vspace{-5mm}
    \label{fig:rmse_ssim_noisy}
\end{figure}

\begin{figure}[t]
\setlength{\gridcolwidth}{.23\linewidth}
\settoheight{\gridrowheight}{\includegraphics[width=\gridcolwidth]{figs/zoomed/synthetic_noise_0/000/image_000_Original_m_40_zoomed_inset.pdf}}%
    \centering
    \hspace{\baselineskip}
    \columnname{$\beta=0$}\hfil%
    \columnname{$\beta=0.01$}\hfil%
    \columnname{$\beta=0.02$}\hfil%
    \columnname{$\beta=0.05$}\\
    \rowname{\textsc{Dct-Lasso-Cv}}
    \begin{subfigure}[b]{\gridcolwidth}
        \includegraphics[trim={0 0 0 10mm},clip,width=\linewidth]{figs/zoomed/synthetic_noise_0/000/image_000_DCT-LASSO_m_40_zoomed_inset.pdf}
    \end{subfigure}\hfil%
    \begin{subfigure}[b]{\gridcolwidth}
        \includegraphics[trim={0 0 0 10mm},clip,width=\linewidth]{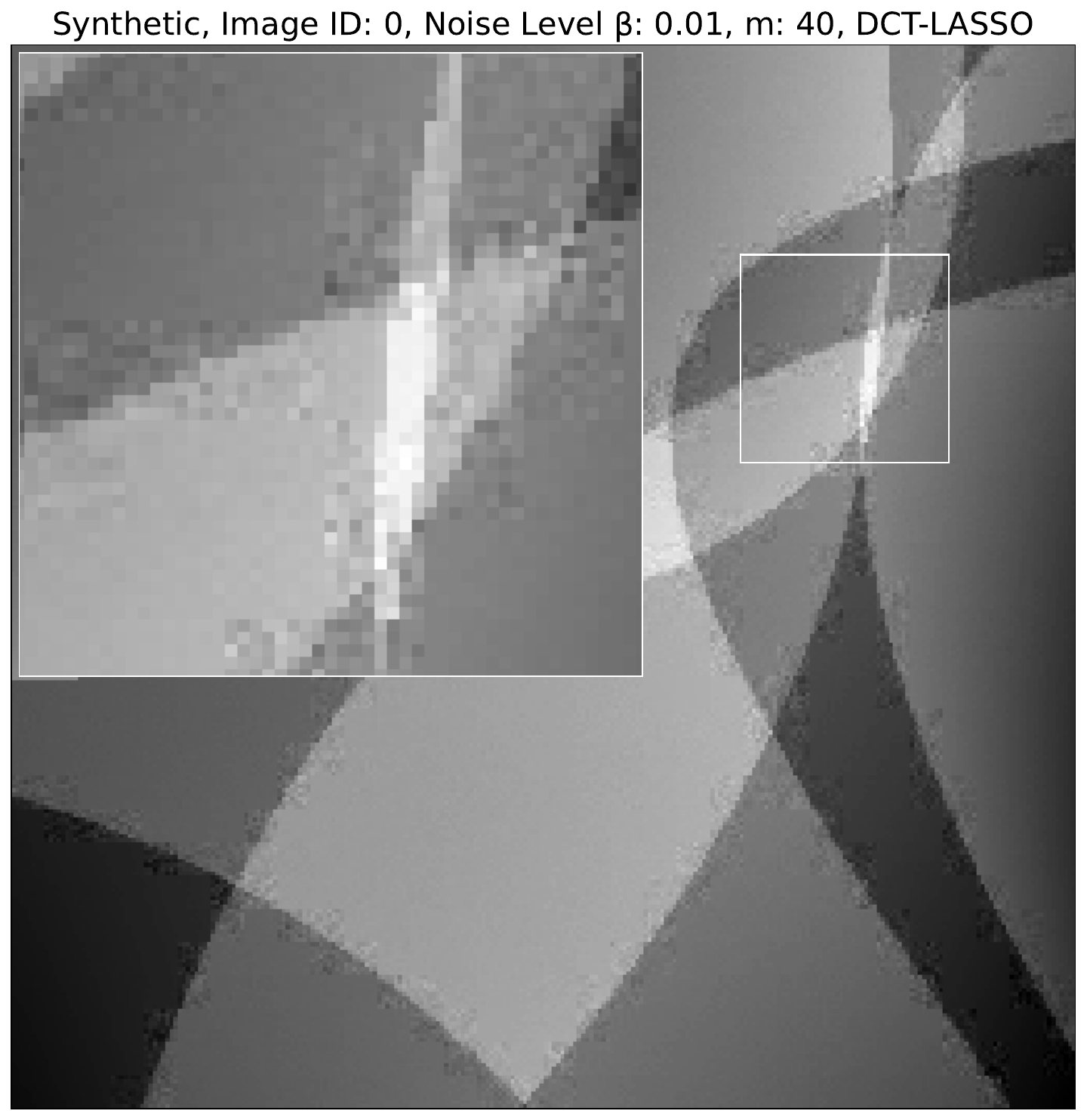}
    \end{subfigure}\hfil%
    \begin{subfigure}[b]{\gridcolwidth}
        \includegraphics[trim={0 0 0 10mm},clip,width=\linewidth]{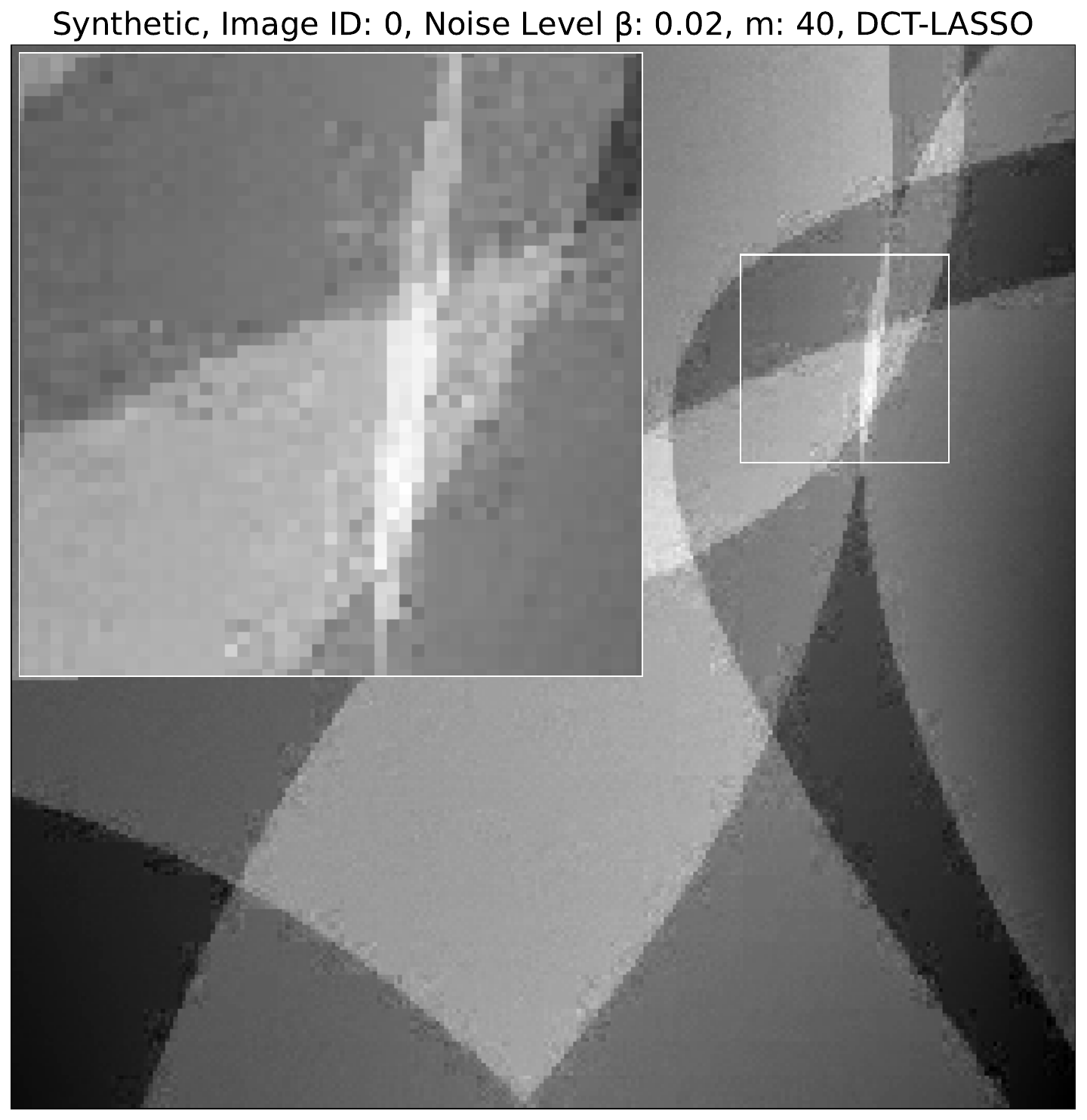}
    \end{subfigure}\hfil%
    \begin{subfigure}[b]{\gridcolwidth}
        \includegraphics[trim={0 0 0 10mm},clip,width=\linewidth]{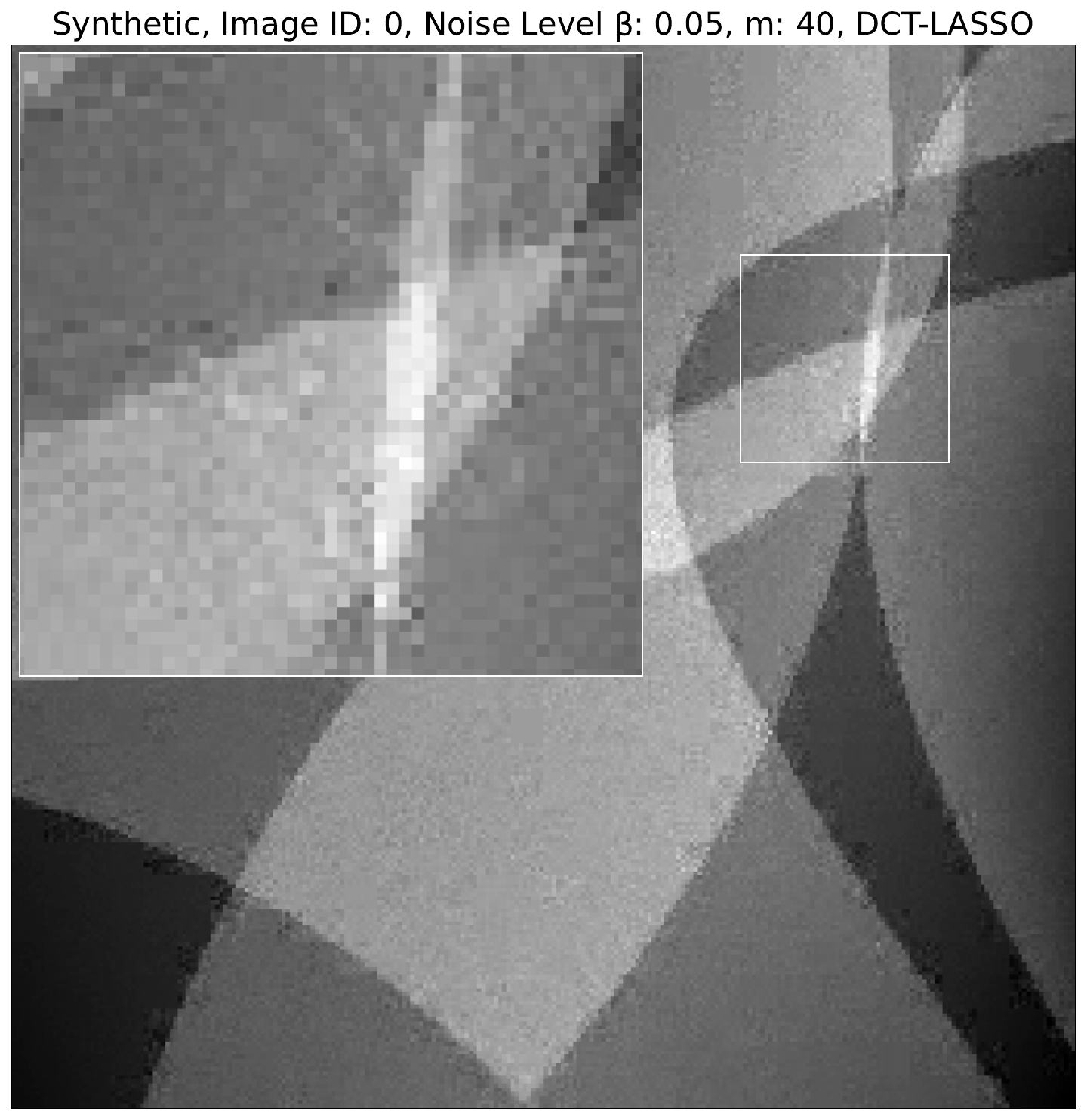}
    \end{subfigure}\\
    \rowname{\textsc{Ilecir}}
    \begin{subfigure}[b]{\gridcolwidth}
        \includegraphics[trim={0 0 0 10mm},clip,width=\linewidth]{figs/zoomed/synthetic_noise_0/000/image_000_ILECIR_m_40_zoomed_inset.pdf}
    \end{subfigure}\hfil%
    \begin{subfigure}[b]{\gridcolwidth}
        \includegraphics[trim={0 0 0 10mm},clip,width=\linewidth]{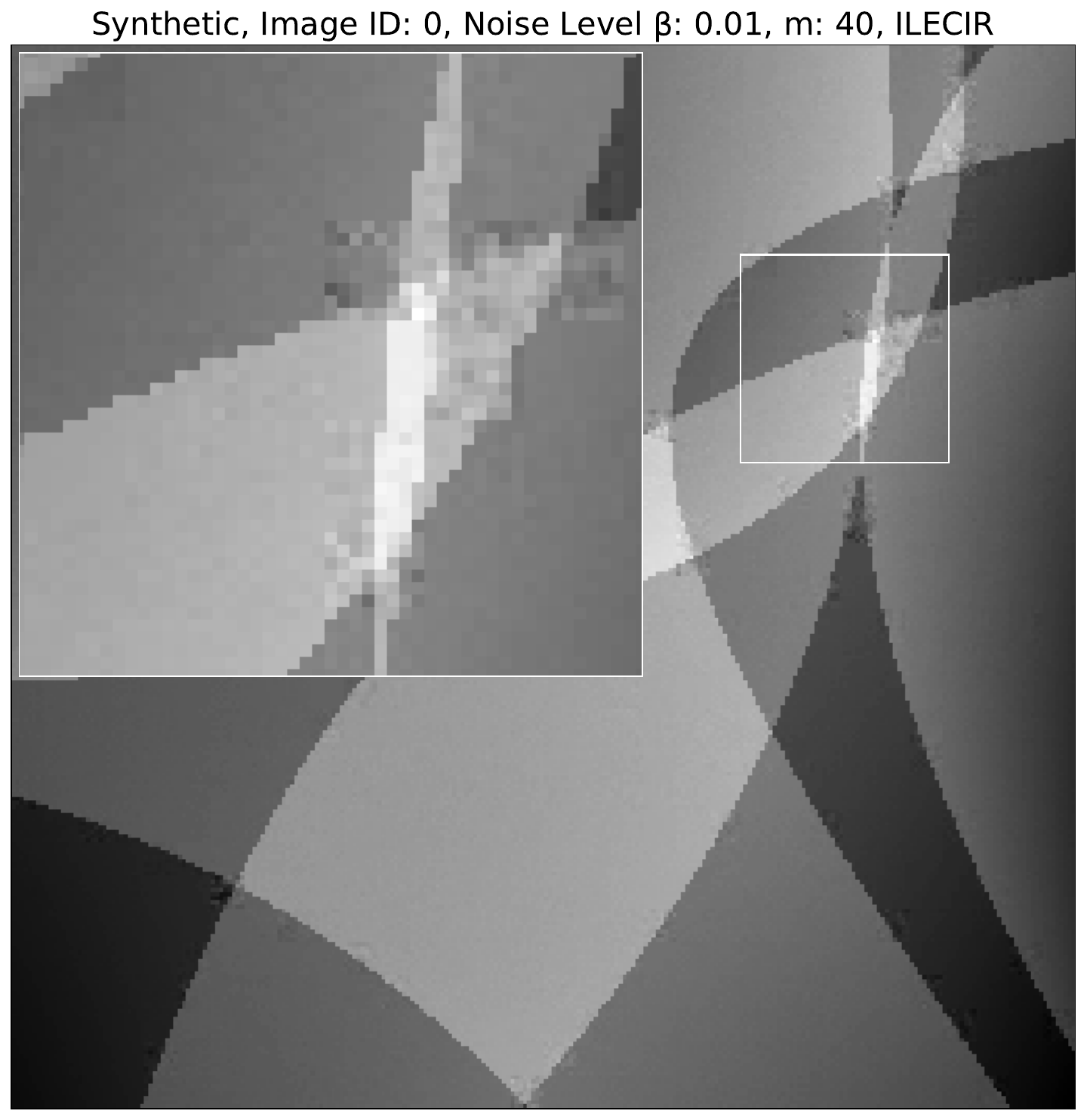}
    \end{subfigure}\hfil%
    \begin{subfigure}[b]{\gridcolwidth}
        \includegraphics[trim={0 0 0 10mm},clip,width=\linewidth]{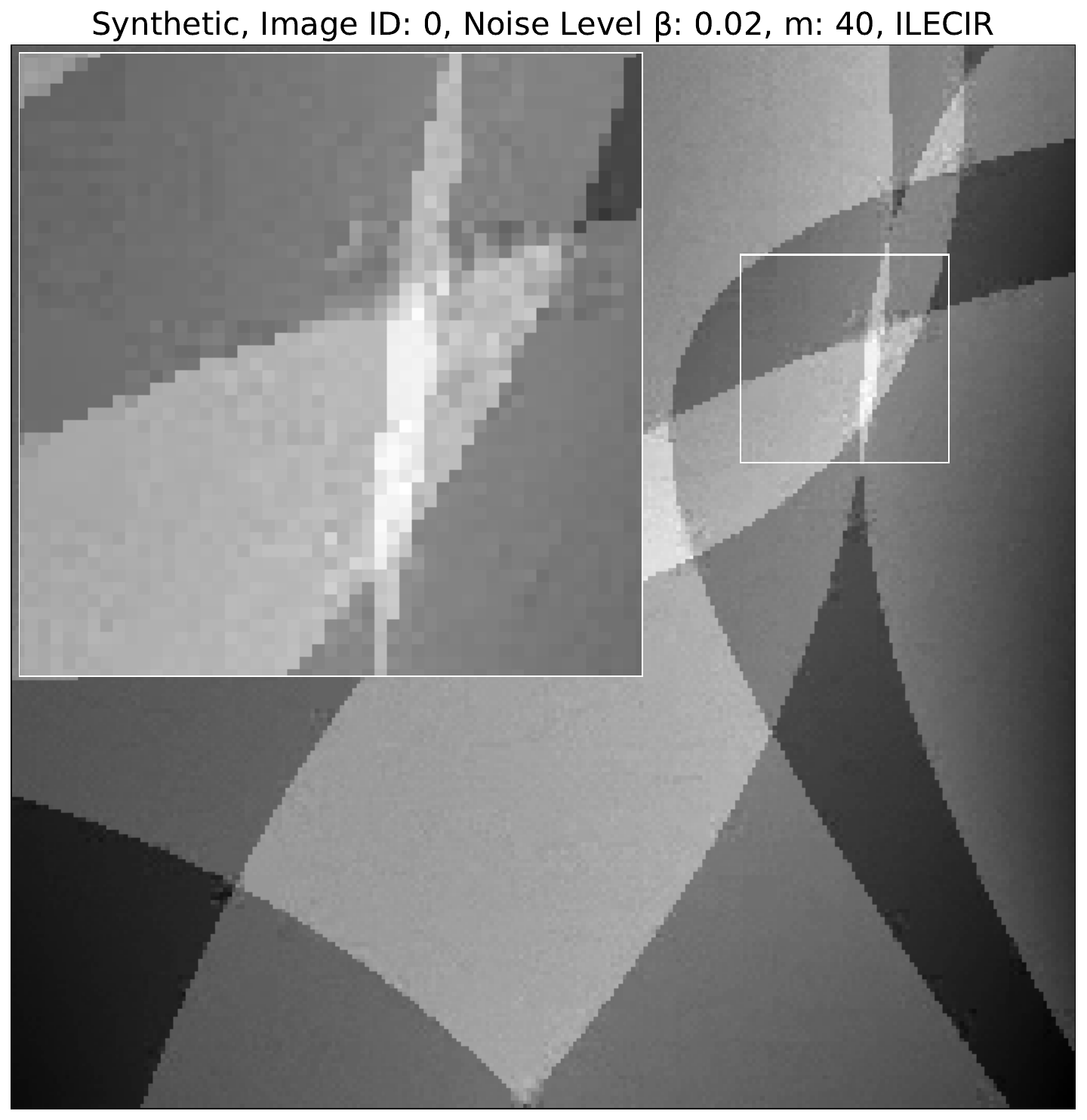}
    \end{subfigure}\hfil%
    \begin{subfigure}[b]{\gridcolwidth}
        \includegraphics[trim={0 0 0 10mm},clip,width=\linewidth]{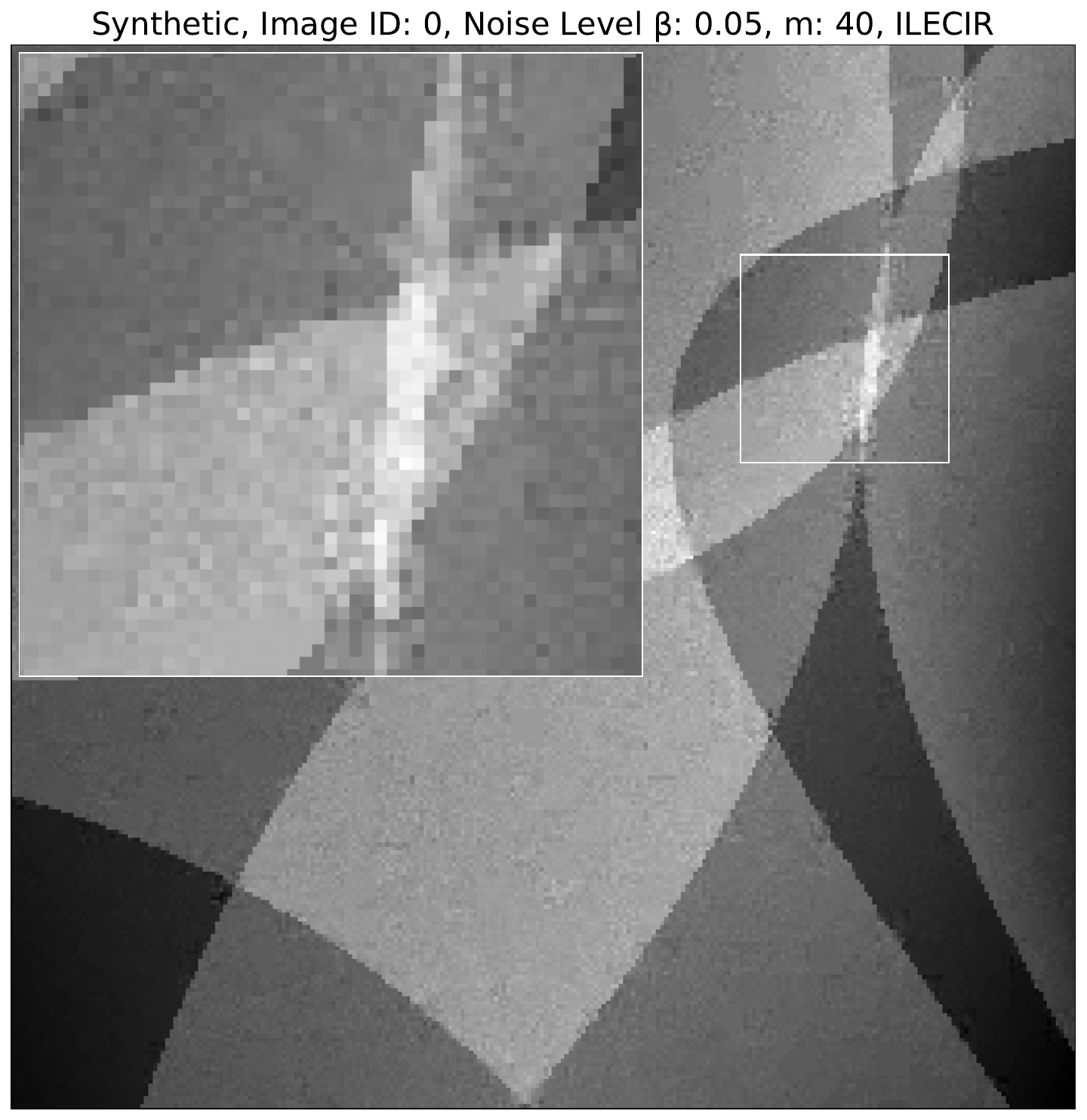}
    \end{subfigure}
    \caption[Images recovered at various noise levels by \textsc{Ilecir}]{Zoomed view of Image ID $0$ of the Synthetic dataset, recovered patch-wise from $m = 40$ compressive measurements per $8\times 8$ patch using the \textsc{Dct-Lasso-Cv} (top row) and the \textsc{Ilecir} (bottom row) algorithms for various noise levels ($\beta \in \{0, 0.01, 0.02, 0.05\}$).
    }
    \vspace{-3mm}
    \label{fig:zoomed_synthetic_noisy}
\end{figure}
\noindent\textbf{Effect of Noise on \textsc{Ilecir} recovery performance:}
\label{subsubsec:noise_ilecir_result}
Fig.~\ref{fig:rmse_ssim_noisy} shows the RRMSE and SSIM of images recovered using \textsc{Ilecir} from noisy measurements of an image of the Synthetic dataset, with different noise levels $\beta \in \{0, 0.01, 0.02, 0.05\}$.
We find that \textsc{Ilecir} performs significantly better than \textsc{Dct-Lasso-Cv} even in the presence of noise.
In fact \textsc{Ilecir} on measurements with $\beta = 0.02$ has better RRMSE and approximately the same SSIM as \textsc{Dct-Lasso-Cv}  with noiseless measurements.
We also find that the RRMSE and SSIM of \textsc{SegGft-Lasso-Cv} worsens significantly with measurement noise. 
Fig.~\ref{fig:zoomed_synthetic_noisy} shows the images recovered using \textsc{Dct-Lasso-Cv} and \textsc{Ilecir} for different noise levels.
We note that the images recovered by \textsc{Ilecir} are of better quality than the corresponding images recovered by \textsc{Dct-Lasso-Cv}, across all noise levels.
The image recovered by \textsc{Ilecir} with noise level $\beta = 0.02$ looks better near the edges in the image than the one recovered by DCT from noiseless measurements.

\section{Conclusion and Future Work}
\label{sec:gsp_conclusion}
This work proposes a new technique of compressive graph signal recovery in cases where the graph topology is partly incorrectly specified.
This is a novel computational problem (to our best knowledge), with no existing literature that targets this specific problem.
At the core of the technique to correct for errors in graph topology, lies the concept of cross validation in compressed sensing.
A novel algorithm is proposed and supportive theoretical results are stated and proved.
A large number of computer simulations are presented, including results on compressive image recovery, where images or image patches are modeled as lattice graphs. Since intensity values at pixels/nodes that lie on opposite sides of an image edge are uncorrelated, the presence of a graph edge that links such nodes is regarded as a perturbation that needs to be corrected.
Thus edge-preserving compressive recovery is cast quite interestingly as a graph perturbation problem.
This specific idea has been used in image compression before \cite{hu2014multiresolution}, where the application innately has access to the complete graph. However its application to compressive image reconstruction is quite novel. Moreover, our technique is naturally applicable to image data that are defined on \emph{arbitrary non-Cartesian domains}, for examples on the vertices of a 3D model. In such cases, our perturbed GFT technique fits in very naturally, even though the eigenvectors of the Laplacian of the actual graph \emph{may not be the DCT}. Furthermore, our technique can be easily extended to handle dropping of edges even in weighted graphs. We also note that the idea of signal sparsity in the GFT for compressive recovery has not been extensively explored in the literature, with most papers focusing on band-limited approximations \cite{zhu2012graph} unlike our approach.
Finally, we show in the supplemental Secs.~\ref*{subsec:graph_tv}, \ref*{subsubsec:graph_tv_ilecir} that our method may be used to recover signals which admit some other structure than sparsity in the GFT, such as being piece-wise constant, by using a different regularizer than the $\ell_1$-norm of GFT, such as Graph Total Variation. 

There are a few major avenues for future work listed below:
(1) Research towards a more computationally efficient strategy to uncover all graph perturbations, over and above our greedy algorithms, and extending the strategy to handle weighted graphs, such as by using gradient descent of the \textsc{Lasso} objective over Laplacian matrices, parameterized in terms of eigenvectors and eigenvalues.
(2) Improving the performance of compressive image recovery possibly using steerable bases such as those in \cite{fracastoro2016steerable} or using notions of image edge smoothness or continuity. 

\nocite{condon2001algorithms,holland1983stochastic,erdos1959onrandomgraphs,penrose2003random,albert2002statistical,zachary1977information,Shi2000}
\vspace{-0.13in}
\bibliographystyle{ieeetr}
\bibliography{References/mybibfile}%

\pagebreak

\title{Supplementary material: Compressive Recovery of Signals Defined on Perturbed Graphs}

\author{
\IEEEauthorblockN{Sabyasachi Ghosh,
Ajit Rajwade} \\
\IEEEauthorblockA{
\{ssgosh,ajitvr.cse.iitb\}@gmail.com\\
Dept. of Computer Science and Engineering,\\
Indian Institute of Technology Bombay, Mumbai, India.}
}
\maketitle

\section{Statement and Proof of Key Theorem}
\label{sec:appendix_proof}
Let $\hat{\bb x}_\mc{G}$ denote the signal recovered using \textsc{Gft-Lasso-Cv}  with the GFT of graph $\mc{G}$. 
We make the assumption that each entry of $\bb \Phi$ is independently drawn from a sub-Gaussian distribution of mean 0 and variance 1.
We restate the following recovery guarantee for the \textsc{Bfgs} algorithm from the main paper:
\begin{theorem}[Brute-Force Algorithm Recovery Guarantee, Theorem~\ref*{thm:brute_force_guarantee} of main paper]
    If in the \textsc{Bfgs} algorithm in Sec.~\ref*{subsec:greedy_edge_selection}, the number of CV measurements $m_\text{cv}$ obeys
    \begin{align}
        m_{\text{cv}} \: \geq \: 4\Big(1+ \frac{2c}{(c-1)^2}\Big) \Big\{ \ln |\Gamma| + \ln{(d_0 + 1)} \nonumber\\ + 2d_0 \ln n + \ln {\frac{1}{\delta}}\Big\}
    \end{align}
    for arbitrary constants $c \in (1, \infty)$ and $\delta \in (0, 1)$, then its recovery error is bounded as
    \begin{align}
        \|\hat{\bb{x}}_{\text{bf}} - \bb x^* \|_2^2 \: < \: c \|\hat{\bb{x}}_{\text{actual}} - \bb x^*\|_2^2 + (c-1)\sigma^2
    \end{align}
    with probability more than $1-\delta$. As a consequence, if 
    \begin{align}
        \|\hat{\bb{x}}_{\mc{G}} - \bb x^* \|_2^2 \: \geq \: c 
        \|\hat{\bb{x}}_{\text{actual}} - \bb x^*\|_2^2 + (c-1)\sigma^2,
    \end{align}
    for all graphs $\mc G \neq \mc G_{\text{actual}}$ which are upto $d_0$ edge perturbations from $\mc G_\text{nominal}$, then with probability more than $1-\delta$, $\hat{\bb x}_{\text{bf}} = \hat{\bb x}_{\text{actual}}$, and the actual graph is recovered.
\end{theorem}

We now prove this theorem. But first, we state the relevant theorem and related concepts from \cite{Zhang2014, zhang2016cross}.
For any recovered signal $\hat {\bb x}^p$ corresponding to a ground truth vector $\boldsymbol{x^*}$, the recovery error vector $\Delta \bb x^p$ and the recovery error $\varepsilon_x^p$ are given by 
\begin{align}
\Delta \bb x^p \triangleq \hat{\bb x}^p - \bb x^*, \: \varepsilon_x^p \triangleq \|\Delta \bb x^p\|_2^2 = \|\hat{\bb x}^p - \bb x^*\|_2^2. \label{eqn:x_p}
\end{align}
They introduce a generalized error vector, which includes the standard deviation of measurement noise, $\sigma$, and generalized error which is the square of its $\ell_2$-norm.
That is, 
\begin{align}
    \Delta \bb x_g^p \triangleq 
    \begin{bmatrix} \Delta \bb x^p \\ \sigma \end{bmatrix},
    \:\varepsilon_g^p \triangleq \|\Delta \bb x_g^p\|_2^2 = \varepsilon_x^p + \sigma^2.\label{eqn:x_g}
\end{align}
The CV error for $\hat{\bb x}^p$ is given by $\epsilon_{\text{cv}}^p = \|\bb y_\text{cv} - \bb \Phi_{\text{cv}} \hat{\bb x}^p\|_2^2$.
The CV comparison theorem from \cite{Zhang2014} states:
\begin{theorem}
\label{thm:cv_comparison}
    \cite[Theorem 2]{Zhang2014}
    Let $\hat{\bb x}^p$ and $\hat{\bb x}^q$ be two recovered signals and assume there are enough measurements for cross-validation. If $\varepsilon_x^p \geq \varepsilon_x^q$, then it holds with probability $\Phi(\lambda_{p, q})$ that $\epsilon_{\text{cv}}^p \geq \epsilon_{\text{cv}}^q$, where $\lambda_{p, q}$ is given by 
    \begin{align}
    \label{eqn:lambda}
        \lambda_{p, q}^2 = \frac{m_{\text{cv}}}{2\Big [ 1 + 2(1-(\rho_g^{p, q})^2) \frac{\varepsilon_g^p \varepsilon_g^q}{(\varepsilon_g^p - \varepsilon_g^q)^2}\Big ]},
    \end{align}
    $\Phi(u) \triangleq \frac{1}{\sqrt{2\pi}} \int_{-\infty}^u e^{-\frac{t^2}{2}} dt$ is the cumulative distribution function of the standard Gaussian distribution, and
    \begin{align}
        \rho_g^{p, q} \triangleq \frac{\langle \Delta\bb x_g^p, \Delta \bb x_g^q \rangle}{\|\Delta \bb x_g^p\|_2\|\Delta \bb x_g^q\|_2}
    \end{align}
    is the correlation coefficient of the two generalized recovery error signals.
\end{theorem}
The condition of `enough number of measurements for cross-validation' is needed for the application of the Central Limit Theorem in the proof of Theorem~\ref{thm:cv_comparison}.
We proceed with the proof of Theorem~\ref*{thm:brute_force_guarantee} under the same assumption.

Each pair $p \triangleq (\mc{P}, \mu)$ of a set of perturbations $\mathcal{P}$ and the \textsc{Lasso} regularization parameter $\mu$ results in a uniquely recovered signal $\hat{\bb x}^p$, with corresponding value of recovery error $\varepsilon_x^p$ and generalized recovery error $\varepsilon_g^p$ as defined earlier.
Let $\mc{P^*}$ denote the set of edges upon perturbing which we recover the actual graph from the nominal graph.
Let $\mu^*$ be the ideal value of the \textsc{Lasso} regularization parameter for the actual graph.
Let $p^* \triangleq (\mc{P^*}, \mu^*)$, and let the corresponding recovery error be $\varepsilon^{p^*}_x$.
From Theorem~\ref{thm:cv_comparison}, for any pair $p$ such that $\varepsilon_x^p \geq \varepsilon_x^{p^*}$, we have 
\begin{align}
   \Pr[\ecv p \geq \ecv {p^*}] = \Pr[u \leq \lambda_{p, p^*}],
\end{align}
where $u$ is a standard Gaussian random variable, and $\lambda_{p, p^*}$ is given by Eqn.~\ref{eqn:lambda}.
Hence the probability of the complement is given by
\begin{align}
\label{eqn:tail_bound}
   \Pr[\ecv p \leq \ecv {p^*}] = \Pr[u \geq \lambda_{p, p^*}] \leq e^{-\frac{\lambda_{p,p^*}^2}{2}}.
\end{align}
The inequality in Eqn.~\ref{eqn:tail_bound} is due to a tail bound on the distribution of $u$, since the standard Gaussian distribution is sub-Gaussian.
Let 
\begin{align}
    h_p \triangleq (1 - (\rho_g^{p, p^*})^2) \frac{\eg{p^*}\eg{p}}{(\eg{p^*} - \eg{p})^2}.
\end{align}
Therefore
\begin{align}
\label{eqn:prob_h_p}
   \Pr[\ecv p \leq \ecv {p^*}] \leq e^{-\frac{m_\text{cv}}
   {4(1 + 2h_p)}
   }.
\end{align}
Since $\rho_g^{p, p^*} \in [-1, 1]$, we have 
\begin{align}
    (1 - (\rho_g^{p, p^*})^2) \leq 1
    \implies  h_p \leq \frac{\eg{p^*}\eg{p}}{(\eg{p^*} - \eg{p})^2} \\
    = \frac{  (\eg{p^*})^2 \frac{\eg{p}}{\eg{p^*}}  }{(\eg{p^*})^2\Big(1  - \frac{\eg{p}}{\eg{p^*}}\Big)^2}
     = \frac{ \frac{\eg{p}}{\eg{p^*}}  }{\Big(\frac{\eg{p}}{\eg{p^*}} - 1\Big)^2}
= f\Big(\frac{\eg{p}}{\eg{p^*}}\Big), \label{eqn:h_p_f}
\end{align}
where we define
\begin{align}
    f(z) \triangleq \frac{z}{(z-1)^2}.
\end{align}
Now, we restrict our consideration to only those $p$ for which $\eg{p} \geq c \eg{p^*}$ for some constant $c$.
We denote this set by $S(c)$, that is:
\begin{align}
    S(c)  \triangleq \{ p : \frac{\eg{p}}{\eg{p^*}} \geq c\}
\label{eqn:sc_defn}
\end{align}
We note that $f(z)$ is a monotonically decreasing function of $z$ for $z > 1$, since
\begin{align}
    f'(z) = \frac{1}{(z-1)^2} - \frac{2z}{(z-1)^3} = - \frac{z+1}{(z-1)^3} < 0 \: \forall {z > 1}.
\end{align}
Hence if $c > 1$, then 
\begin{align}
    f\Big(\frac{\eg{p}}{\eg{p^*}}\Big) &\leq f(c) \:\: \forall p \in S(c) \\
    \implies h_p &\leq f(c) \:\: \forall p \in S(c). \text{\:\:[using Eqn.~\ref{eqn:h_p_f}]}
\end{align}
We build on this to get an inequality for the R.H.S. of Eqn.~\ref{eqn:prob_h_p}:
\begin{align}
    \forall p \in S(c), & \nonumber \\
     h_p &\leq f(c) \\
    \implies  4(1 + 2h_p) &\leq 4(1+2f(c)) \\
    \implies  \frac{m_\text{cv}}{4(1 + 2h_p)} &\geq \frac{m_\text{cv}}{4(1 + 2f(c))} \nonumber \\
    &\text{ [}\because  h_p \text{ and } f(c) \text{ are both } > 0 \text{] } \\
    \implies   - \frac{m_\text{cv}}{4(1 + 2h_p)} &\leq - \frac{m_\text{cv}}{4(1 + 2f(c))} & \\
    \implies   e^{- \frac{m_\text{cv}}{4(1 + 2h_p)}} &\leq e^{- \frac{m_\text{cv}}{4(1 + 2f(c))}} \\
    \implies  \Pr[\ecv p \leq \ecv {p^*}] &\leq e^{- \frac{m_\text{cv}}{4(1 + 2f(c))}}. \text{ [using Eqn.~\ref{eqn:prob_h_p}]} \label{eqn:prob_f_c}
\end{align}
Eqn.~\ref{eqn:prob_f_c} bounds the probability of the events $\ecv p \leq \ecv {p^*}$ separately for each $p \in S(c)$.
However, we want $\ecv {p^*}$ to be less than $\ecv {p}$ for all $p \in S(c)$ at the same time, i.e. we want the probability of the event $\bigcap\limits_{p\in S(c)} (\ecv {p^*} \leq \ecv {p})$ to be large. Hence, we find an upper bound on the probability of the complement event $\bigcup\limits_{p\in S(c)}^{} (\ecv p \leq \ecv {p^*})$ using the union bound:
\begin{align}
    \Pr\Big[\bigcup_{p\in S(c)}^{} (\ecv p \leq \ecv {p^*})\Big]  \leq \sum_{p \in S(c)}\Pr[\ecv p \leq \ecv {p^*}]&  \\
     \leq \sum_{p \in S(c)}  e^{- \frac{m_\text{cv}}{4(1 + 2f(c))}}  \text{ [from Eqn.~\ref{eqn:prob_f_c}]}& \\
     = |S(c)|e^{- \frac{m_\text{cv}}{4(1 + 2f(c))}}.& \nonumber
    \\\text{[ } \because \text{term inside the sum is a constant]}& \label{eqn:union_sc}
\end{align}
In Eqn.~\ref{eqn:union_sc}, $|S(c)|$ denotes the cardinality of the set $S(c)$.
This number is less than the total number of pairs of perturbations and \textsc{Lasso} regularization parameter values used for grid search (the pair $p^*$ is not in $S(c)$ for $c > 1$).
Hence,
\begin{align}
    |S(c)| < |\Gamma| N, \label{eqn:bound_Sc}
\end{align}
where $N$ is the number of perturbations of size $\leq d_0$.
Since there are $n\choose2$ possible edges, of which $s$ are chosen for perturbation, for all $s \in \{0, \dots, d_0\}$, hence,
\begin{align}
   &\:\:\:\:\:\:\:\:\:\:\:\:N = {{n\choose2}\choose 0} + {{n\choose2}\choose 1} + \dots + {{n\choose2}\choose d_0}. \\
   & \implies N < (d_0 + 1)n^{2d_0}. \nonumber\\
    &\text{ [}\because {n\choose 2}  < n^2 \text{ \& } {{n^2} \choose k} \leq  {{n^2} \choose d_0} < n^{2d_0} \text { for } k \leq d_0 \text{]} \\
   & \implies |S(c)| < |\Gamma| (d_0 + 1)n^{2d_0}. \text{ [using Eqn.~\ref{eqn:bound_Sc}]} \label{eqn:bound_Sc_n}
\end{align}
Putting this in Eqn.~\ref{eqn:union_sc},
\begin{align}
 \Pr\Big[\bigcup_{p\in S(c)}^{} (\ecv p \leq \ecv {p^*})\Big]  < |\Gamma|(d_0+1)n^{2d_0}   e^{- \frac{m_\text{cv}}{4(1 + 2f(c))}}
\end{align}
We may upper bound the R.H.S. by a constant $\delta \in (0, 1)$ of our choosing:
\begin{align}
    |\Gamma|(d_0+1)n^{2d_0}   e^{- \frac{m_\text{cv}}{4(1 + 2f(c))}} \leq \delta \\
    \implies \ln(|\Gamma|(d_0+1)n^{2d_0}) - \frac{m_\text{cv}}{4(1 + 2f(c))} \leq \ln \delta \\
    \implies m_{\text{cv}} \geq 4\Big(1+ \frac{2c}{(c-1)^2}\Big) \Big\{ \ln |\Gamma| + \ln{(d_0 + 1)}\nonumber\\
    +~2d_0 \ln n + \ln {\frac{1}{\delta}}\Big\}. \label{eqn:mcv_final}
\end{align}
Hence, if the condition of Eqn.~\ref{eqn:mcv_final} is true, then 
\begin{align}
    \Pr\Big[\bigcup_{p\in S(c)}^{} (\ecv p \leq \ecv {p^*})\Big]  < \delta.
\end{align}
Equivalently,
\begin{align}
    \Pr\Big[\bigcap_{p\in S(c)}^{} (\ecv {p^*} \leq \ecv {p})\Big]  > 1 - \delta. \label{eqn:ecv_all}
\end{align}
Hence, with an arbitrarily high probability (more than $1-\delta$), no pair in $S(c)$ gets chosen by the cross-validation procedure -- either $p^*$ gets chosen or some other pair $p'$ for which $\eg {p'}$ is small gets chosen -- provided the condition in Eqn.~\ref{eqn:mcv_final} is true.
In such a case, for the chosen pair $\hat{p}_{\text{bf}}$, since $\hat{p}_{\text{bf}} \notin S(c)$, it must be true that
\begin{align}
     \eg{\hat{p}_{\text{bf}}} < c \eg{p^*}. & \zerotext[49mm]{\text{[from defn. of $S(c)$ in Eqn.~\ref{eqn:sc_defn}]}}   \label{eqn:not_sc}
\end{align}
Using defintion of $\eg{p}$ and $\ex p$ (Eqns.~\ref{eqn:x_g} and ~\ref{eqn:x_p}) and putting in Eqn.~\ref{eqn:not_sc},
\begin{align}
    &\ex{\hat{p}_{\text{bf}}} + \sigma^2 < c \ex{p^*} + c\sigma^2 \\
    \implies &\ex{\hat{p}_{\text{bf}}}  < c \ex{p^*} + (c-1)\sigma^2 \\
    \implies &||\bb x^* - \hat{\bb x}_{\text{bf}}||_2^2 < c ||\bb x^* - \hat{\bb x}_{\text{actual}}||_2^2 + (c-1)\sigma^2. \label{eqn:error_bound_appendix}
\end{align}
Eqns.~\ref{eqn:mcv_final}, ~\ref{eqn:ecv_all} and ~\ref{eqn:error_bound_appendix} together prove Theorem ~\ref*{thm:brute_force_guarantee}. \qed

\section{Details Regarding Random Graph Models}
\label{sec:details_RGM}
We test \textsc{Ges} on the following graphs commonly uses in the Network Science literature:  Planted Partition Model (PPM, \cite{condon2001algorithms}), Stochastic Block Model (SBM, \cite{holland1983stochastic}), Erd\H{o}s–R\'{e}nyi  Graph (ERG \cite{erdos1959onrandomgraphs}), Random Geometric Graph (RGG, \cite{penrose2003random}), Barabasi-Albert Graph (BAG \cite{albert2002statistical}), and Karate Club Graph (KCG, \cite{zachary1977information}) -- all are random graph models, other than the Karate Club graph.
In a PPM, modelling community structure, there are $n=lr$ nodes, with $l$ communities (clusters) and and $r$ nodes per community, with each edge being independently present with probability $p$ (for intra-cluster edge) or $q$ (for inter-cluster edge).
The SBM is a generalization of the PPM in which the community sizes may be different, and the edge presence probabilities depend on the communities of the edge endpoints, and are given by a $l\times l$ symmetric matrix.
In an ER Graph, modelling random graphs with no particular structure, each edge is independently present with a probability $p$.
In a RGG, modelling real-world networks on a plane, the $n$ nodes are points selected uniformly at random in a unit square, with points within a radius $r$ from each other being connected.
The BA graph, modelling preferential attachment, is generated by adding new nodes one at a time to a star graph with $r+1$ nodes, with each new node connecting to $r$ existing nodes chosen with a probability proportional to their current degree. The Karate Club graph is a two-community social network having $n=34$ nodes, observed in \cite{zachary1977information}.

\section{Recovery Analysis of Edge Recovery} 
\label{sec:recovery_analysis_edges}

\begin{figure}[ht]
    \centering
    \includegraphics[width=0.36\linewidth]{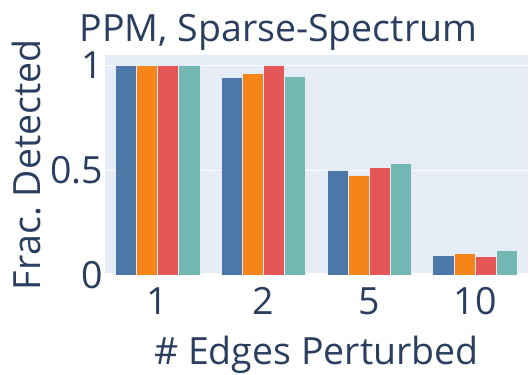}\hfill%
    \includegraphics[width=0.63\linewidth]{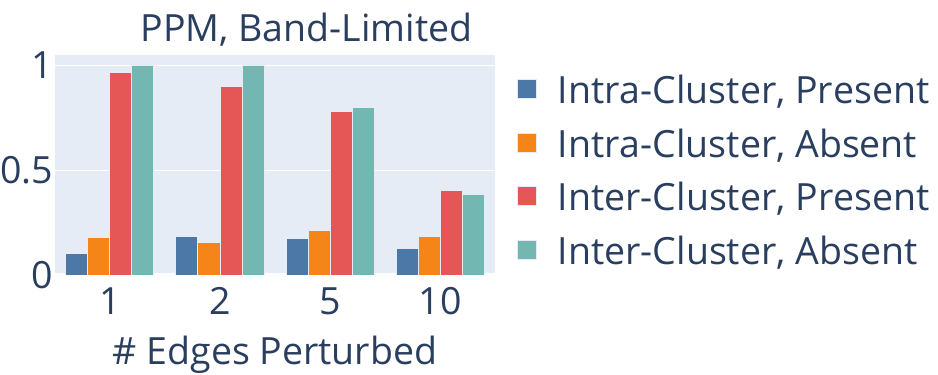}
    \caption[\textsc{Ges} edge perturbation recovery performance by edge type]{Fraction of Edge Perturbations of each type recovered via \textsc{Ges} for the Planted Partition Model with sparse-spectrum (left) and band-limited (right) signals.
    }
    \label{fig:ges_analysis_edge_type}
\end{figure}

Fig.~\ref{fig:ges_analysis_edge_type} shows the fraction of edge perturbations which were detected by \textsc{Ges} on the Planted Partition Model, categorized by edge type, for the sparse-spectrum (left) and the band-limited (right) signals.
We note that for band-limited signals, inter-cluster edges are highly likely to be recovered by \textsc{Ges}.
For sparse-spectrum signals, all edge types are equally likely to be recovered.
In general, an edge perturbation is more likely to be detected by \textsc{Ges} if perturbing that edge significantly perturbs a least one of the eigenvectors on which the original signal $\bb x^*$ has a non-zero component.
From the formula in Eqn.~\ref*{eqn:eigvec_perturbation_formula} (Sec.~\ref*{subsec:ges_ceci_barbarossa}), we note that an edge perturbs an eigenvector significantly only if the absolute difference of the values of that eigenvector on the two nodes of the edge is large.
It is well-known that the first $k$ low-frequency eigenvectors do not change much within a community for a community-structured graph with $k$ communities, a fact that is used in graph spectral clustering and image segmentation \cite{Shi2000}.
Hence adding or removing an intra-cluster edge does not significantly perturb the first $k$ eigenvectors of the PPM. On the other hand, the value of the first $k$ eigenvectors of such PPMs vary a lot across clusters. Hence, adding or removing an inter-cluster edge causes a significant perturbation in the first $k$ eigenvectors, which is easily detected by \textsc{Ges}. For high-frequency eigenvectors, both inter-cluster and intra-cluster edges are highly likely to have different values at the endpoints. Hence for signals which are sparse-spectrum but not band-limited, \textsc{Ges} is able to recover both inter-cluster and intra-cluster edges.

\section{Alternatives to Graph Fourier Transform for Regularization}
\label{subsec:graph_tv}
The methods presented in this section can be easily extended to settings where some other graph-based regularizer may be more appropriate than the $\ell_1$-norm of the Graph Fourier Transform of the graph signal.
For example, if the graph signal under consideration is known to be piece-wise constant 
(i.e. the graph can be partitioned into sets of topologically close nodes each having the same value of the graph signal),
it may be more appropriate to use the Graph Total Variation of the signal as the regularization term.
The graph total variation of a signal is the sum of the absolute differences of the value of the graph signal at the endpoints of all the edges of the graph. 
That is, we have:
\begin{equation}
    \label{eq:graph_tv}
    \mathtt{TV}_{\mc G}(\bb x) = \sum_{\{a, b\} \in \mathcal{E}} |x_a - x_b|.
\end{equation}
It is straightforward to see that piece-wise constant signals will have very small graph total variation.

For recovery of an unknown graph signal $\bb x^*$ from compressive measurements $\bb y$ (see Eqn.~\ref*{eq:cs_measurement}) using an arbitrary regularizer $\mathtt{R}_{\mc G}:\mathbb R^n \rightarrow \mathbb R_{\geq 0}$ on some graph $\mc G$, one may use:
\begin{equation}
    \label{eq:recovery_generalized_lasso}
    {\hat{\bb x}}_{\mathtt{R}_{\mc G}} = \underset{\bb x}{\arg\min} \:\:\: \|\bb y - \bb \Phi \bb x\|_2^2 + \mu \mathtt{R}_{{\mc G}}(\bb x).
\end{equation}
Accordingly, for a problem setting similar to Sec.~\ref*{subsec:problem_statement}, when only a nominal graph $\nom G$ is known,
and the graph signal $\bb x^*$ has a small value of the regularizer $\mathtt R_{\mc G}(\bb x^*)$ for $\mc G = \mc G_{\text{actual}}$,
the \textsc{Lasso} steps in the brute-force algorithm (from Sec.~\ref*{subsec:greedy_edge_selection}), cross-validation (Eqn.~\ref*{eqn:cv_error_formula}),
greedy edge selection (Alg.~\ref*{alg:greedy_edge_selection}), or \textsc{Ilecir} (Alg.~\ref*{alg:image_reconstruction}) may be replaced appropriately by Eqn.~\ref{eq:recovery_generalized_lasso}. When the underlying signal is known to be piece-wise constant, one may use the graph total variation for the respective graphs in each of these algorithms, i.e. with $\mathtt R_{\mc G}(\bb x^*) := \mathtt{TV}_{\mc G}(\bb x^*)$.

\section{Performance of \textsc{Ilecir} with Graph Total Variation regularization}
\label{subsubsec:graph_tv_ilecir}
\begin{figure}[htp]
    \centering
    \begin{subfigure}[b]{0.64\linewidth}
    \centering
    \includegraphics[width=\linewidth]{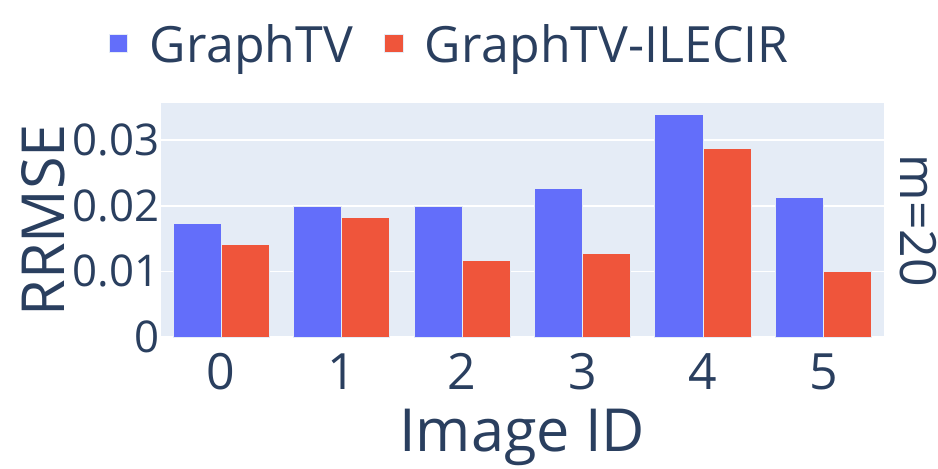}
    \subcaption{RRMSE, Synthetic dataset}
    \end{subfigure}\hfill%
    \begin{subfigure}[b]{0.35\linewidth}
    \centering
    \includegraphics[trim={0 0 0 10mm},clip,width=\linewidth]{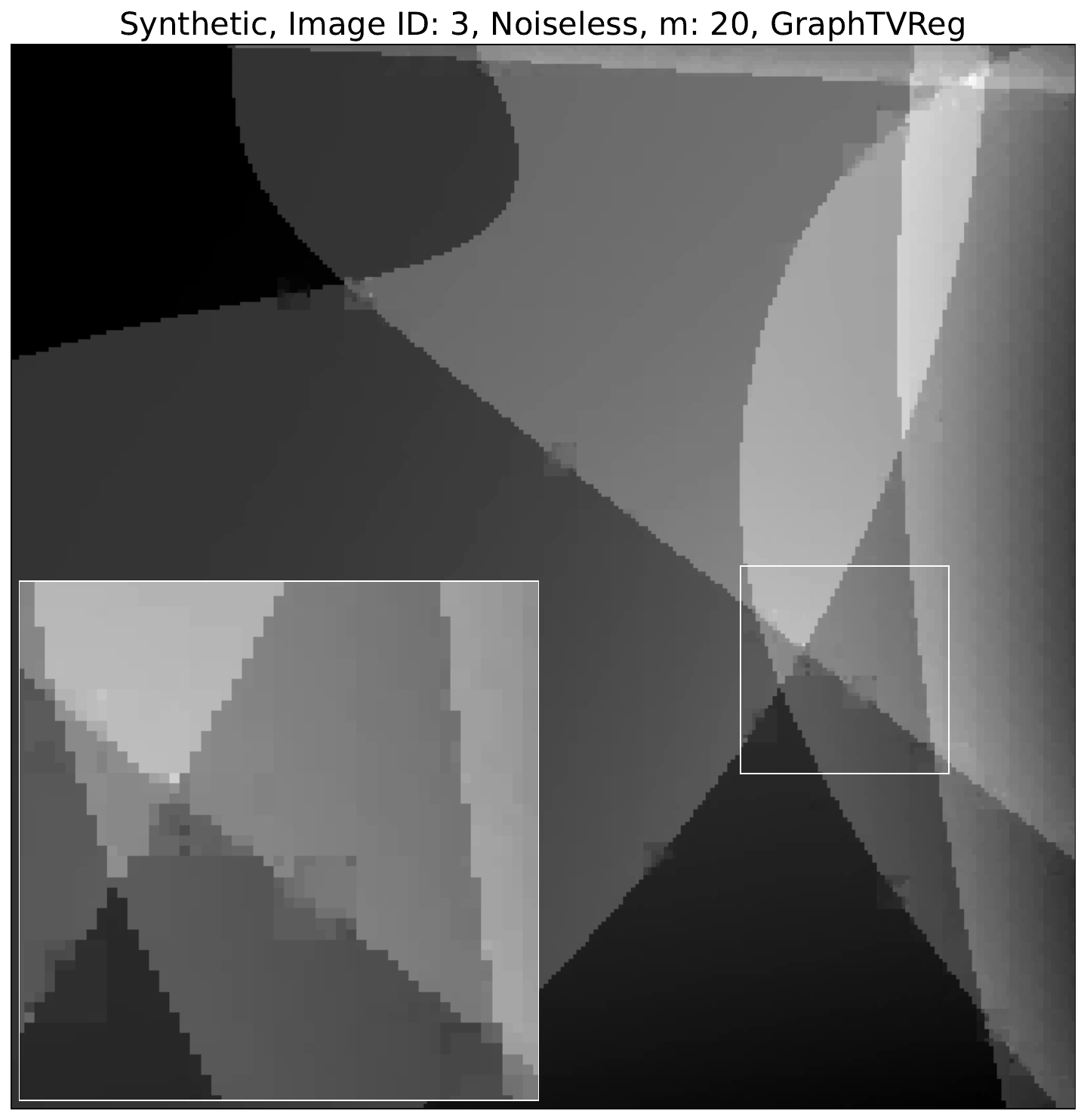}\\
    \subcaption{\textsc{GraphTvReg-Cv}}
    \end{subfigure}\\
    \begin{subfigure}[b]{0.64\linewidth}
    \centering
    \includegraphics[width=\linewidth]{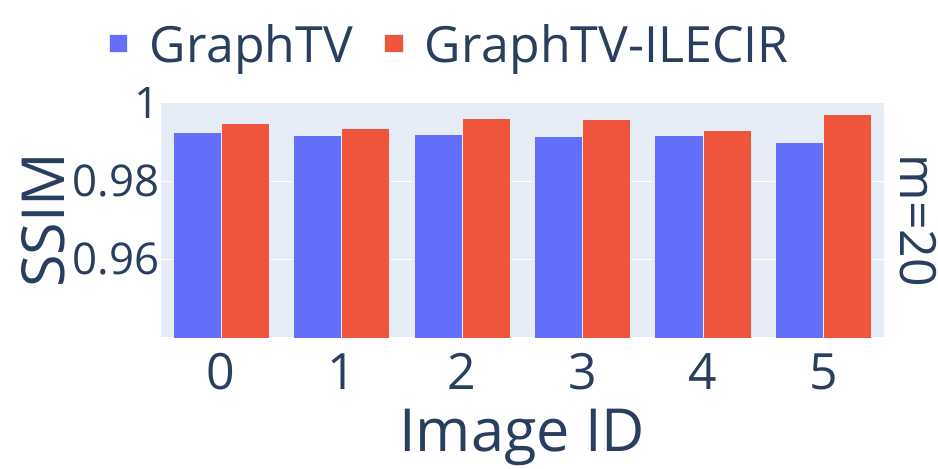}
    \subcaption{SSIM, Synthetic dataset}
    \end{subfigure}\hfill%
    \begin{subfigure}[b]{0.35\linewidth}
    \centering
    \includegraphics[trim={0 0 0 10mm},clip,width=\linewidth]{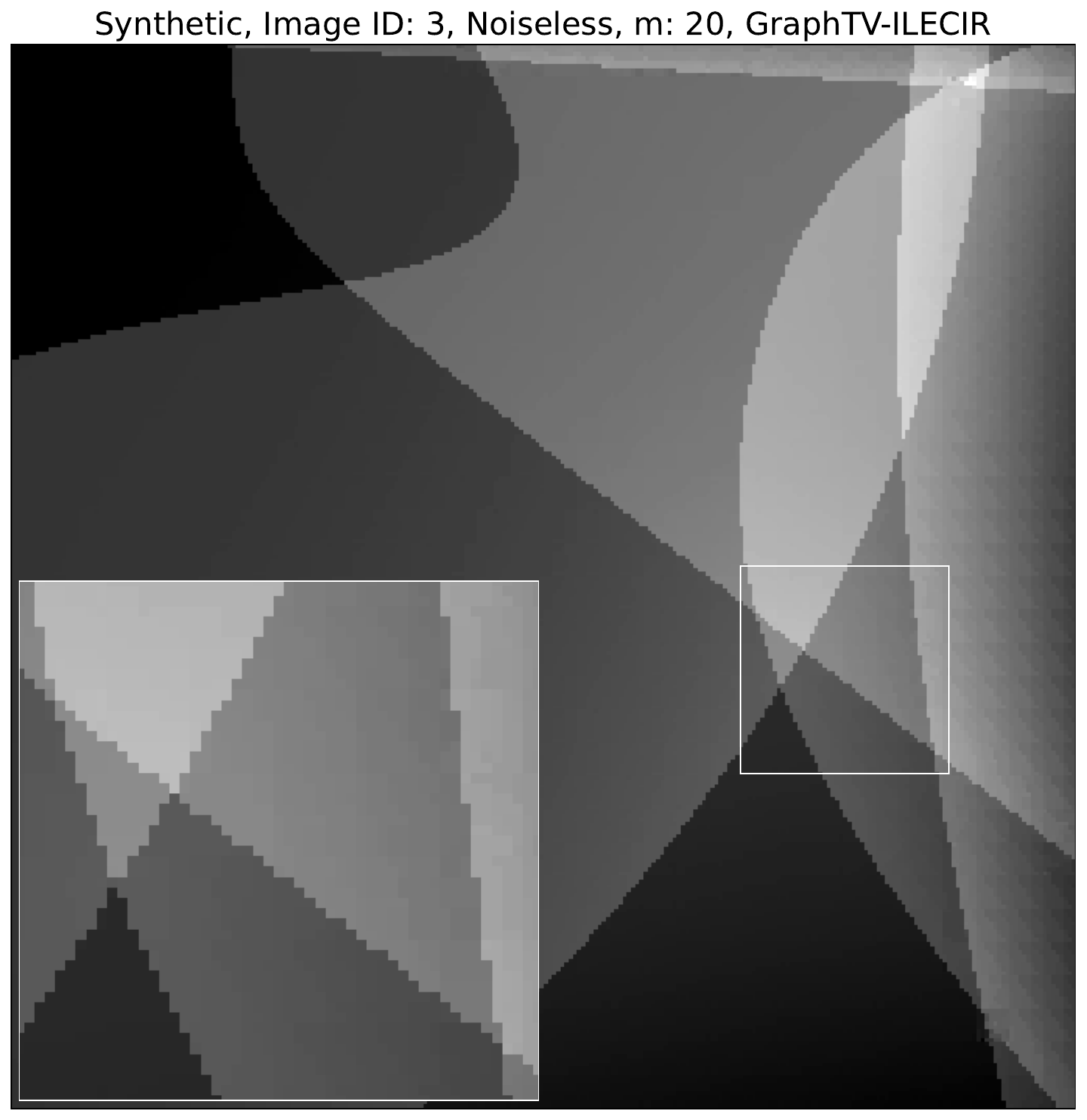}
    \subcaption{\textsc{GraphTv-Ilecir}}
    \end{subfigure}
    \caption[Performance of \textsc{GraphTv-Ilecir}]{Column 1: RRMSE and SSIM of $6$ images from the Synthetic dataset, recovered patch-wise from $m=20$ compressive measurements per $8\times 8$ patch using the \textsc{GraphTvReg-Cv} and \textsc{GraphTv-Ilecir} algorithms in the noiseless setting;
    Column 2: Zoomed view of Image ID $3$ of the Synthetic dataset, recovered patch-wise from $m = 20$ compressive measurements per $8\times 8$ patch using the \textsc{GraphTvReg-Cv} and the \textsc{GraphTV-Ilecir} algorithms in the noiseless setting.}
    \label{fig:ilecir_graph_tv}
\end{figure}
We demonstrate the effectiveness of \textsc{Ilecir} when used with the Graph Total Variation regularizer.
Fig.~\ref{fig:ilecir_graph_tv} (first column) shows the RRMSE and SSIM of six images from the Synthetic dataset recovered using \textsc{GraphTv-Ilecir}, and the baseline method of \textsc{GraphTvReg-Cv}.
We see that for all images, \textsc{GraphTv-Ilecir} performs better than \textsc{GraphTvReg-Cv}.
We also compare the images recovered by the two algorithms for Image ID $3$ of the Synthetic dataset (Fig.~\ref{fig:ilecir_graph_tv}, second column).
Careful inspection of the two images reveals that there are fewer artifacts along the edges and corners in the image recovered by \textsc{GraphTv-Ilecir} than in the one recovered by \textsc{GraphTvReg-Cv}.

\end{document}

%% file: preamble.tex
\usepackage{todonotes}
\usepackage{lineno}
\usepackage{longtable}
\usepackage{subfiles}
\usepackage{comment}
\usepackage{amsmath,amssymb,amsthm,amsfonts,graphicx,enumitem}
\usepackage{multirow}
\usepackage{algorithm}
\usepackage{siunitx}
\usepackage[mathscr]{euscript}
\usepackage{xspace}
\usepackage[noEnd]{algpseudocodex}
\usepackage{soul}
\usepackage{scalerel}
\usepackage{afterpage}
\usepackage[section]{placeins}
\usepackage[font=small,labelfont=bf]{caption}
\usepackage[labelformat=simple]{subcaption}

\newcommand{\bb}[1]{\boldsymbol{#1}}
\newcommand{\act}[1]{{#1}_{\text{actual}}}
\newcommand{\nom}[1]{{#1}_{\text{nominal}}}
\newcommand{\can}[1]{{#1}_{\text{candidate}}}
\newcommand{\E}{\mathcal{E}}
\newcommand{\mc}[1]{\mathcal{#1}}

\newcommand{\ecv}[1]{\epsilon_{\text{cv}}^{#1}}
\newcommand{\eg}[1]{\varepsilon_{\text{g}}^{#1}}
\newcommand{\ex}[1]{\varepsilon_{\text{x}}^{#1}}
\newcommand{\zerotext}[2][0pt]{\makebox[#1][l]{\qquad#2}}

\makeatletter
\def\blfootnote{\xdef\@thefnmark{}\@footnotetext}
\makeatother

\newlength{\gridrowheight}
\newlength{\gridcolwidth}

\newcommand{\rowname}[1]
{\rotatebox{90}{\makebox[\gridrowheight][c]{\footnotesize #1}}}

\newcommand{\columnname}[1]
{\makebox[\gridcolwidth][c]{\footnotesize #1}}

%% file: preamble1.tex
\usepackage{times}
\usepackage[T1]{fontenc}
\usepackage[utf8]{inputenc}
\usepackage{amssymb}
\usepackage{amsmath, amsfonts, amsthm}
\usepackage{graphicx,graphics}
\usepackage{xurl}

\usepackage[bookmarks,pdfencoding=auto,pagebackref,hidelinks]{hyperref}

\usepackage[capitalise]{cleveref}
\usepackage{textcomp}
\usepackage{enumitem}
\usepackage[sort, numbers]{natbib}
\usepackage{pdflscape}

\usepackage{titlesec}
\usepackage{bm,bbm}
\usepackage{float,lscape}
\usepackage{tabu}
\usepackage{multirow}
\usepackage{array}
\usepackage{booktabs}
\usepackage{graphicx}
\usepackage{lettrine}
\usepackage{enumitem}
\usepackage{lscape}
\usepackage{rotating}
\usepackage{booktabs}
\usepackage{longtable}
\usepackage{anyfontsize}
\usepackage{t1enc}
\usepackage{relsize}
\usepackage{nomencl}
\usepackage{rotating}
\usepackage{glossaries}
\usepackage{scrlayer}
\usepackage{enumitem}

\usepackage[nottoc]{tocbibind}

\usepackage{mathtools}

\usepackage{stackengine}

\usepackage[Export]{adjustbox}
\usepackage{tabularx}
\usepackage{cellspace}

\newtheorem{theorem}{Theorem}[]

\newtheorem{problem}[]{Problem}

%% file: preamble_xr.tex
\usepackage{xr}

